\documentclass[sn-mathphys-ay]{sn-jnl}

\usepackage{graphicx}
\usepackage{multirow}
\usepackage{amsmath,amssymb,amsfonts}
\usepackage{amsthm}
\usepackage{mathrsfs}
\usepackage[title]{appendix}
\usepackage{xcolor}
\usepackage{textcomp}
\usepackage{manyfoot}
\usepackage{booktabs}
\usepackage{algorithm}
\usepackage{algorithmicx}
\usepackage{algpseudocode}
\usepackage{listings}

\begin{document}

\title[3D analytical theory of the perturbed single-synchronous state. Application to the post-impact Didymos-Dimorphos system]{3D analytical theory of the perturbed single-synchronous state. Application to the post-impact Didymos-Dimorphos system}

\author*[1]{\fnm{Michalis} \sur{Gaitanas}}\email{mgaitana@physics.auth.gr}
\author[2]{\fnm{Christos} \sur{Efthymiopoulos}}\email{cefthym@math.unipd.it}
\author[1]{\fnm{Ioannis} \sur{Gkolias}}\email{igkoli@physics.auth.gr}
\author[1]{\fnm{George} \sur{Voyatzis}}\email{voyatzis@auth.gr}
\author[1]{\fnm{Kleomenis} \sur{Tsiganis}}\email{tsiganis@auth.gr}

\affil[1]{\orgdiv{Department of Physics}, \orgname{Aristotle University of Thessaloniki}, \orgaddress{\city{Thessaloniki}, \postcode{54124}, \country{Greece}}}

\affil[2]{\orgdiv{Department of Mathematics}, \orgname{Universit\`a degli Studi di Padova}, \orgaddress{\street{Via Trieste 63}, \city{Padova}, \postcode{35121},  \country{Italy}}}

\abstract{{\small We develop the 3D generalization of the planar analytical theory presented in \citep{Michelangelo2024}, which deals with states slightly perturbed from the exact `single-synchronous equilibrium state' (SSES) of the full two-body problem. The SSES corresponds to two non-spherical gravitationally interacting bodies, settled in nearly circular relative orbit, with rotation axes normal to the orbital plane, rapid rotation of the primary and synchronous rotation of the secondary. In the present paper we remove all simplifying assumptions of our previous work \citep{Michelangelo2024}, and show how to compute analytical solutions describing a 3-dimensional perturbation of the system from the SSES in the framework of two distinct theories, called `linear' and `nonlinear'. Linear theory stems from averaging the equations of motion over the primary's rapid rotation angle. This maps the SSES to an equilibrium point of the averaged system, around which analytical solutions can be computed by linearization of the equations of motion. In nonlinear theory, instead, we compute a high order normal form for the Hamiltonian of motion through a sequence of canonical transformations in the form of series. Resonances between the basic system's frequencies appear in the nonlinear theory as small divisors. We show that, close to resonances, the nonlinear theory leads to a partially integrable model, sufficient to analytically describe the evolution of the relative orbit, but only of some of the Euler angles of the system. As a basic application, we compute analytical solutions representing various possible Didymos-Dimorphos post-impact orbital and rotational states. In this case, all analytical formulas here proposed are of direct utility in fitting algorithms exploiting available time series of post-impact observational data.}}


\maketitle

\section{Introduction}
\label{sec:Intro}
The present paper generalizes in three dimensions the methods and results discussed in our previous work \citep{Michelangelo2024}, referring to the analytical study, in the planar case, of the orbital and rotational dynamics in a system of two non-spherical bodies interacting through gravitational forces. Our focus is on the case when the system gets `perturbed' (by natural or artificial causes) with respect to a basic equilibrium state, hereafter referred to as the `single synchronous equilibrium state' (SSES). 

The SSES and nearby states are among the most frequently occurring endstates in astronomical applications of the full two-body problem. Leading examples are planets with a satellite (e.g. Earth-Moon), or binary asteroids with a secondary in synchronous rotation. In the context of the present paper we adopt the following four requirements of definition of an exact SSES: (i) the primary (more massive) body rotates rapidly with respect to the orbital period of the companion and has axisymmetric (typically oblate) shape, (ii) the secondary is triaxial and rotates synchronously with the orbit, (iii) the rotation axes of both bodies are normal to the orbital plane, (iv) the relative orbit is circular. 

By `perturbation' to the SSES we mean a small variation of any of the above conditions (i) to (iv). Such variations can be caused by altering one or more of the system's parameters with respect to a parameter set for which the SSES exists, or by assuming a small variation in the orbital or rotational state of the system with respect to an initial SSES. Here, as in \cite{Michelangelo2024}, we focus on a main  example of current interest, namely the change of the dynamical state of the binary asteroid system 65803 Didymos after its impact with DART. In this example, the `perturbation' consists in that condition (i) is not precisely fulfilled (the primary is triaxial), while conditions (iii) and (iv) are no longer fulfilled after the impact. 

We call the perturbation `planar' if the vectors of the orbital and spin angular momenta of the two bodies undergo no change in orientation in the perturbed state, respective to the considered exact SSES (condition (iii) maintained). Planar perturbations were the subject of our previous paper, in which we developed two kinds of analytical theories, called `linear'and `nonlinear', to describe the evolution of all generalized coordinates of the problem in the perturbed state, given the latter's initial conditions. In the Hamiltonian context, the planar problem has four degrees of freedom (length and azimuth of the relative orbit's radius vector, and one Euler angle per body). Averaging over the fast rotation of the primary, and exploring the invariance of the total angular momentum, the number of degrees of freedom can be reduced to two. 

In the present paper we deal with the same problem in three dimensions. We now have nine degrees of freedom, the coordinates $(X_R,Y_R,Z_R)$ of the relative radius vector, and three angles (for example the `roll-pitch-yaw' angles) per body. By reduction to the relative rotation angles (see \cite{Macie1996}) the number of degrees of freedom can be reduced to six. However, in what follows we avoid making this reduction, since it leads to cumbersome formulas in what regards the application of perturbation theory, while our perturbative scheme based on canonical transformations guarantees invariance of the non-reduced system in all three components of the total angular momentum at any order of perturbation theory. Averaging with respect to the primary's fast rotation is only required in the theory called below `linear': in the averaged system the primary's effect is virtually reduced to that of an axisymmetric body, for which the SSES is mathematically represented as an equilibrium point. We then compute approximative analytical solutions for initial conditions near this point through the linearized equations of motion. No simplifications are made, instead, in nonlinear theory, which yields analytical (series) solutions in time for all nine generalized coordinates of the problem as a function of the initial conditions.

To outline the key steps in our theories, after stating the mathematical formulation of the problem in \textit{Section \ref{sec:ProbState}}, we initially develop linear perturbation theory (\textit{Section \ref{sec:LinearTheory}}). If the primary has triaxial shape, our adopted definition of the SSES is not strictly fulfilled (condition (i) is violated). However, by `scissor'-averaging the  Hamiltonian with respect to the fast-rotating angle of the primary, we effectively recover a SSES as equilibrium point of the averaged Hamiltonian. Physically, the process of averaging mimics the effect of approximating the shape of the primary as axisymmetric by rotation. Nonetheless, the information on the primary's true triaxial shape is retained in the solutions obtained by linear theory after averaging, since the variational matrix of the linearized equations of motion around the equilibrium depends on the values of all three primary's principal moments of inertia. In summary, we give below the linear solution in explicit form, i.e., by a set of formulas exhibiting explicit dependence on all the physical parameters of the problem. 

In \textit{Section \ref{sec:CanonicalTheory}}, a more precise, nonlinear theory is developed through series obtained by the method of the Lie canonical transformations. In this we implement a `book‐keeping' method introduced in \cite{Efthymiopoulos2011}. A key difference between the two theories is that the nonlinear theory is based on expansions taken not with respect to the SSES of the averaged problem, but around a basic state of spherical rotating bodies in Keplerian orbit. The latter turns to be a mathematically more convenient starting point for nonlinear series expansions. On the other hand, SSES-like equilibrium states of the normalized equations of motion are recovered by this approach `on the go', i.e., computed by iteration of the normal form algorithm, together with their nearby perturbed states. 

In \textit{Section \ref{sec:dart}} we apply the above theories to the Didymos-Dimorphos problem, thus obtaining 3D analytical solutions for the post-impact spin-orbit state of the system under the assumption that the impactor's velocity vector is slightly inclined with respect to the pre-impact orbital plane. The momentum enhancement parameter $\beta$, which encapsulates ejecta-related effects (see \cite{Rivkin2021}), remains central to our analysis of this problem: $\beta$ appears symbolically in all equations of motion and all obtained analytical solutions. As a result, the formulas presented below are of direct utility as entry data in algorithms seeking to compute a best-fit value for $\beta$, or of the orbital and rotational parameters of the system post-impact, on the basis of detailed time-series observations. In the case of the Didymos system, one of the effects observed numerically even by mild excitations of the system around a starting SSES is the destabilization of the roll and pitch angles, which, for our assumed bodies' physical and collision parameters, are found to enter into chaotic motion after impact with DART for a value of $\beta$ beyond $\beta \gtrsim 3.6$. Since estimates from Earth observations render such a state a real possibility, we here pay some attention to it. \textit{Section \ref{sec:dart}} makes a comparison of the analytical with the numerical solutions for the post-impact state of Didymos-Dimorphos under various assumptions on the value of $\beta$ and on the mass and ellipsoidal axial ratios of the secondary. From these comparisons we find that the onset of this partially chaotic regime is connected with a particular set of resonances between the basic orbital and librational frequencies of the problem (\cite{Agrusa2021}). The instability marginally affects the relative orbit, which can be described very accurately as a Keplerian ellipse with precessing longitudes of the pericenter and of the nodes, with the nodal line taken as the line of intersection of the two orbital planes before and after the impact. Fixing the orbit by an ad hoc model of (precessing) Keplerian ellipse, we can derive two separate restricted spin-orbit problems (one for the primary and another for the secondary) which approximate the full problem. Then, we show that the instability in the secondary's rotational state observed under the full problem corresponds to a parametric resonance observed in the respective restricted problem. 

Notwithstanding our particular interest in the Didymos-DART problem, we stress that the analytical theories here presented are generic, i.e., applicable to any two-body system of astronomical or astrodynamical interest settled close to  single-synchronous equilibrium state. \textit{Section \ref{sec:Conclusions}} summarizes, in respect, our main results and conclusions.

\section{Problem statement and equations of motion}
\label{sec:ProbState}

\subsection{Coordinate system}
\label{subsec:CoordSys}
\begin{figure}[h]
    \centering
	\includegraphics[scale=0.35]{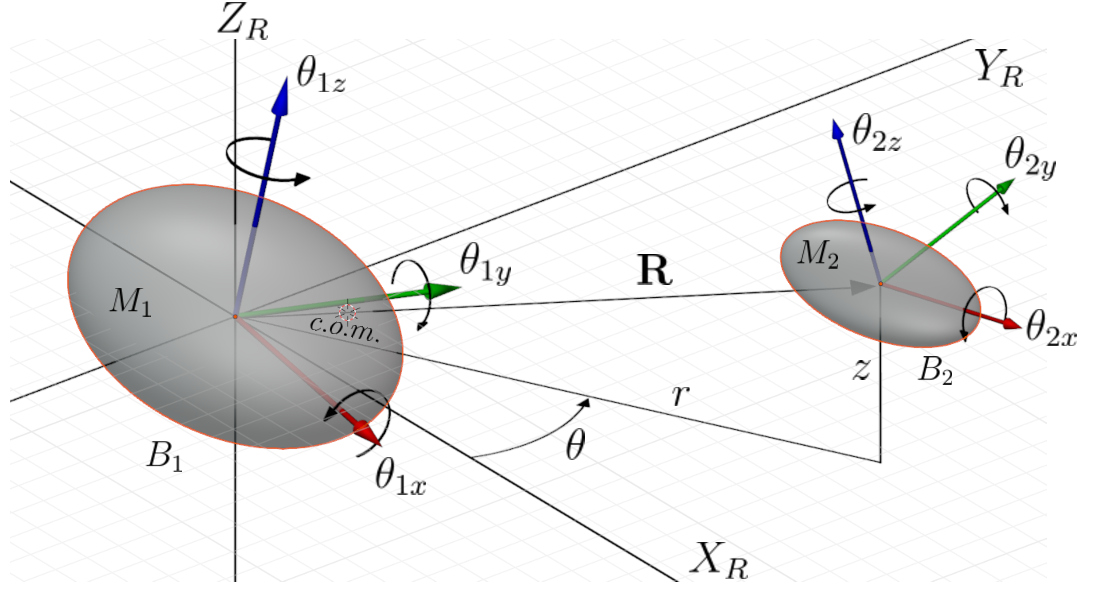}
    \caption{Coordinate system setup of the F2BP. Both bodies can move along 3 directions and spin around 3 axes, while gravitationally interacting.}
    \label{fig:F2BP3D}
\end{figure}

We consider two rigid bodies $B_1,B_2$ in space, of masses $M_1$, $M_2$, under their mutual gravitational influence, as depicted in Figure \ref{fig:F2BP3D}. We assume $M_1>M_2$ and call $B_1$ the `primary' and $B_2$ the `secondary'. Both bodies are allowed arbitrary rotation, described relative to their principal axes of inertia. We adopt the Tait-Bryan angle set 1-2-3 (commonly referred to as Euler angles, i.e. roll, pitch, yaw) for each body, denoted hereafter by the symbols $\theta_{ix}, \theta_{iy}, \theta_{iz}$, $i=1,2$, where the indices $(x,y,z)$ respectively refer to each body's long, intermediate, and short principal axis of inertia. For the description of the relative orbit we employ cylindrical coordinates $(r, \theta, z)$ such that the relative position vector $\boldsymbol{R}$ with origin at the center of mass of the primary and endpoint at the center of mass of the secondary has relative coordinates $\boldsymbol{R}=(X_R,Y_R,Z_R)$ with $X_R=r\cos\theta$, $Y_R=r\sin\theta$, $Z_R=z$. Assuming an inertial reference frame $O(XYZ)$ with origin at the center of mass of the system we recover the barycentric radius vectors $\boldsymbol{R}_1=\boldsymbol{OB_1}$, $\boldsymbol{R}_2=\boldsymbol{OB_2}$ through the relations $\boldsymbol{R_1}=-M_2\boldsymbol{R}/(M_1+M_2)$, $\boldsymbol{R}_2=M_1\boldsymbol{R}/(M_1+M_2)$. 

The bodies' gravitational interaction is expressed through a `second-moment' approximation of the mutual potential function (\cite{Scheeres2009})
\begin{equation}\label{Vord2}
\begin{split}
V(\boldsymbol{R}, \boldsymbol{A}_1, \boldsymbol{A}_2) &= -\frac{GM_1M_2}{R} \\
        & - \frac{G}{2 R^3}
      \big[M_2 \text{tr}(\boldsymbol{\mathrm{I}}_1) + M_1 \text{tr}(\boldsymbol{\mathrm{I}}_2)\big] \\
      & + \frac{3G}{2R^5} \boldsymbol{R}\cdot(M_2 \boldsymbol{A}_1 \boldsymbol{\mathrm{I}}_1 \boldsymbol{A}_1^T + 
                                     M_1 \boldsymbol{A}_2 \boldsymbol{\mathrm{I}}_2 \boldsymbol{A}_2^T)
                                     \cdot \boldsymbol{R}~~~,
\end{split}
\end{equation}
where $G$ is the gravitational constant, $\boldsymbol{A}_i$ is the rotation matrix of the body $B_i$ around its center of mass and $\boldsymbol{I}_i = \text{diag}(I_{ix}, I_{iy}, I_{iz})$, with $I_{ix} < I_{iy} < I_{iz}$, is the (constant) inertia matrix of the body $B_i$ expressed in its own solidal frame. In terms of the Euler angles $\theta_{ix}, \theta_{iy}$ and $\theta_{iz}$, the matrix $\boldsymbol{A}_i$ is expressed as 
\begin{equation}
    \boldsymbol{A}_i = \boldsymbol{A}_{iz}\boldsymbol{A}_{iy}\boldsymbol{A}_{ix}
\end{equation}
where
\begin{equation}\label{Aiz}
\boldsymbol{A}_{iz} = 
\begin{bmatrix}
\cos{\theta_{iz}} & -\sin{\theta_{iz}} & 0 \\
\sin{\theta_{iz}} &  \cos{\theta_{iz}} & 0 \\
       0       &         0       & 1
\end{bmatrix}
\end{equation}
\begin{equation}\label{Aiy}
\boldsymbol{A}_{iy} = 
\begin{bmatrix}
 \cos{\theta_{iy}} & 0 & \sin{\theta_{iy}} \\
       0        & 1 &        0       \\
-\sin{\theta_{iy}} & 0 & \cos{\theta_{iy}}
\end{bmatrix}
\end{equation}
\begin{equation}\label{Aix}
\boldsymbol{A}_{ix} =
\begin{bmatrix}
1 &        0       &         0       \\
0 & \cos{\theta_{ix}} & -\sin{\theta_{ix}} \\
0 & \sin{\theta_{ix}} &  \cos{\theta_{ix}}
\end{bmatrix}
\end{equation}
are the matrices corresponding to rotations about the bodies solidal axes $z$, $y$, and $x$  respectively.

For dynamical states close to the SESS (see below), the solidal $z-$axis of both bodies remains always nearly normal to the orbital plane. It then turns convenient to replace the generalized coordinates $\theta_{iz}$ (yaw angles) in the equations of motion with the `relative yaw angles' $\phi_i$, $i=1,2$ given by
\begin{equation}\label{InertialYawToRelativeYaw}
\phi_i = \theta_{iz} - \theta~~.
\end{equation}
When the coordinate $z$ of the relative orbit, as well as all four angles $\theta_{ix},\theta_{iy}$, are small quantities, the angles $\phi_i$ become nearly equal (with an error of second order in the small quantities) to the geometrical angles formed between the lines containing the relative radius vector $\boldsymbol{R}$ and each of the bodies' major inertial axis $x_i$. 

\subsection{Hamiltonian}
\label{subsec:HamFull}
For our study, we employ the Hamiltonian formalism. The total kinetic energy of the system is
\begin{equation}\label{TKinetic}
T  = \frac{1}{2}m \boldsymbol{v}\cdot\boldsymbol{v} +
      \frac{1}{2}\boldsymbol{\omega}_1 \cdot \boldsymbol{\mathrm{I}}_1 \cdot \boldsymbol{\omega}_1 +
      \frac{1}{2}\boldsymbol{\omega}_2 \cdot \boldsymbol{\mathrm{I}}_2 \cdot \boldsymbol{\omega}_2~~~,
\end{equation}
where
\begin{equation}
    m = \frac{M_1M_2}{M_1 + M_2}
\end{equation}
is the reduced mass of the binary,
\begin{equation}
    \boldsymbol{v} \equiv \dot{\boldsymbol{R}} =  
    \begin{bmatrix}
    \dot{r}\cos{\theta} - r\dot{\theta}\sin{\theta} \\
    \dot{r}\sin{\theta} + r\dot{\theta}\cos{\theta} \\
    \dot{z}
    \end{bmatrix}
\end{equation}
is the linear mutual velocity vector in the cylindrical coordinates and $\boldsymbol{\omega}_i$ is the angular velocity of $B_i$, expressed in the corresponding body frame. Denoting $\boldsymbol{\theta}_i=(\theta_{ix},\theta_{iy},\theta_{iz})$, for $i=1,2$, and using the formula 
\begin{equation}\label{OmegasAsFuncOfThetas}
\boldsymbol{\omega}_i(\boldsymbol{\theta}_i, \dot{\boldsymbol{\theta}}_i) = 
\begin{bmatrix}
\dot{\theta}_{ix} - \dot{\theta}_{iz}\sin{\theta_{iy}}                             \\
\dot{\theta}_{iy}\cos{\theta_{ix}} + \dot{\theta}_{iz}\cos{\theta_{iy}}\sin{\theta_{ix}} \\
 \dot{\theta}_{iz}\cos{\theta_{ix}}\cos{\theta_{iy}} - \dot{\theta}_{iy}\sin{\theta_{ix}}
\end{bmatrix}~~~,
\end{equation}
the Lagrangian is expressed in terms of the generalized coordinates and velocities ($\boldsymbol{R},\boldsymbol{\theta}_1,\boldsymbol{\theta}_2,\dot{\boldsymbol{R}},\dot{\boldsymbol{\theta}}_1,\dot{\boldsymbol{\theta}}_2$) as 
\begin{equation}\label{Lagrangian}
L  = T(\boldsymbol{R},\boldsymbol{\theta}_1,\boldsymbol{\theta}_2,\dot{\boldsymbol{R}},\dot{\boldsymbol{\theta}}_1,\dot{\boldsymbol{\theta}}_2) - V(\boldsymbol{R},\boldsymbol{\theta}_1,\boldsymbol{\theta}_2)~~,
\end{equation}
from which we derive the Hamiltonian
\begin{equation}\label{Hfull}
    H = \dot{r}p_r +
        \dot{\theta}p_\theta +
        \dot{z}p_z +
        \dot{\theta}_{1x}p_{\theta_{1x}} +
        \dot{\theta}_{1y}p_{\theta_{1y}} +
        \dot{\phi}_{1}p_{\phi_{1}} +
        \dot{\theta}_{2x}p_{\theta_{2x}} +
        \dot{\theta}_{2y}p_{\theta_{2y}} +
        \dot{\phi}_{2}p_{\phi_{2}} - L~~,
\end{equation}
where
\begin{equation}\label{HfullMomenta}
\begin{gathered}
p_r  = \frac{\partial L }{\partial \dot{r}} \hspace{1cm}
p_\theta  = \frac{\partial L }{\partial \dot{\theta}} \hspace{1cm}
p_z  = \frac{\partial L }{\partial \dot{z}} \\
p_{\theta_{1x}}  = \frac{\partial L }{\partial \dot{\theta}_{1x}} \hspace{1cm}
p_{\theta_{1y}}  = \frac{\partial L }{\partial \dot{\theta}_{1y}} \hspace{1cm}
p_{\phi_{1}}  = \frac{\partial L }{\partial \dot{\phi}_{1}} \\
p_{\theta_{2x}}  = \frac{\partial L }{\partial \dot{\theta}_{2x}} \hspace{1cm}
p_{\theta_{2y}}  = \frac{\partial L }{\partial \dot{\theta}_{2y}} \hspace{1cm}
p_{\phi_2}  = \frac{\partial L }{\partial \dot{\phi}_2}
\end{gathered} 
\end{equation}
are the canonical momenta, corresponding to the Hamiltonian (\ref{Hfull}). The functional form of $H(\boldsymbol{q},\boldsymbol{p})$, with $\boldsymbol{q}=$ $(r,\theta,z,\theta_{1x},\theta_{1y},\phi_1,\theta_{2x},\theta_{2y},\phi_2)$, $\boldsymbol{p}=$ $(p_r,p_\theta,p_z,p_{\theta_{1x}},p_{\theta_{1y}},p_{\phi_1},p_{\theta_{2x}},p_{\theta_{2y}},p_{\phi_2})$ is
\begin{align}\label{Hfullform}
H&={1\over 2m}(p_r^2+p_z^2)
  +{1\over 2mr^2}(p_\theta-p_{\phi_1}-p_{\phi_2})^2 \nonumber\\
&+\sum_{i=1}^{2}
\Bigg\{
\Bigg(
{1\over 2I_{ix}}
+{\sin^2(\theta_{ix})\tan^2(\theta_{iy})\over 2I_{iy}}
+{\cos^2(\theta_{ix})\tan^2(\theta_{iy})\over 2I_{iz}}
\Bigg)p_{\theta_{ix}}^2 \nonumber\\
&~~~~~~~+
\Bigg(
 {\cos^2(\theta_{ix})\over 2I_{iy}}
+{\sin^2(\theta_{ix})\over 2I_{iz}}
\Bigg)p_{\theta_{iy}}^2 
+
\Bigg(
 {\sin^2(\theta_{ix})\over 2\cos^2(\theta_{iy})I_{iy}}
+{\cos^2(\theta_{ix})\over 2\cos^2(\theta_{iy})I_{iz}}
\Bigg)p_{\phi_{i}}^2 \nonumber\\
&~~~~~~~+
\Bigg(
 {\sin(2\theta_{ix})\tan(\theta_{iy})\over 2I_{iy}}
-{\sin(2\theta_{ix})\tan(\theta_{iy})\over 2I_{iz}}
\Bigg)p_{\theta_{ix}}p_{\theta_{iy}} \\
&~~~~~~~+
\Bigg(
 {\sin^2(\theta_{ix})\sin(\theta_{iy})\over \cos^2(\theta_{iy})I_{iy}}
+{\cos^2(\theta_{ix})\sin(\theta_{iy})\over \cos^2(\theta_{iy})I_{iz}}
\Bigg)p_{\theta_{ix}}p_{\phi_{i}} \nonumber\\
&~~~~~~~+
\Bigg(
 {\sin(2\theta_{ix})\over 2\cos(\theta_{iy})I_{iy}}
-{\sin(2\theta_{ix})\over 2\cos(\theta_{iy})I_{iz}}
\Bigg)p_{\theta_{iy}}p_{\phi_{i}} \Bigg\}\nonumber\\
~&+V(r,\theta,z,\theta_{1x},\theta_{1y},\phi_1,
\theta_{2x},\theta_{2y},\phi_2) \nonumber~~,
\end{align}
with the potential $V$ given as
$$
V=V_{kep}+V_{2,0}+V_{2,1}+V_{2,2}~~,
$$
with
\begin{align*}
V_{kep}&= -{G M_1 M_2\over(r^2+z^2)^{1/2}} \\
V_{2,0} &= -{G\left(
(I_{1x}+I_{1y}+I_{1z})M_2+(I_{2x}+I_{2y}+I_{2z})M_1
\right)\over 2(r^2+z^2)^{3/2}} \\
V_{2,1} &= {3 G M_2\over 2(r^2 + z^2)^{
 5/2}} \times \\
~&\times\Bigg(
 I_{1x}(r\cos(\theta_{1y})\cos(\phi_1) - z\sin(\theta_{1y}))^2 \\
&+I_{1y}\left(
        z\cos(\theta_{1y})\sin(\theta_{1x}) 
      + r\cos(\phi_1)\sin(\theta_{1x})\sin(\theta_{1y}) 
      - r\sin(\phi_1)\cos(\theta_{1x})
       \right)^2 \\
&+I_{1z}\left(
        z\cos(\theta_{1x})\cos(\theta_{1y}) 
      + r\cos(\phi_1)\cos(\theta_{1x})\sin(\theta_{1y}) 
      + r\sin(\phi_1)\sin(\theta_{1x})
       \right)^2 \Bigg) \\
V_{2,2} &= {3 G M_1\over 2(r^2 + z^2)^{
 5/2}} \times \\
~&\times\Bigg(
 I_{2x}(r\cos(\theta_{2y})\cos(\phi_2) - z\sin(\theta_{2y}))^2 \\
&+I_{2y}\left(
        z\cos(\theta_{2y})\sin(\theta_{2x}) 
      + r\cos(\phi_2)\sin(\theta_{2x})\sin(\theta_{2y}) 
      - r\sin(\phi_2)\cos(\theta_{2x})
       \right)^2 \\
&+I_{2z}\left(
        z\cos(\theta_{2x})\cos(\theta_{2y}) 
      + r\cos(\phi_2)\cos(\theta_{2x})\sin(\theta_{2y}) 
      + r\sin(\phi_2)\sin(\theta_{2x})
       \right)^2 \Bigg)~.
\end{align*}
Note that, Hamilton's equations for the canonical pair $(\theta,p_\theta)$ read
\begin{equation}\label{Hfulltheta}
\dot{\theta}={1\over mr^2}(p_\theta-p_{\phi_1}-p_{\phi_2}),~~\dot{p}_\theta=0~~.
\end{equation}
The second of the above equations comes from the fact that the gravitational interaction between the bodies depends only on the relative angles $\phi_i$, hence $\theta$ is ignorable, and it expresses the conservation of the total angular momentum magnitude of the system, equal in modulus to $p_\theta$. Note also that by the chosen set of Euler angles, the Hamiltonian (\ref{Hfullform}) exhibits a virtual singularity when the pitch angle becomes equal to $\theta_{iy} = k\pi \pm \pi/2$, $k \in \mathbb{Z}$. However, our perturbation theories below provide analytical expressions valid around the state $\theta_{iy} = 0$, i.e., `far' from the singularity. 

\subsection{Single-synchronous equilibrium state}
\label{subsec:StateInterest}
\begin{figure}
  \centering
    \includegraphics[scale=0.35]{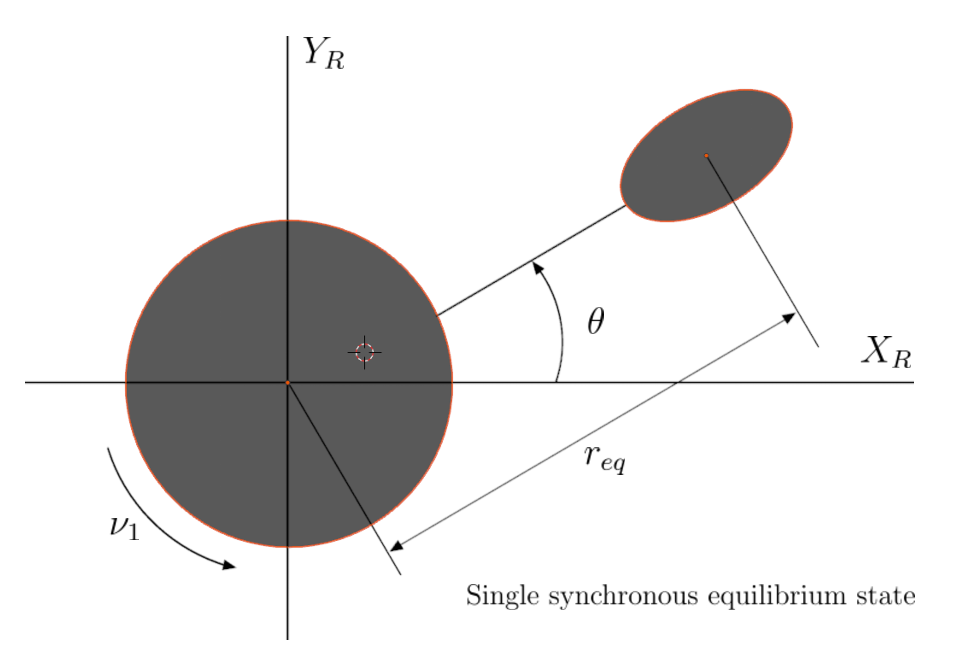}
    \includegraphics[scale=0.35]{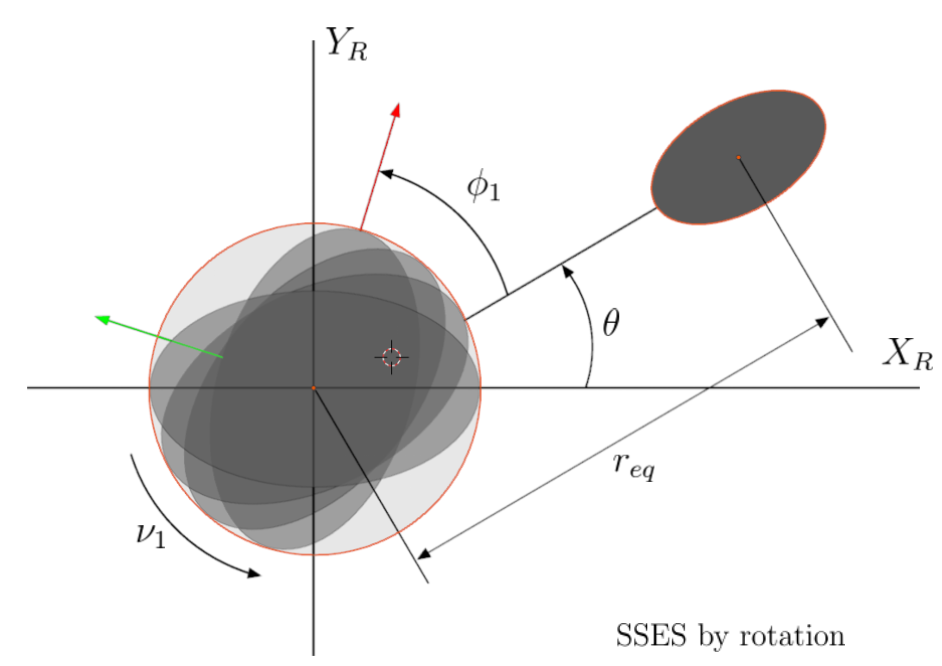}
    \caption{Top: schematic representation of the SSES. The primary has two of its principal axes lying in the orbital plane and rotates around its third principal axis with spin frequency equal to $\dot{\phi}_1+\dot{\theta}=\nu_1$. The relative orbit is circular, with frequency $\dot{\theta}=\nu_\theta$. Bottom: an SSES created `by rotation', i.e., by averaging the Hamiltonian (\ref{Hfullform}) with respect to the `fast' angle $\phi_1$.}
    \label{fig:SSES}
\end{figure}

Assume a binary system with axisymmetric primary $I_{1x}=I_{1y}=I_s$. We can readily show that, for any observed value of the primary's spin frequency $\nu_{1,obs}$, and of the system's orbital period $T_{\theta,obs}=2\pi/\nu_{\theta,obs}$, Hamilton's equations under the Hamiltonian (\ref{Hfullform}) admit the following exact solution, called the `single-synchronous equilibrium state' (SSES):
\begin{align}\label{sses}
r(t)&=r_{eq}, 
&z(t)&=0,
&\theta(t)&=~~\theta(t_0)+\nu_{\theta,obs} (t-t_0), 
\nonumber\\
\theta_{1x}(t)&=0, 
&\theta_{1y}(t)&=0,
&\phi_{1}(t)&=\phi_{1}(t_0)+(\nu_{1,obs}-\nu_{\theta,obs})(t-t_0), 
\nonumber\\
\theta_{2x}(t)&=0, 
&\theta_{2y}(t)&=0,
&\phi_{2}(t)&=0, \\
p_r(t)&=0,
&p_z(t)&=0,
&p_\theta(t)&=p_{\theta,eq}=(mr_{eq}^2+I_{2z})\nu_{\theta,obs}+I_{1z}\nu_{1,obs},
\nonumber\\
p_{\theta_{1x}}(t)&=0, 
&p_{\theta_{1y}}(t)&=0,
&p_{\phi_{1}}(t)&=p_{\phi_1,eq}=\nu_{1,obs}I_{1z}, 
\nonumber\\
p_{\theta_{2x}}(t)&=0, 
&p_{\theta_{2y}}(t)&=0,
&p_{\phi_{2}}(t)&=p_{\phi_2,eq}=\nu_{\theta,obs}I_{2z}.
\nonumber
\end{align} 
In Eq. (\ref{sses}), $\theta(t_0)$, $\phi_1(t_0)$ are arbitrary initial conditions, while $r_{eq}$ is the positive algebraic root of the equation (\cite{Scheeres2009})
\begin{equation}\label{req}
\nu_{\theta,obs}=\sqrt{\frac{G(M_1+M_2)}{r_{eq}^3}\bigg[
    1 + \frac{3}{2r_{eq}^2} \bigg( \frac{I_{1z}-I_{s}}{M_1} + \frac{-2I_{2x} + I_{2y} + I_{2z}}{M_2} \bigg)
    \bigg]}~,
\end{equation}
closest to the Keplerian estimate $r_{eq}\simeq r_{eq,kep}=[G(M_1+M_2)]^{1/3}/\nu_{\theta,obs}^{2/3}$. Physically, the system's relative orbit is circular at radius $r=r_{eq}$ with uniform angular frequency $\nu_{\theta,obs}$ (see Fig. \ref{fig:SSES}, top panel).  

Consider, now, the case in which the primary body is triaxial $I_{1x}\neq I_{1y}$. Strictly, an SSES of the form of Eq. (\ref{sses}) does not exist for Hamilton's equations under the full Hamiltonian (\ref{Hfullform}). However, we readily find that an SSES still can be defined for Hamilton's equations in all nine degrees of freedom under the averaged Hamiltonian
\begin{equation}\label{Havg}
    H_{\text{avg}}(r,z,\theta_{1x},\theta_{1y},\theta_{2x},\theta_{2y},\phi_2) = \frac{1}{2\pi}\int_{0}^{2\pi}H(r,z,\theta_{1x},\theta_{1y},\phi_1,\theta_{2x},\theta_{2y},\phi_2)d\phi_1~~.
\end{equation}
The equilibrium is given by the same formulas as in Eq. (\ref{sses}), with the equilibrium radius $r_{eq}$ now given as the root of
\begin{equation}\label{reqave}
\nu_{\theta,obs}=\sqrt{\frac{G(M_1+M_2)}{r_{eq}^3}\bigg[
    1 + \frac{3}{2r_{eq}^2} \bigg( \frac{2I_{1z}-I_{1x}-I_{1y}}{2M_1} + \frac{-2I_{2x} + I_{2y} + I_{2z}}{M_2} \bigg)
    \bigg]}~,
\end{equation}
which closely follows the formulation of \cite{Scheeres2009} in Eq. (\ref{req}), extended here to the triaxial case where $I_{1x}\neq I_{1y}$, but for the SSES case. Physically, the averaging of Eq. (\ref{Havg}) mimics approximating the triaxial primary as axisymmetric by rotation, yet it remains more general than directly setting $I_{1x}=I_{1y}=I_s=(I_{1x}+I_{1y})/2$, since both $I_{1x}$ and $I_{1y}$ are kept as symbolic parameters in the averaged Hamiltonian (see Fig. \ref{fig:SSES}, bottom panel). 

\begin{figure}
  \centering
    \includegraphics[scale=0.3]{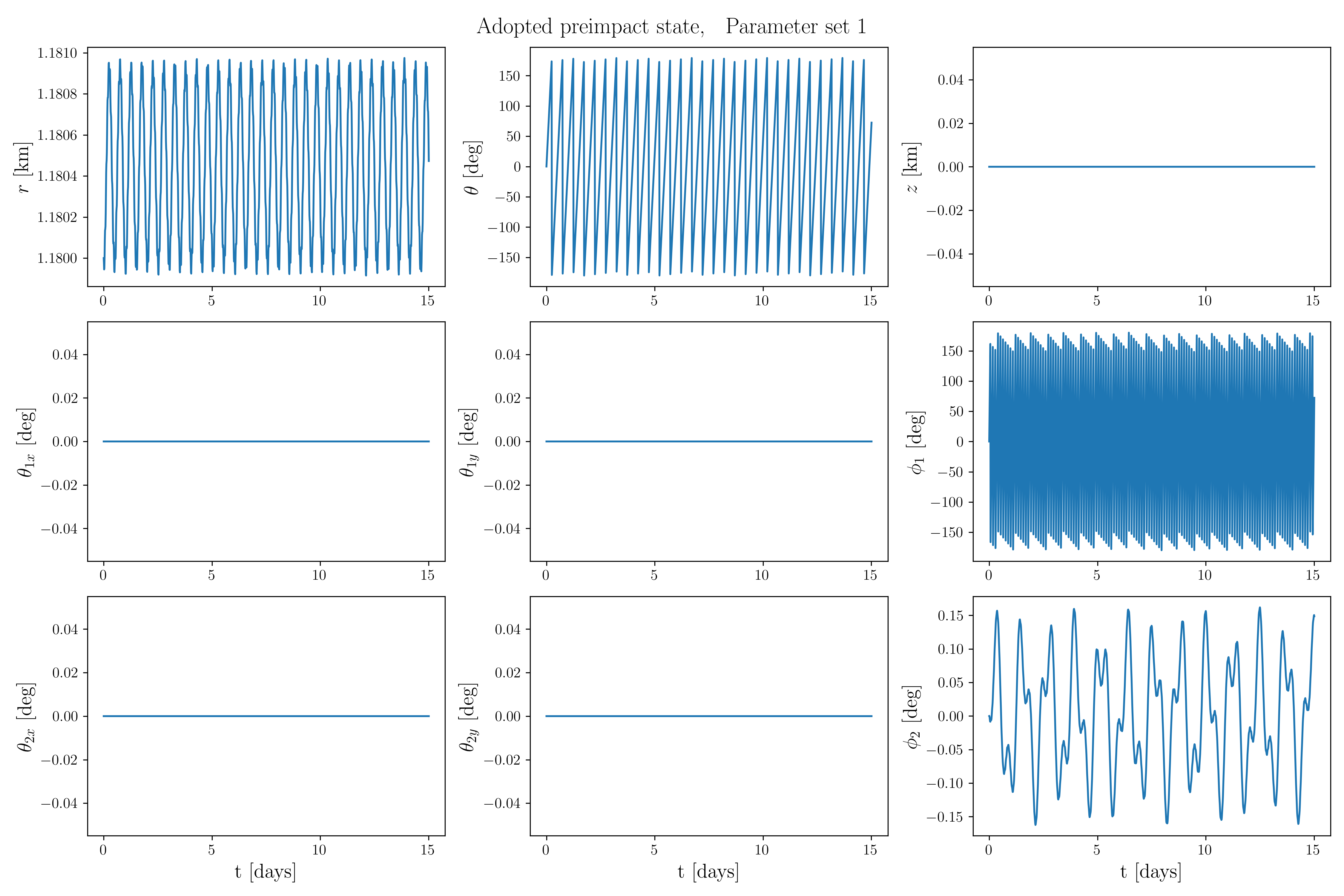}
    \caption{Our adopted pre-impact kinetic state of the binary system 65803 Didymos. The functions $r(t),\theta(t),z(t)$ (upper row), $\theta_{1x}(t),\theta_{1y}(t),\phi_{1}(t)$ (mid row) and $\theta_{2x}(t), \theta_{2y}(t), \phi_2(t)$ (lower row) are numerically integrated in time for $G = 0.0864989$, $M_1 = 5.150418$, $M_2 = 0.0392693$, $I_{1x} = 0.260108$, $I_{1y} = 0.267618$, $I_{1z} = 0.337960$, $I_{2x} = 8.20454 \cdot 10^{-5}$, $I_{2y} = 8.88782 \cdot 10^{-5}$, $I_{2z} = 1.1899 \cdot 10^{-4}$, which we refer to as `parameter set 1' in Section \ref{sec:dart}. The units of measurement are in [hr], [km], [kg$^{\ast}$], where 1 [kg$^{\ast}] = 10^{11}$[kg]. The system is very close to an exact SSES.}
    \label{fig:Observed_state_set1}
\end{figure}
Numerically, we find that the SSES recovered by averaging the equations of motion with respect to $\phi_1$ is justified if the primary is a fast rotator $\nu_{1,obs}\gg\nu_{\theta,obs}$, and the ratio $I_{1y}/I_{1x}$ is a number of the order of unity. As an example consider the orbit of the 65803 Didymos system prior to DART impact, with initial conditions at $t_0=0$ equal to those corresponding to Eq. (\ref{sses}) and $r_{eq}$ computed through Eq. (\ref{reqave}). Fig. \ref{fig:Observed_state_set1} shows the result in time of the numerical integration of the equations of motion under the exact Hamiltonian (\ref{Hfullform}), with physical parameters as in the caption of the figure, and estimated observed values $\nu_{\theta,obs} = 2\pi/T_{\theta,obs}$, $T_{\theta,orb} = 11.92149$ [hr], $\nu_{1,obs} = 2\pi/T_{1,obs}$, $T_{1,obs} = 2.26$ [hr] and $r_{obs} = 1.18$ [km]. As shown in Fig. \ref{fig:Observed_state_set1}, the SSES is reproduced by the real pre-impact observed motion of the system except for a fluctuation of about one meter with respect to the equilibrium radius, and a small libration of the angle $\phi_2$ of about $0.15$ degrees amplitude with respect to the exact alignment of the SSES. Note, finally, that the non-planar degrees of freedom $(z,\theta_{1x},\theta_{1y},\theta_{2x},\theta_{2y})$ satisfy the SSES condition exactly. This occurs because the assumed potential (of second multipole order) preserves any initial planar symmetry by confining forces strictly to the $x-y$ plane and torques to the $z-$axis. In contrast, a potential of higher multipole order, along with the influence of possible shape irregularities, would introduce a coupling of in-plane and out-of-plane motions, thereby rendering the SSES approximate for the out-of-plane degrees of freedom as well.

\section{Linear perturbation theory}
\label{sec:LinearTheory}

\subsection{Linearized equations of motion around the SSES}
\label{subsec:lineqmo}
The angle $\phi_1$ is ignorable in the averaged Hamiltonian (\ref{Havg}), hence its conjugate momentum $p_{\phi_1}$ is a constant of motion of the flow under Hamilton's equations of $H_{\text{avg}}$, additional to the constant $p_\theta$. Consider the equilibrium solution of the SSES type defined for a particular choice of parameters $\nu_1=\nu_{1,obs}$, $\nu_\theta=\nu_{\theta,obs}$ and $r_{eq}$ given by Eq. (\ref{reqave}). The constant quantities $p_\theta=p_{\theta,eq}$ and $p_{\phi_1}=p_{\phi_1,eq}$ with values as in Eq. (\ref{sses}) can be considered as parameters in the Hamiltonian $H_{\text{avg}}$, effectively reducing the latter to a Hamiltonian with seven degrees of freedom. In the so-reduced phase space, Hamilton's equations read
\begin{equation}\label{AvgEOMs}
\begin{aligned}
    \dot{r} &= \frac{\partial H_{\text{avg}} }{\partial p_r} & \quad \dot{p}_r  &= -\frac{\partial H_{\text{avg}} }{\partial r} \\
    \dot{z} &= \frac{\partial H_{\text{avg}} }{\partial p_z} & \quad \dot{p}_z  &= -\frac{\partial H_{\text{avg}} }{\partial z} \\
    \dot{\theta}_{1x} &= \frac{\partial H_{\text{avg}} }{\partial p_{\theta_{1x}}} & \quad \dot{p}_{\theta_{1x}}  &= -\frac{\partial H_{\text{avg}} }{\partial \theta_{1x}} \\
    \dot{\theta}_{1y} &= \frac{\partial H_{\text{avg}} }{\partial p_{\theta_{1y}}} & \quad \dot{p}_{\theta_{1y}}  &= -\frac{\partial H_{\text{avg}} }{\partial \theta_{1y}} \\
    \dot{\theta}_{2x} &= \frac{\partial H_{\text{avg}} }{\partial p_{\theta_{2x}}} & \quad \dot{p}_{\theta_{2x}}  &= -\frac{\partial H_{\text{avg}} }{\partial \theta_{2x}} \\
    \dot{\theta}_{2y} &= \frac{\partial H_{\text{avg}} }{\partial p_{\theta_{2y}}} & \quad \dot{p}_{\theta_{2y}}  &= -\frac{\partial H_{\text{avg}} }{\partial \theta_{2y}} \\
    \dot{\phi}_{2} &= \frac{\partial H_{\text{avg}} }{\partial p_{\phi_{2}}} & \quad \dot{p}_{\phi_{2}}  &= -\frac{\partial H_{\text{avg}} }{\partial \phi_{2}}~~.
\end{aligned}
\end{equation}
Introducing a small disturbance to the SESS
\begin{equation}\label{LinDisturb}
\begin{aligned}
r(t) &= r_{eq} + \delta r(t) & \quad p_r(t) &= \delta p_r(t) \\
z(t) &= \delta z(t) & \quad p_z(t) &= \delta p_z(t) \\
\theta_{1x}(t) &= \delta \theta_{1x}(t) & \quad p_{\theta_{1x}}(t) &= \delta p_{\theta_{1x}}(t) \\
\theta_{1y}(t) &= \delta \theta_{1y}(t) & \quad p_{\theta_{1y}}(t) &= \delta p_{\theta_{1y}}(t) \\
\theta_{2x}(t) &= \delta \theta_{2x}(t) & \quad p_{\theta_{2x}}(t) &= \delta p_{\theta_{2x}}(t) \\
\theta_{2y}(t) &= \delta \theta_{2y}(t) & \quad p_{\theta_{2y}}(t) &= \delta p_{\theta_{2y}}(t) \\
\phi_2(t) &= \delta \phi_2(t) & \quad p_{\phi_2}(t) &= p_{\phi_{2eq}} + \delta \phi_2(t)
\end{aligned}
\end{equation}
and linearizing the averaged equations of motion (\ref{AvgEOMs}) around the equilibrium state, leads to the linear system

\begin{equation}\label{LinSysDelta}
\dot{\boldsymbol{\xi}} = \boldsymbol{J} \cdot \boldsymbol{\xi}~,
\end{equation}
where 
$$
\boldsymbol{\xi} = 
\big[\delta r,\delta z,
\delta\theta_{1x},\delta\theta_{1y},
\delta\theta_{2x},\delta\theta_{2y},\delta\phi_2,
\delta p_r,\delta p_z,
\delta p_{\theta_{1x}},\delta p_{\theta_{1y}},
\delta p_{\theta_{2x}},\delta p_{\theta_{2y}},\delta p_{\phi_2}\big]^T~~
$$
and the variational Jacobian matrix $\boldsymbol{J}$ has the structure 
\begin{equation*}
\boldsymbol{J} =\left[
\begin{array}{cccccccccccccc}
   0    &    0    &    0     &    0     &    0     &    0     &    0     & j_{1,8} &    0    &    0     &    0     &    0     &    0     &   0      \\
   0    &    0    &    0     &    0     &    0     &    0     &    0     &    0    & j_{1,8} &    0     &    0     &    0     &    0     &   0      \\
   0    &    0    &    0     & j_{3,4}  &    0     &    0     &    0     &    0    &    0    & j_{3,10} &    0     &    0     &    0     &   0      \\
   0    &    0    & j_{4,3}  &    0     &    0     &    0     &    0     &    0    &    0    &    0     & j_{4,11} &    0     &    0     &   0      \\
   0    &    0    &    0     &    0     &    0     & j_{5,6}  &    0     &    0    &    0    &    0     &    0     & j_{5,12} &    0     &   0      \\
   0    &    0    &    0     &    0     & j_{6,5}  &    0     &    0     &    0    &    0    &    0     &    0     &    0     & j_{6,13} &   0      \\
j_{7,1} &    0    &    0     &    0     &    0     &    0     &    0     &    0    &    0    &    0     &    0     &    0     &    0     & j_{7,14} \\
j_{8,1} &    0    &    0     &    0     &    0     &    0     &    0     &    0    &    0    &    0     &    0     &    0     &    0     & -j_{7,1} \\
   0    & j_{9,2} &    0     &    0     &    0     & j_{9,6}  &    0     &    0    &    0    &    0     &    0     &    0     &    0     &    0     \\
   0    &    0    & j_{10,3} &    0     &    0     &    0     &    0     &    0    &    0    &    0     & -j_{4,3} &    0     &    0     &    0     \\
   0    &    0    &    0     & j_{11,4} &    0     &    0     &    0     &    0    &    0    & -j_{3,4} &    0     &    0     &    0     &    0     \\
   0    &    0    &    0     &    0     & j_{12,5} &    0     &    0     &    0    &    0    &    0     &    0     &    0     & -j_{6,5} &    0     \\
   0    & j_{9,6} &    0     &    0     &    0     & j_{13,6} &    0     &    0    &    0    &    0     &    0     & -j_{5,6} &    0     &    0     \\
   0    &    0    &    0     &    0     &    0     &    0     & j_{14,7} &    0    &    0    &    0     &    0     &    0     &    0     &    0
\end{array}
\right]~.
\end{equation*}
The non-zero elements of the Jacobian, $j_{n,m}$, evaluated at the equilibrium point, depend only on the chosen values of the system's physical parameters and SSES. Their mathematical form is provided in Appendix \ref{app:JacElements}. 

The block structure of the Jacobian matrix $\boldsymbol{J}$ implies a natural decoupling of the linearized equations of motion (\ref{LinSysDelta}) in three non-interacting subsystems, namely: i) the (planar) variables $(\delta r, \delta {\phi_2}, \delta p_r, \delta p_{\phi_2})$, are grouped together into a 4x4 linear subsystem, ii) the (non-planar) variables $(\delta z, \delta \theta_{2x}, \delta \theta_{2y}, \delta p_z, \delta p_{\theta_{2x}}, \delta p_{\theta_{2y}})$ form a 6x6 linear subsystem, and iii) the primary's (non-planar) orientation variables $(\delta \theta_{1x}, \delta \theta_{1y}, \delta p_{\theta_{1x}}, \delta p_{\theta_{1y}})$ form a 4x4 linear subsystem. We have
\renewcommand{\arraystretch}{1.2}
\begin{equation}\label{LinSys1}
\begin{bmatrix}
 \delta \dot{r} \\
 \dot{\delta \phi_2} \\
 \dot{\delta p_r} \\
 \dot{\delta p_{\phi_2}}
\end{bmatrix} = 
\begin{bmatrix}
 j_{1,8}{\delta p_r} \\
 j_{7,1}\delta r + j_{7,14}\delta p_{\phi_2} \\
 j_{8,1}\delta r - j_{7,1}\delta p_{\phi_2} \\
 j_{14,7}\delta \phi_2
\end{bmatrix}
\end{equation}
\begin{equation}\label{LinSys2}
\begin{bmatrix}
 \delta \dot{z} \\
 \dot{\delta \theta_{2x}} \\
 \dot{\delta \theta_{2y}} \\
 \dot{\delta p_z} \\
 \dot{\delta p_{\theta_{2x}}} \\
 \dot{\delta p_{\theta_{2y}}}
\end{bmatrix} = 
\begin{bmatrix}
 j_{1,8}{\delta p_z} \\
 j_{5,6}\delta \theta_{2y} + j_{5,12}\delta p_{\theta_{2x}} \\
 j_{6,5}\delta \theta_{2x} + j_{6,13}\delta p_{\theta_{2y}} \\
 j_{9,2}\delta z + j_{9,6}\delta \theta_{2y} \\
 j_{12,5}\delta \theta_{2x} - j_{6,5}\delta p_{\theta_{2y}} \\
 j_{9,6}\delta z + j_{13,6}\delta \theta_{2y} - j_{5,6}\delta p_{\theta_{2x}}
\end{bmatrix}
\end{equation}
\begin{equation}\label{LinSys3}
\begin{bmatrix}
 \dot{\delta \theta_{1x}} \\
 \dot{\delta \theta_{1y}} \\
 \dot{\delta p_{\theta_{1x}}} \\
 \dot{\delta p_{\theta_{1y}}}
\end{bmatrix} = 
\begin{bmatrix}
 j_{3,4}\delta \theta_{1y} + j_{3,10}\delta p_{\theta_{1x}} \\
 j_{4,3}\delta \theta_{1x} + j_{4,11}\delta p_{\theta_{1y}} \\
 j_{10,3}\delta \theta_{1x} - j_{4,3}\delta p_{\theta_{1y}} \\
 j_{11,4}\delta \theta_{1y} - j_{3,4}\delta p_{\theta_{1x}}
\end{bmatrix}
\end{equation}
\renewcommand{\arraystretch}{1.0}
Note that the above block structure can be understood substituting Eqs. (\ref{LinDisturb}) to the system of Hamilton's equations under the \textit{full} (non-averaged) equations of motion (\ref{Hfullform}) and expanding up to terms of second degree in the small disturbance. As regards the primary body, we then find that the expanded Hamiltonian contains: leading terms involving pairs of variables within the set $(\delta\theta_{1x},\delta\theta_{1y}, \delta p_{\theta_{1x}}, \delta p_{\theta_{1y}})$ that are quadratic (class (i)), pairs involving the variable $\delta z$ and one of the variables in the previous set (class (ii)). However, the pairs of class (i) come from the linearization of the torque-free Euler's equations for the primary body, hence they are not related to the averaging process. On the other hand, the pairs of class (ii) are spin-orbit terms with coefficients depending trigonometrically on the fast angle $\phi_1$, and hence vanishing after the averaging. The same 2 classes of terms appear also for the variables describing the secondary, but the class (ii) terms are not averaged out due to libration. In physical terms, this difference represents the difference between the primary (fast rotator in $\varphi_1$) and secondary (librator in $\varphi_2$). Finally, the decoupling of the linear block (\ref{LinSys1}) from the rest of the linearized equations of motion represents the fact that the problem is assumed to remain essentially planar, i.e., all off-plane disturbances are considered small quantities. 

\subsection{Solution of the linearized system}\label{subsec:LinSol}
The 4x4 subsystem of equation (\ref{LinSys1}) is identical to the linear system discussed in \cite{Michelangelo2024} when $I_{1x}=I_{1y}=I_s$ (primary axisymmetric). If, instead, $(I_{1x}\neq I_{1y})$, the entries $j_{n,m}$ depend on both $I_{1x}$ and $I_{1y}$, and hence the general solution resumes a slightly different form. The subsystem (\ref{LinSys1}) has the characteristic polynomial
\begin{equation}\label{P1Morph}
P_1(\lambda) = \lambda^4 + u_2\lambda^2 + u_0~,
\end{equation}
with parameters $u_0,u_2$ whose explicit form is given in the Appendix \ref{app:SysLinSol}. The roots are $\lambda_{1,2} = \mp i\omega_1$, $\lambda_{3,4} = \mp i\omega_2$, where 
\begin{equation}\label{P1Roots}
    \omega_{1,2} = \frac{\sqrt{u_2 \pm \sqrt{u_2 ^2-4 u_0 } }}{\sqrt{2}}~.
\end{equation}
Taking into account the inequalities $I_{ix} < I_{iy} < I_{iz}$, $i=1,2$, and by a leading term expression of the coefficients $u_0,u_2$ in the small quantities $M_2/M_1$, as well as $I_{ix}/mr_{eq}^2$, $I_{iy}/mr_{eq}^2$, $I_{iz}/mr_{eq}^2$, we find that $\omega_1, \omega_2$ are real and positive, hence they represent the fundamental set of linear frequencies of libration around the SSES for planar perturbations. For a generic initial perturbation $\boldsymbol{\xi}(0)$, we obtain the general solution of the 4x4 system in the form
\begin{equation}\label{P1LinVargen}
\begin{aligned}
\delta r(t) & = \sum_{k=1}^2
  C_{r,k}(\omega_1,\omega_2;\boldsymbol{\xi}(0))\cos{(\omega_k t)} 
+ S_{r,k}(\omega_1,\omega_2;\boldsymbol{\xi}(0))\sin{(\omega_k t)} \\
\delta \phi_2(t) & = \sum_{k=1}^2
  C_{\phi_2,k}(\omega_1,\omega_2;\boldsymbol{\xi}(0))\cos{(\omega_k t)} 
+ S_{\phi_2,k}(\omega_1,\omega_2;\boldsymbol{\xi}(0))\sin{(\omega_k t)} \\
\delta p_r(t) & = \sum_{k=1}^2
  C_{p_r,k}(\omega_1,\omega_2;\boldsymbol{\xi}(0))\cos{(\omega_k t)} 
+ S_{p_r,k}(\omega_1,\omega_2;\boldsymbol{\xi}(0))\sin{(\omega_k t)} \\
\delta p_{\phi_2}(t) & = \sum_{k=1}^2
  C_{p_{\phi_2}}(\omega_1,\omega_2;\boldsymbol{\xi}(0))\cos{(\omega_k t)} 
+ S_{p_{\phi_2}}(\omega_1,\omega_2;\boldsymbol{\xi}(0))\sin{(\omega_k t)}~,
\end{aligned}
\end{equation}
where the coefficients $C_{r,k}$, $S_{r,k}$, etc. depend on the the eigenvectors of the 4x4 submatrix of the system. Generic explicit expressions for these coefficients are cumbersome to obtain. However, they obtain a simple form in particular cases, such as the case of an oblique DART-like impact in the secondary asteroid in the case of the system 65803 Didymos (see Subsection \ref{subsec:dartlin}). 

Next, we examine the 6x6 linear subsystem (\ref{LinSys2}), which introduces a coupling of only off-plane coordinates and momenta $(\delta z,\delta \theta_{2x},\delta \theta_{2y}, \delta p_z,\delta p_{\theta_{2x}},\delta p_{\theta_{2y}})$. The corresponding characteristic polynomial has the form
\begin{equation}\label{P2Morph}
P_2(\lambda) = \lambda ^6 + \mathrm{v}_4\lambda ^4 + \mathrm{v}_2\lambda ^2 + \mathrm{v}_0~.
\end{equation}
Its roots are $\lambda_{5,6} = \mp i\omega_3$, $\lambda_{7,8} = \mp i\omega_4$ and $\lambda_{9,10} = \mp i\omega_5$, where
\begin{equation}\label{P2RootOm3}
    \omega_3 = \frac{\sqrt{-2^{2/3}\kappa^{1/3} + 2\mathrm{v}_4 - \frac{2^{4/3}(-3\mathrm{v}_2 + \mathrm{v}_4^2)}{\kappa^{1/3}}}}{\sqrt{6}}
\end{equation}
\begin{equation}\label{P2RootOm4}
    \omega_4 = \frac{\sqrt{2^{2/3}(1 + i\sqrt{3})\kappa^{1/3} + 4\mathrm{v}_4 - \frac{2^{4/3}(1-i\sqrt{3})(3\mathrm{v}_2-\mathrm{v}_4^2)}{\kappa^{1/3}}}}{2\sqrt{3}}
\end{equation}
\begin{equation}\label{P2RootOm5}
    \omega_5 = \frac{\sqrt{2^{2/3}(1 - i\sqrt{3})\kappa^{1/3} + 4\mathrm{v}_4 - \frac{2^{4/3}(1+i\sqrt{3})(3\mathrm{v}_2-\mathrm{v}_4^2)}{\kappa^{1/3}}}}{2\sqrt{3}}~~~.
\end{equation}
Explicit expressions for the polynomial's coefficients $\mathrm{v}_0,\mathrm{v}_2,\mathrm{v}_4$, along with the parameter $\kappa$, are given in the Appendix \ref{app:SysLinSol}. Again, by a leading terms analysis in the small quantities $M_2/M_1$, $I_{ix}/mr_{eq}^2$, $I_{iy}/mr_{eq}^2$, $I_{iz}/mr_{eq}^2$, we find that $\omega_3$, $\omega_4$, and $\omega_5$ are real and positive, hence they represent the fundamental set of linear frequencies of the off-plane librations of the secondary around the SSES. For a generic initial perturbation $\boldsymbol{\xi}(0)$, the corresponding linear solution can be expressed in terms of the cosines and sines $\cos(\omega_k t)$, $\sin(\omega_k t)$, $k=3,4,5$, with amplitudes $C_{z,k}$, $S_{z,k}$, etc., similar as in Eq. (\ref{P1LinVargen}) for the planar perturbations. Explicit expressions in the case of a DART-like perturbation are deferred to Subsection \ref{subsec:dartlin}.

The final 4x4 subsystem (\ref{LinSys3}) contains a coupling only in the variables $(\theta_{1x}, \theta_{1y}, p_{\theta_{1x}}, p_{\theta_{1y}})$, which correspond to the angular motion of the primary body. The characteristic polynomial has the form
\begin{equation}\label{P3Morph}
P_3(\lambda) = \lambda ^4 + \mathrm{w}_2\lambda ^2 + \mathrm{w}_0~~~,
\end{equation}
with explicit expressions for the coefficients $\mathrm{w}_0, \mathrm{w}_2$ given in the Appendix \ref{app:SysLinSol}. The polynomial's roots are $\lambda_{11,12} = \mp i\omega_6$, $\lambda_{13,14} = \mp i\omega_7$, and the two linear frequencies, are
\begin{equation}\label{P3Roots}
    \omega_{6,7} = \frac{\sqrt{\mathrm{w}_2 \pm \sqrt{\mathrm{w}_2 ^2-4 \mathrm{w}_0 } }}{\sqrt{2}}~~.
\end{equation}
As before, the generic solution can be expressed in terms of the cosines and sines $\cos(\omega_k t)$, $\sin(\omega_k t)$, $k=6,7$. Explicit expressions in the case of a DART-like perturbation are again deferred to Subsection \ref{subsec:dartlin}.

\section{Nonlinear perturbation theory}\label{sec:CanonicalTheory}
The linear theory developed in the previous section yields a good first approximation to the librational dynamics of the secondary. However, as shown by numerical examples in the next section, the linear theory fails to encapsulate several relevant effects on the dynamics, which manifest themselves when i) the amplitudes of librations are large, and/or ii) the system approaches a \textit{resonance condition} between the fundamental frequencies of the system. In the present section, we develop a nonlinear perturbation theory based on a Hamiltonian normal form. Our treatment is similar as in \cite{Michelangelo2024}, but extended to three dimensions. In particular, our series are based on a non-resonant normal form, which yields some approximate integrals of motion. Far from resonances, we check the accuracy of the method by quantifying the variations of these integrals. Close to resonances, the method partially fails, but we still find that the series are able to partially describe a subset of librational degrees of freedom. We attribute this partial (near-)integrability to the multiplicity of the resonance close to which the initial conditions are taken, as shown by numerical examples in Section \ref{sec:dart}. 

\subsection{Hamiltonian expansion}\label{subsec:HamExp}
As in our previous work \citep{Michelangelo2024}, the point of departure of our non-linear perturbation theory is an expansion of the Hamiltonian (\ref{Hfullform}) around the SSES of two spheres (so, different from before) of masses $M_1$, $M_2$ in Keplerian circular relative orbit, defined as:
\begin{align}\label{TwoSpheresEquil}
r(t)&=r^\ast,~
&p_r(t)&=p_r^\ast=0 
\nonumber\\
\theta(t)&=\theta(t_0)+\nu_{\theta}^\ast(t-t_0),
&p_\theta(t)&=p_\theta^\ast=p_{\phi_1}^\ast + p_{\phi_2}^\ast + mr^{\ast 2}\nu_{\theta}^{\ast}
\nonumber\\
z(t)&=z^\ast=0, 
&p_z(t)&=p_z^\ast=0
\nonumber\\
\theta_{1x}(t)&=\theta_{1x}^\ast=0, 
&p_{\theta_{1x}}(t)&=p_{\theta_{1x}}^\ast=0
\\
\theta_{1y}(t)&=\theta_{1y}^\ast=0, 
&p_{\theta_{1y}}(t)&=p_{\theta_{1y}}^\ast=0
\nonumber\\
\phi_1(t)&=\phi_1(t_0) + (\nu_1-\nu_\theta^\ast)(t-t_0), 
&p_{\phi_1}(t)&=p_{\phi_1}=\nu_1 I_{1z}
\nonumber\\
\theta_{2x}(t)&=\theta_{2x}^\ast=0, 
&p_{\theta_{2x}}(t)&=p_{\theta_{2x}}^\ast=0
\nonumber\\
\theta_{2y}(t)&=\theta_{2y}^\ast=0, 
&p_{\theta_{2y}}(t)&=p_{\theta_{2y}}^\ast=0
\nonumber\\
\phi_2(t)&=\phi_2^\ast=0, 
&p_{\phi_2}(t)&=p_{\phi_2}^\ast=\nu_{\theta}^{\ast} I_{2z}~,
\nonumber
\end{align}
where $\nu_1 \equiv \nu_{1,obs} = \dot{\theta}_{1z}(0)$ and $\nu_{\theta}^{\ast} = \sqrt{G(M_1+M_2)/r^{\ast 3}}$ is the Keplerian orbital frequency. We now perform an expansion of the full (non-averaged) Hamiltonian (\ref{Hfullform}) up to a chosen truncation order $N+2$, where $N \geq 1$, around the above SSES, suitable for the computation of a normal form. To this end, consider first the shift-of-center canonical transformation of 1) all \textit{librational} variables
\begin{equation}\label{CanDisturb}
\begin{aligned}
    r                 &= r^{\ast} + \varepsilon\delta r & \quad p_r             &= \varepsilon\delta p_r                            \\
    z                 &=  \varepsilon\delta z           & \quad p_z             &= \varepsilon\delta p_z                            \\
    \theta_{1x}       &= \varepsilon\delta \theta_{1x}  & \quad p_{\theta_{1x}} &= \varepsilon\delta p_{\theta_{1x}}                \\
    \theta_{1y}       &= \varepsilon\delta \theta_{1y}  & \quad p_{\theta_{1y}} &= \varepsilon\delta p_{\theta_{1y}}                \\
    \theta_{2x}       &= \varepsilon\delta \theta_{2x}  & \quad p_{\theta_{2x}} &= \varepsilon\delta p_{\theta_{2x}}                \\
    \theta_{2y}       &= \varepsilon\delta \theta_{2y}  & \quad p_{\theta_{2y}} &= \varepsilon\delta p_{\theta_{2y}}                \\
    \phi_{2}          &= \varepsilon\delta \phi_{2}     & \quad p_{\phi_{2}}    &= p_{\phi_2}^\ast + \varepsilon\delta p_{\phi_{2}} \\
\end{aligned} 
\end{equation}
and 2) all \textit{rotational} variables
\begin{equation}
p_\theta = p_\theta^{\ast} + \varepsilon^2\delta p_\theta~~~~~~p_{\phi_1} = p_{\phi_1}^{\ast} + \varepsilon^2\delta p_{\phi_1}~~.
\end{equation}
The rotational variables are scaled with $\varepsilon^2$ to reflect their larger characteristic magnitude compared to the librational ones. Let 
\begin{equation}\label{NonLinDeltaVars}
\resizebox{12.0cm}{!}{$
\boldsymbol{\zeta} = \big(\delta r,
                        \delta z,
                        \delta \theta_{1x},
                        \delta\theta_{1y},
                        \delta \theta_{2x},
                        \delta \theta_{2y},
                        \delta \phi_2,
                        \delta p_r,
                        \delta p_z,
                        \delta p_{\theta_{1x}},
                        \delta p_{\theta_{1y}},
                        \delta p_{\theta_{2x}},
                        \delta p_{\theta_{2y}},
                        \delta p_{\phi_2}\big)$}
\end{equation}
be the vector of all new librational variables. The Hamiltonian then takes the form $H=H(\boldsymbol{\zeta},\phi_1,\delta p_{\phi_1},\delta p_\theta)$. Through the use the `book-keeping' assignment (\cite{Efthymiopoulos2011}), we separate the above Hamiltonian in powers of the `book-keeping symbol' $\varepsilon$, whose numerical value is equal to $\varepsilon=1$. The expansion then, obtains the form

\begin{equation}\label{HfullExpanded}
H = \sum_{j=0}^{N+2} \varepsilon^j H_j(\boldsymbol{\zeta},\phi_1, \delta p_{\phi_1} \delta p_\theta)~~~.
\end{equation}

\noindent
Because $\varepsilon = 1$, we can freely lower/raise the exponents of $\varepsilon$ at any term of the Hamiltonian. In our case we want to isolate the quadratic part of $H$ in order to enable diagonalization. Therefore we lower the book-keeping exponents in the Hamiltonian (\ref{HfullExpanded}) by two, but we do so only for $j \geq 2$ and then apply the substitution

\begin{equation}\label{TrigPush}
\begin{aligned} 
\cos{(k\phi_1)} & \leftarrow \varepsilon \cos{(k\phi_1)} \\
\sin{(k\phi_1)} & \leftarrow \varepsilon \sin{(k\phi_1)}~~,
\end{aligned} 
\end{equation}

\noindent
where $k \in \mathbb{Z}$. The Hamiltonian then becomes

\begin{equation}\label{HfullExpandedTrigPushed}
H = Z_0(\boldsymbol{\zeta}, \delta p_{\phi_1}\delta p_\theta) + \sum_{j=1}^{N+1} \varepsilon^j H_{j+2}(\boldsymbol{\zeta},\phi_1, \delta p_{\phi_1} \delta p_\theta)~~~,
\end{equation}

\noindent
where $Z_0$ is a sign-definite quadratic expression in the variables $\boldsymbol{\zeta}$. The exact form of the function $Z_0$ is given in Appendix \ref{app:Z0}. Physically, this function represents the integrable dynamics of two rotators (variables $(\theta, p_\theta))$, $(\phi_1,p_{\phi_1})$ $\times$ seven coupled linear oscillators (variables $\boldsymbol{\zeta}$). Furthermore, by a diagonalizing canonical transformation we switch from the variables $\boldsymbol{\zeta}$ to Birkhoff complex canonical variables $(Q_j,P_j)$, $j = 1,...,7$. The transformation is given by
\begin{equation}\label{delta2QP57}
\begin{bmatrix}
\delta r \\
\delta \phi_2 \\
\delta p_r \\
\delta p_{\phi_2}
\end{bmatrix} =
\boldsymbol{A}_{4\text{x}4}
\begin{bmatrix}
Q_5 \\
Q_7 \\
P_5 \\
P_7
\end{bmatrix}
\end{equation}
\begin{equation}\label{delta2QP346}
\begin{bmatrix}
\delta z \\
\delta \theta_{2x} \\
\delta \theta_{2y} \\
\delta p_z \\
\delta p_{\theta_{2x}} \\
\delta p_{\theta_{2y}} \\
\end{bmatrix} =
\boldsymbol{B}_{6\text{x}6}
\begin{bmatrix}
Q_3 \\
Q_4 \\
Q_6 \\
P_3 \\
P_4 \\
P_6 \\
\end{bmatrix}
\end{equation}
\begin{equation}\label{delta2QP12}
\begin{bmatrix}
\delta \theta_{1x} \\
\delta \theta_{1y} \\
\delta p_{\theta_{1x}} \\
\delta p_{\theta_{1y}} \\
\end{bmatrix} =
\boldsymbol{C}_{4\text{x}4}
\begin{bmatrix}
Q_1 \\
Q_2 \\
P_1 \\
P_2 \\
\end{bmatrix}
\end{equation}
where the matrices $\boldsymbol{A}_{4\text{x}4}$, $\boldsymbol{B}_{6\text{x}6}$, $\boldsymbol{C}_{4\text{x}4}$ contain as columns the eigenvectors of the matrix $J_7\nabla_{\boldsymbol{\zeta}}Z_0$, where $J_7$ is the 14x14 standard symplectic matrix. By substituting equations (\ref{delta2QP57})-(\ref{delta2QP12}) into (\ref{HfullExpandedTrigPushed}), we finally obtain the Hamiltonian
\begin{equation}\label{HQPInit}
    H \equiv H^{(0)} = Z_0(Q_k,P_k,\phi_1,\delta p_{\phi_1},\delta p_\theta) + \sum_{j=1}^{N+1} \epsilon^j h_j(Q_k,P_k,\phi_1,\delta p_{\phi_1},\delta p_\theta),~~
\end{equation}
for $k=1,\ldots,7$, where (omitting the constant term)
\begin{equation}\label{Z0Init}
    Z_0 = \nu_{\theta}^{\ast} \delta p_\theta + (\nu_1 - \nu_\theta^{\ast}) \delta p_{\phi_1} - \sum_{j=1}^{7} i\omega_{j,\text{ker}} Q_jP_j~.
\end{equation}
The quantity $Z_0$ is called the `kernel' of the Hamiltonian (\ref{HQPInit}) and contains the fundamental information regarding the system's decomposition into 7 oscillators with (kernel) frequencies $\omega_{j,\text{ker}}$ and 2 rotators with frequencies $\nu_\theta^{\ast}$ and $\nu_1-\nu_\theta^{\ast}$. 

\subsection{Hamiltonian normal form and solution}\label{subsec:HamNormForm}
To obtain a nonlinear theory of analytical solutions of the equations of motion under the Hamiltonian (\ref{HQPInit}), we apply Birkhoff canonical normalization, based on the Lie series approach \citep{Hori1966, Deprit1969}, carrying out a recursive normalization process for a maximum of $N$ iterations (described in detail by \cite{Efthymiopoulos2011}). In summary, after $n$ near-to-identity canonical transformations, the Hamiltonian in new canonical variables reads:
\begin{equation}\label{hamn}
H^{(n)}=Z_0+\varepsilon Z_1+\ldots+\varepsilon^n Z_n +\sum_{s=n+1}^N\varepsilon^s h^{(n)}_s~~.
\end{equation}
Spliting the term $h^{(n)}_{n+1}$ in a normal form part $Z_{n+1}$ and a part to normalize $\tilde{h}^{(n)}_{n+1}$, the homological equation
\begin{equation}\label{HomoloEq}
  \{Z_0, \chi_{n+1}\} + \varepsilon^{n+1}\tilde{h}_{n+1}^{(n)} = 0
\end{equation}
is then solved (where $\{,\}$ is the Poisson bracket binary operator), allowing to determine the Lie generating function $\chi_{n+1}$. In our case, setting
\begin{equation}\label{htilde}
\tilde{h}^{(n)}_{n+1} = \sum_{|\boldsymbol{k}| \neq 0}^{}\tilde{h}^{(n)}_{n+1,\boldsymbol{k}}(\delta p_{\phi_1})
Q_1^{k_1}\cdots Q_7^{k_7}P_1^{\ell_1}\cdots P_7^{\ell_7}e^{i k_{\phi_1}\phi_1}
\end{equation}
with $|\boldsymbol{k}|=\left(\sum_{j=1,\ldots 7}|k_j-\ell_j|\right)+|k_{\phi_1}|$, the solution of the homological equation (\ref{HomoloEq}) reads
\begin{equation}\label{EvalChi}
\chi_{n+1} = \varepsilon^{n+1}\sum_{|\boldsymbol{k}| \neq 0}^{}
\frac{\tilde{h}^{(n)}_{n+1,\boldsymbol{k}}(\delta p_{\phi_1})
Q_1^{k_1}\cdots Q_7^{k_7}P_1^{\ell_1}\cdots P_7^{\ell_7}e^{i k_{\phi_1}\phi_1}}
{i(\omega_1(k_1-\ell_1) + \cdots + \omega_7(k_7-\ell_7) +\omega_{\phi_1} k_{\phi_1})}~.
\end{equation}
The Hamiltonian in the next step is then expressed as
\begin{equation}\label{HStepN}
H^{(n+1)} = \exp(\mathcal{L}_{\chi_{n+1}} )H^{(n)}~~,
\end{equation}
truncated at the maximum order (in book-keeping) equal to $N$. 

The Hamiltonian $H^{(n+1)}$ depends on the new canonical variables $(Q^{(n+1)}_j,P^{(n+1)}_j)$, $j=1,...,7$, as well as on the canonical pair $(\phi_1^{(n+1)}\equiv\phi_1,\delta p_{\phi_1}^{(n+1)})$ and on the (constant) $\delta p_\theta$. By the structure of the generating function (\ref{EvalChi}), the new variables in all nine degrees of freedom are near to identity transformations of the original variables, except for the angles $\theta,\phi_1$, which do not transform, and the momentum $\delta p_\theta$, which is an invariant of the Hamiltonian flow. Thus, after $N$ normalization steps, the composition of all near-to-identity transformations leads to the new canonical variables
\begin{equation}\label{nonlintra}
\begin{aligned}
Q_j^{(N)}&=Q_j+\Delta Q_j^{(N)}(\boldsymbol{\zeta},\phi_1,\delta p_{\phi_1}),~\\
P_j^{(N)}&=P_j+\Delta P_j^{(N)}(\boldsymbol{\zeta},\phi_1,\delta p_{\phi_1}),~\\
\delta p_{\phi_1}^{(N)}&=\delta p_{\phi_1}+
\Delta p_{\phi_1}^{(N)}(\boldsymbol{\zeta},\phi_1,\delta p_{\phi_1}),~\\
\phi_1^{(N)}&=\phi_1+\Delta \phi_1^{(N)}(\boldsymbol{\zeta},\phi_1,\delta p_{\phi_1}),~\\
\theta^{(N)}&=\theta,~~\delta p_{\theta}^{(N)}=\delta p_{\theta}~.
\end{aligned}
\end{equation}
After $N$ normalization steps, we obtain the final normal form Hamiltonian
\begin{equation}\label{ZNormalForm}
Z^{(N)} = \sum_{n_1=0}^{N_1 \leq N}\cdots\sum_{n_7=0}^{N_7 \leq N}\Big[C_{n_1\dots n_7}^{(N)}(\delta p_\theta,\delta p_{\phi_1}^{(N)})
\prod_{j=1}^{7}(Q_j^{(N)}P_j^{(N)})^{n_j}\Big]~~~.
\end{equation}
Since  all dependence on the fast angle $\phi_{1}$ is eliminated in the normal form $Z^{(N)}$, the conjugate momentum $\delta p_{\phi_1}^{(N)}$ becomes an integral of motion in the new canonical variables. Physically, this is equivalent to averaging out the primary's rapid spin from the equations of motion. Nonetheless, the averaging is now achieved through proper  canonical normalization, instead of the non-invertible `scissor' averaging adopted in Eq. (\ref{Havg}). Furthermore, in (\ref{ZNormalForm}) the quantities $C_{n_1\dots n_7}^{(N)}$ are constants that depend on the physical parameters, as well as on the quantities $\delta p_\theta, \delta p_{\phi_1}$, which are integrals of motion. The variables $Q_j^{(N)},P_j^{(N)}$ appear only through their product $Q_j^{(N)}P_j^{(N)}$. Hence, the real quantities 
\begin{equation}\label{integrals}
I_j^{(N)}=iQ_j^{(N)}P_j^{(N)}
\end{equation}
are integrals of motion under the flow generated by $Z^{(N)}$. Finally, the equations of motion, corresponding to the normal form Hamiltonian (\ref{ZNormalForm}) are
\begin{equation}\label{ZNormalFormEquationsDeriv}
\begin{aligned}
  \dot{Q}_j^{(N)} &= \frac{\partial Z^{(N)}}{\partial P_j^{(N)}} = i\omega_{jc}Q_j^{(N)} \\
  \dot{P}_j^{(N)} &= -\frac{\partial Z^{(N)}}{\partial Q_j^{(N)}} = -i\omega_{jc}P_j^{(N)}\\
  \dot{\theta} &= \frac{\partial Z^{(N)}}{\partial \delta p_\theta} = \omega_{\theta c}\\
  \dot{\phi}_1 &= \frac{\partial Z^{(N)}}{\partial \delta p_{\phi_1}} = \omega_{\phi_1 c}~~~,
\end{aligned}
\end{equation}
with $\omega_{jc},\omega_{\theta c},\omega_{\phi_1 c}$ being the canonical frequencies, containing only products of the form $Q_j^{(N)}P_j^{(N)}$. As a result, equations (\ref{ZNormalFormEquationsDeriv}) are directly integrable and yield the solution in time in the new canonical coordinates, namely
\begin{equation}\label{ZNormalFormSol}
\begin{aligned}
  Q_j^{(N)}(t)    &= Q_j^{(N)}(0)e^{ i\omega_{jc}t }\\
  P_j^{(N)}(t)    &= P_j^{(N)}(0)e^{-i\omega_{jc}t }\\
  \theta(t) &= \theta(0) + \omega_{\theta c}t \\
  \phi_1^{(N)}(t) &= \phi_1^{(N)}(0) + \omega_{\phi_1 c}t ~~~,
\end{aligned} 
\end{equation}
where $( Q_j^{(N)}(0), P_j^{(N)}(0), \theta(0), \phi_1^{(N)}(0) )$ are the initial conditions of the problem, computed from the initial conditions in the original variables through Eqs. (\ref{delta2QP57}) - (\ref{delta2QP12}) and (\ref{nonlintra}). Using (\ref{ZNormalFormSol}), and by the inverse of the transformations (\ref{nonlintra}) and (\ref{delta2QP57}) - (\ref{delta2QP12}), we finally retrieve the analytical solution in time expressed in our original variables (see Subsection \ref{subsec:dartnonlin}).

\section{Application to the post impact state of the 65803 Didymos system}
\label{sec:dart}

\subsection{Oblique impact and new equilibrium}
\label{subsec:ImpactAndNewEq}
\begin{figure}[H]
  \centering
    \includegraphics[scale=0.3]{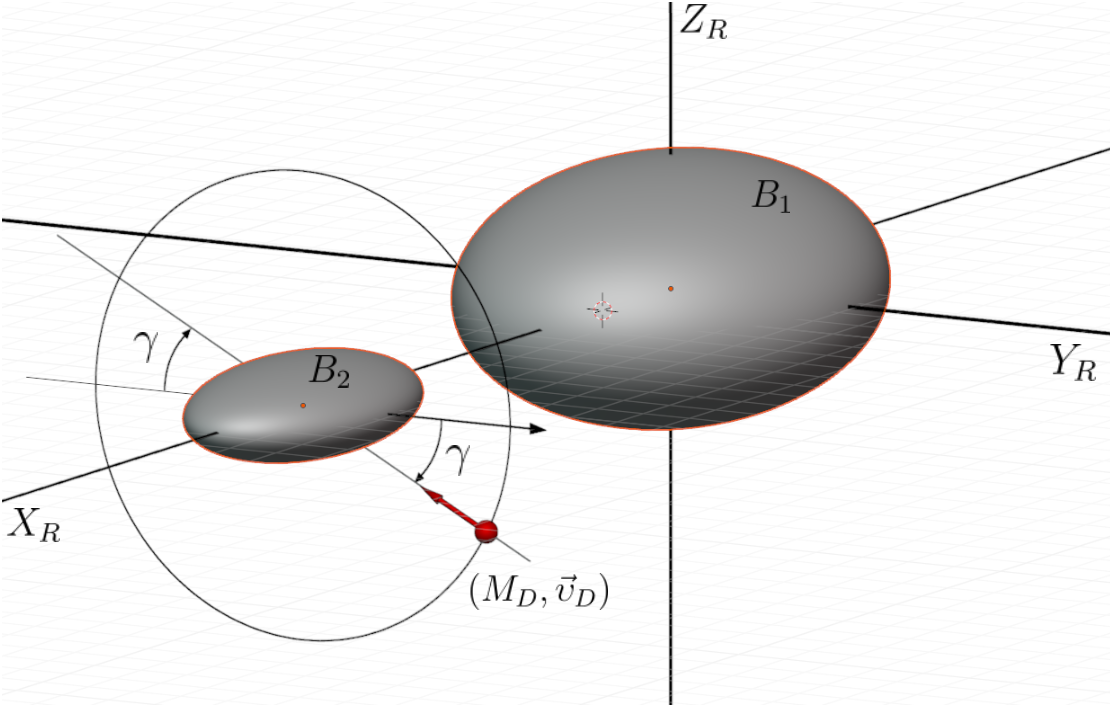}
    \caption{Schematic representation of an oblique collision between the impactor and the secondary asteroid. In the chosen coordinate system, the impactor's velocity $\vec{\upsilon}_D$ is decomposed into a tangential component (lying in the $X_R-Y_R$ plane) and a normal component (along the $Z_R$ axis), with an inclination $\gamma$. During the collision it is assumed zero net torque on the secondary.}
    \label{fig:DARThit}
\end{figure}

We assume an oblique DART-like impact upon the secondary asteroid (Dimorphos) of the 65803 Didymos system, as shown in Figure \ref{fig:DARThit}. The impactor’s velocity $\vec{\upsilon}_D$ has two components in the relative coordinate system: the first is tangent to the orbital plane, parallel to the velocity vector of the secondary asteroid, aiming at its center of mass. The second is perpendicular to the orbital plane, also aiming at the secondary's center of mass. This second component is quantified by the inclination $\gamma$ between $\vec{\upsilon}_D$ and the initial orbital plane. During the impact, it is assumed that no direct torque is induced upon the secondary asteroid. The resulting change in the velocity $\Delta \vec{\upsilon}$ imparted to the secondary can be described using the momentum enhancement factor $\beta$, via the equation \citep{Rivkin2021}
\begin{equation}
    \Delta \vec{\upsilon} = \frac{\beta M_D\vec{\upsilon}_D}{M_2}, \hspace{0.4cm} \beta \geq 1
\end{equation}
where, in cylindrical coordinates
\begin{equation}
    \vec{\upsilon}_D = -(\upsilon_D\cos{\gamma}) \hat{\theta} + (\upsilon_D\sin{\gamma}) \hat{z}~~~.
\end{equation}

Once the collision occurs, the binary moves away from its pre‐impact single-synchronous state. The momentum transfer from the impactor, which is oblique in this case, modifies both the integral of motion $p_\theta$ and the initial momentum $p_z$. Setting $t_0=0$ at the time of the impact in the approximation of the averaged Hamiltonian (\ref{Havg}), the post‐impact state has initial conditions such that all the generalized position variables remain identical to those given in Eq. (\ref{sses}) for $t=t_0=0$ and $r_{eq}$ computed by Eq. (\ref{reqave}), while the initial velocities and momenta are
\begin{equation}\label{PostimpactICVel}
\begin{gathered}
   \dot{r}(0) = 0, \hspace{0.2cm} \dot{\theta}(0) = \dot{\theta}_{imp} = \nu_{\theta,obs} - \frac{M_D\upsilon_D\beta\cos{\gamma}}{r_{eq}M_2}, \hspace{0.2cm} \dot{z}(0) = \dot{z}_{imp} = \frac{M_D\upsilon_D\beta\sin{\gamma}}{M_2} \\
   \dot{\theta}_{1x}(0) = 0, \hspace{0.2cm} \dot{\theta}_{1y}(0) = 0, \hspace{0.2cm} \dot{\phi}_{1}(0) =  \dot{\phi}_{1,imp} = \nu_{1,obs} - \dot{\theta}_{imp} \\
   \dot{\theta}_{2x}(0) = 0, \hspace{0.2cm} \dot{\theta}_{2y}(0) = 0, \hspace{0.2cm} \dot{\phi}_{2}(0) =  \dot{\phi}_{2,imp} = \frac{M_D\upsilon_D\beta\cos{\gamma}}{r_{eq}M_2}
\end{gathered}
\end{equation}
\begin{equation}\label{AvgPostmpactICMom}
\begin{gathered}
     p_r(0) = 0, \hspace{0.2cm} p_{\theta}(0) = p_{\theta eq} - \frac{m r_{eq}M_D\upsilon_D\beta\cos{\gamma}}{M_2}, \hspace{0.2cm} p_z(0) = m\dot{z}_{imp} \\
    p_{\theta_{1x}}(0) = 0, \hspace{0.2cm} p_{\theta_{1y}}(0) = 0, \hspace{0.2cm} p_{\phi_{1}}(0) = \nu_1I_{1z} \\
    \hspace{0.6cm} p_{\theta_{2x}}(0) = 0, \hspace{0.2cm} p_{\theta_{2y}}(0) = 0, \hspace{0.2cm} p_{\phi_{2}}(0) = \nu_{\theta,obs}I_{2z}~.
\end{gathered}
\end{equation}
To the new (post-impact) value of the integral $p_{\theta}'=p_\theta(0)$, corresponds a new theoretical SSES, given again by Eqs. ({\ref{sses}}), but with a new equilibrium radius $r_{eq}'$ and orbital frequency $\nu_{\theta}'$ given by 
\begin{equation}\label{NewEqState}
\resizebox{12.0cm}{!}{$
\begin{split}
 \dot{\theta}_{eq}' & = \nu_{\theta}'=
 \frac{(I_{2z} + mr_{eq}^2)\dot{\theta}_{eq}}{I_{2z} + mr^{2'}_{eq}} -
 \frac{mr_{eq}M_D\upsilon_D\beta\cos{\gamma}}{M_2(I_{2z} + mr^{2'}_{eq})} \\
 & = \sqrt{\frac{G(M_1+M_2)}{r_{eq}^{3'}}\bigg[
    1 + \frac{3}{2r_{eq}^{2'}} \bigg( \frac{2I_{1z} - I_{1x} - I_{1y}}{2M_1} + \frac{-2I_{2x} + I_{2y} + I_{2z}}{M_2} \bigg)
    \bigg]}~~~.
\end{split}$}
\end{equation}
Solving for $r_{eq}'$, yields the equilibrium state around which the linear theory is to be applied, post-impact. The radius $r_{eq}'$ can be expressed as an analytical function of the physical parameters and of the impact parameter $\beta$. Taylor-expanding Eq. (\ref{NewEqState}) in powers of the small quantity $\delta r = r_{eq} - r_{eq}'$, results in 
\begin{equation}\label{reqnewLin}
r_{eq}' = r_{eq} + C_{r_{eq}}\beta + O(\delta r^2)~~~,
\end{equation}
where
\begin{equation}\label{reqnewLinCoeff}
\begin{gathered}
    C_{r_{eq}} =
\frac{ 2mr_{eq}^{9/2}\sqrt{6C_I+4r_{eq}^2}M_D\upsilon_D\cos{\gamma} }
     { M_2\sqrt{G(M_1+M_2)}\big( 6I_{2z}r_{eq}^2 - 2mr_{eq}^4 +3C_I(5I_{2z} + mr_{eq}^2) \big)} \\
    C_I = \frac{2I_{1z} -I_{1x} - I_{1y}}{2M_1} + \frac{-2I_{2x} + I_{2y} + I_{2z}}{M_2}~.
\end{gathered} 
\end{equation}
In the following, we adopt two distinct physical configurations for the Didymos-Dimorphos system, hereafter referred to as `parameter set 1' and `parameter set 2'. Each set specifies a different combination of the system's physical parameters, namely the individual masses $M_i$ and the principal moments of inertia $I_{ix},I_{iy},I_{iz}$, $i=1,2$. The adopted values are listed in Table \ref{tab:ParamSets}.
\begin{table}[htbp]
\centering
\begin{tabular}{lcc}
\hline
\textbf{Parameter} & \textbf{Parameter set 1} & \textbf{Parameter set 2} \\
\hline
$M_1$ [kg$^\ast$] & 5.150418  &  5.159394  \\
$M_2$ [kg$^\ast$] & 0.0392693 &  0.0287669 \\
$I_{1x}$ [kg$^\ast \cdot$km$^2$] & 0.260108 & 0.260562 \\
$I_{1y}$ [kg$^\ast \cdot$km$^2$] & 0.267618 & 0.268085 \\
$I_{1z}$ [kg$^\ast \cdot$km$^2$] & 0.337960 & 0.338550 \\
$I_{2x}$ [kg$^\ast \cdot$km$^2$] & $8.20454 \cdot 10^{-5}$  & $4.33115 \cdot 10^{-5}$ \\
$I_{2y}$ [kg$^\ast \cdot$km$^2$] & $8.88782 \cdot 10^{-5}$  & $6.07052 \cdot 10^{-5}$ \\
$I_{2z}$ [kg$^\ast \cdot$km$^2$] & $1.18990 \cdot 10^{-4}$  & $6.59724 \cdot 10^{-5}$ \\
\hline
\end{tabular}
\caption{Physical parameters adopted for each configuration of the Didymos-Dimorphos system. 1 [kg$^{\ast}] = 10^{11}$[kg].}
\label{tab:ParamSets}
\end{table}
Our motivation for checking two parameter regimes stems from the current uncertainty surrounding the precise post-impact configuration of the system. Observational and dynamical analysis indicate that the DART impact likely induced reshaping of Dimorphos \citep{Naidu2024}, altering its inertia tensor and orbital response (`parameter set 2'). At the same time, pre-impact determinations of Dimorphos' ellipsoidal dimensions, as obtained from DRACO imaging \citep{Daly2024}, provide a natural baseline for `parameter set 1'. The adopted masses and inertia values are therefore constructed with nominal observational constraints on bulk densities and volumes, such that the ellipsoid semi-axes and the resulting inertia tensors are compatible with the reported figures. Importantly, rather than inferring the system's total mass from a purely Keplerian fit to the binary orbit, we acknowledge that the Didymos - Dimorphos system is not strictly Keplerian and therefore, the orbital mean motion is governed by Eq. (\ref{reqave}).

Notably, for a fixed value of momentum enhancement factor $\beta$, we find that different choices of the system's physical parameters, each consistent with current observational uncertainties, can lead to qualitatively different dynamical outcomes. In particular, while some parameter values yield regular, quasi-periodic evolution following the impact, others give rise to chaotic behavior under the same perturbation. This sensitivity underscores the importance of exploring multiple plausible configurations, as even slight variations in the parameters, such as in $I_{2x}$, may suffice to induce a transition from order to chaos. 

\subsection{Comparison of the linear and non-linear theories}
\label{subsec:linnonlincomp}

\subsubsection{Solution of the linear theory}
\label{subsec:dartlin}
From the particular DART-like impact initial conditions of Eq. (\ref{AvgPostmpactICMom}), we have $\delta r(0)=r_{eq}-r_{eq}'$, $\delta p_r=0$, $\delta\phi_2=0$, $\delta p_{\phi_2}(0)=(\nu_\theta-\nu_\theta')I_{2z}$. Then, we readily find that only some of the sine and cosine coefficients of the general solution (\ref{P1LinVargen}) of the 4x4 planar linear subsystem acquire non-zero values. The solution (\ref{P1LinVargen}) then takes the form
\begin{equation}\label{P1LinVar}
\begin{aligned}
\delta r(t) & = A_r(\omega_1,\omega_2)\cos{(\omega_1 t)} + A_r(\omega_2,\omega_1)\cos{(\omega_2 t)} \\
\delta \phi_2(t) & =
A_{\phi_2}(\omega_1,\omega_2)\sin{(\omega_1t)} + A_{\phi_2}(\omega_2,\omega_1)\sin{(\omega_2t)}  \\ 
\delta p_r(t) & =
A_{p_r}(\omega_1,\omega_2)\sin{(\omega_1 t)} + A_{p_r}(\omega_2,\omega_1)\sin{(\omega_2 t)} \\ 
\delta p_{\phi_2}(t) & =
A_{p_{\phi_2}}(\omega_1,\omega_2)\cos{(\omega_1t)} + A_{p_{\phi_2}}(\omega_2,\omega_1)\cos{(\omega_2t)}~~~,
\end{aligned}
\end{equation}
with coefficients $A_r$,$A_{\phi_2}$,$A_{p_r}$,$A_{p_{\phi_2}}$ as given in the Appendix \ref{app:SysLinSol}. Note that these coefficients depend on the differences $\delta r(0)=r_{eq}-r_{eq}'$, and $\nu_\theta-\nu_\theta'$. Using Eq. (\ref{reqnewLin}), we can then obtain a linear solution of the form 
\begin{equation}\label{LinSolPlan}
\begin{gathered}   
    r_{lin}(t;\beta) = r_{eq}'(\beta) + \delta r(t;\beta) \\
    \phi_{2,lin}(t;\beta) = \delta \phi_2(t;\beta) \\
    p_{r,lin}(t;\beta) = \delta p_r(t;\beta) \\
    p_{{\phi_2},lin}(t;\beta) = p_{\phi_2}(r_{eq}'(\beta)) + \delta p_{\phi_2}(t;\beta)~~~.
\end{gathered} 
\end{equation}
i.e., depending explicitly on the impact parameter $\beta$.

Working in the same way, we obtain the linear solution of the 6x6 system (\ref{LinSys2}) for a DART-like oblique impact as
\begin{equation}\label{LinSolz}
\begin{aligned}   
    z_{lin}(t) & = \delta z(t) \\
    \theta_{2x,lin}(t) & = \delta \theta_{2x}(t) \\
    \theta_{2y,lin}(t) & = \delta \theta_{2y}(t) \\
    p_{z,lin}(t) & = \delta p_z(t) \\
    p_{\theta_{2x,lin}}(t) & = \delta p_{\theta_{2x}}(t) \\
    p_{\theta_{2y,lin}}(t) & = \delta p_{\theta_{2y}}(t)~~,
\end{aligned}
\end{equation}
with
\begin{equation}\label{P2LinVar}
\begin{aligned}
\delta z(t) & = A_z(\omega_3, \omega_4, \omega_5)\sin{(\omega_3 t)} \\
            & + A_z(\omega_4, \omega_5, \omega_3)\sin{(\omega_4 t)} \\
            & + A_z(\omega_5, \omega_3, \omega_4)\sin{(\omega_5 t)}
\end{aligned}
\quad\quad\quad
\begin{aligned}
\delta p_z(t) & = A_{p_z}(\omega_3, \omega_4, \omega_5)\cos{(\omega_3 t)} \\
              & + A_{p_z}(\omega_4, \omega_3, \omega_5)\cos{(\omega_4 t)} \\
              & + A_{p_z}(\omega_5, \omega_3, \omega_4)\cos{(\omega_5 t)}
\end{aligned}
\end{equation}
\begin{equation}
\begin{aligned}
\delta \theta_{2x}(t) & = A_{\theta_{2x}}(\omega_4, \omega_3, \omega_5)\cos{(\omega_3 t)} \\
                      & + A_{\theta_{2x}}(\omega_3, \omega_4, \omega_5)\cos{(\omega_4 t)} \\
                      & + A_{\theta_{2x}}(\omega_4, \omega_5, \omega_3)\cos{(\omega_5 t)}
\end{aligned}
\quad\quad\quad
\begin{aligned}
\delta p_{\theta_{2x}}(t) & = A_{p_{\theta_{2x}}}(\omega_3, \omega_4, \omega_5)\sin{(\omega_3 t)} \\
                          & + A_{p_{\theta_{2x}}}(\omega_4, \omega_3, \omega_5)\sin{(\omega_4 t)} \\
                          & + A_{p_{\theta_{2x}}}(\omega_5, \omega_3, \omega_4)\sin{(\omega_5 t)}
\end{aligned}
\end{equation}
\begin{equation}
\begin{aligned}
\delta \theta_{2y}(t) & = A_{\theta_{2y}}(\omega_3, \omega_4, \omega_5)\sin{(\omega_3 t)} \\
                      & + A_{\theta_{2y}}(\omega_4, \omega_3, \omega_5)\sin{(\omega_4 t)} \\
                      & + A_{\theta_{2y}}(\omega_5, \omega_3, \omega_4)\sin{(\omega_5 t)}
\end{aligned}
\quad\quad\quad
\begin{aligned}
\delta p_{\theta_{2y}}(t) & = A_{p_{\theta_{2y}}}(\omega_3, \omega_4, \omega_5)\cos{(\omega_3 t)} \\
                          & + A_{p_{\theta_{2y}}}(\omega_4, \omega_3, \omega_5)\cos{(\omega_4 t)} \\
                          & + A_{p_{\theta_{2y}}}(\omega_5, \omega_3, \omega_4)\cos{(\omega_5 t)}~,
\end{aligned}
\end{equation}
where the linear amplitudes $A_z(\omega_\ell,\omega_m,\omega_n)$, $A_{\theta_{2x}}(\omega_\ell,\omega_m,\omega_n)$, $A_{\theta_{2y}}(\omega_\ell,\omega_m,\omega_n)$, $A_{p_z}(\omega_\ell,\omega_m,\omega_n)$, $A_{p_{\theta_{2x}}}(\omega_\ell,\omega_m,\omega_n)$, $A_{p_{\theta_{2y}}}(\omega_\ell,\omega_m,\omega_n)$ are provided explicitly in the Appendix \ref{app:SysLinSol}. 

Finally, since we assume that the impactor strikes at the secondary asteroid, the primary exhibits no direct initial disturbance in its angular momenta, resulting in $\delta p_{\theta_{1x}}(0) = \delta p_{\theta_{1y}}(0) = 0$. In such a scenario, the corresponding linear solution for the primary's variables $\theta_{1x}(t),\theta_{1y}(t)$ is trivial ($\theta_{1x}(t)=\theta_{1y}(t)=0$). In order to obtain a non-trivial solution for these variables, we then need assign an initial disturbance (offset) in the orientation itself of the primary at the moment of impact, namely $\delta \theta_{1x}(0) = \delta \theta_{1x0} \neq 0$ and $\delta \theta_{1y}(0) = \delta \theta_{1y0} \neq 0$. By doing so, both the angular positions and the momenta get excited, rendering a non-zero linear solution of the 4x4 subsystem (\ref{LinSys3}). We find:
\begin{equation}\label{LinSolth1xy}
\begin{aligned}   
    \theta_{1x,lin}(t) & = \delta \theta_{1x}(t) \\
    \theta_{1y,lin}(t) & = \delta \theta_{1y}(t) \\
    p_{\theta_{1x,lin}}(t) & = \delta p_{\theta_{1x}}(t) \\
    p_{\theta_{1y,lin}}(t) & = \delta p_{\theta_{1y}}(t)~
\end{aligned}
\end{equation}
with
\begin{equation}\label{P3LinVar}
\begin{aligned}
\delta \theta_{1x}(t) & = A_{1_{\theta_{1x}}}(\omega_7, \omega_6)\cos{(\omega_6t)} \\
                      & + A_{1_{\theta_{1x}}}(\omega_6, \omega_7)\cos{(\omega_7t)} \\
                      & + A_{2_{\theta_{1x}}}(\omega_6, \omega_7)\sin{(\omega_6t)} \\
                      & + A_{2_{\theta_{1x}}}(\omega_7, \omega_6)\sin{(\omega_7t)}
\end{aligned}
\quad\quad\quad
\begin{aligned}
\delta p_{\theta_{1x}}(t) & = A_{2p_{\theta_{1x}}}(\omega_6, \omega_7)\cos{(\omega_6t)} \\
                          & + A_{2p_{\theta_{1x}}}(\omega_7, \omega_6)\cos{(\omega_7t)} \\
                          & + A_{1p_{\theta_{1x}}}(\omega_7, \omega_6)\sin{(\omega_6t)} \\
                          & + A_{1p_{\theta_{1x}}}(\omega_6, \omega_7)\sin{(\omega_7t)}
\end{aligned}
\end{equation}
\begin{equation}
\begin{aligned}
\delta \theta_{1y}(t) & = A_{2_{\theta_{1y}}}(\omega_7, \omega_6)\cos{(\omega_6t)} \\
                      & + A_{2_{\theta_{1y}}}(\omega_6, \omega_7)\cos{(\omega_7t)} \\
                      & + A_{1_{\theta_{1y}}}(\omega_7, \omega_6)\sin{(\omega_6t)} \\
                      & + A_{1_{\theta_{1y}}}(\omega_6, \omega_7)\sin{(\omega_7t)}
\end{aligned}
\quad\quad\quad
\begin{aligned}
\delta p_{\theta_{1y}}(t) & = A_{1p_{\theta_{1y}}}(\omega_7, \omega_6)\cos{(\omega_6t)} \\
                          & + A_{1p_{\theta_{1y}}}(\omega_6, \omega_7)\cos{(\omega_7t)} \\
                          & + A_{2p_{\theta_{1y}}}(\omega_6, \omega_7)\sin{(\omega_6t)} \\
                          & + A_{2p_{\theta_{1y}}}(\omega_7, \omega_6)\sin{(\omega_7t)}~.
\end{aligned}
\end{equation}
Again, explicit expressions for the amplitudes $A_{1_{\theta_{1x}}}(\omega_\ell, \omega_m)$, $A_{2_{\theta_{1x}}}(\omega_\ell, \omega_m)$, $A_{1_{\theta_{1y}}}(\omega_\ell, \omega_m)$, $A_{2_{\theta_{1y}}}(\omega_\ell, \omega_m)$, $A_{1p_{\theta_{1x}}}(\omega_\ell, \omega_m)$, $A_{2p_{\theta_{1x}}}(\omega_\ell, \omega_m)$, $A_{1p_{\theta_{1y}}}(\omega_\ell, \omega_m)$ and $A_{2p_{\theta_{1y}}}(\omega_\ell, \omega_m)$ are given in the Appendix \ref{app:SysLinSol}. \\
\\

\begin{figure}
  \centering
    \includegraphics[scale=0.35]{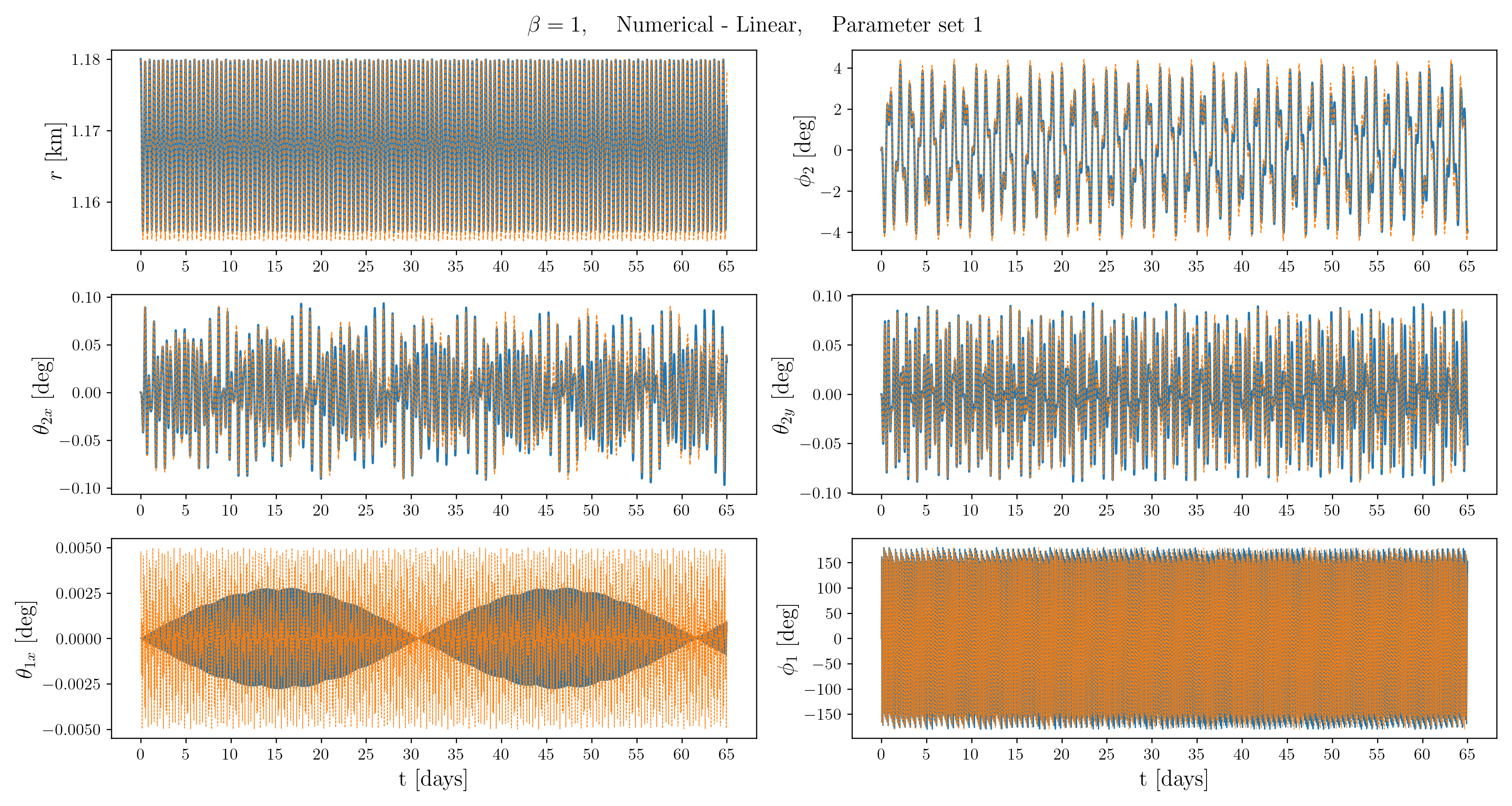}
    \caption{Comparison between the numerical (blue) and the linear (orange) solution for the parameter set 1, post DART impact for $\beta = 1$. The impactor's inclination, relative velocity and mass are $\gamma = 9^o$, $\upsilon_D = 22121.6$ [km/hr] and $M_D = 5.79434 \cdot 10^{-9}$ [kg$^{\ast}$] respectively. The primary's initial orientation is set to $\delta \theta_{1x0} = \delta \theta_{1y0} = 0.0025^o$, corresponding to the maximum amplitude resulted by the numerical solution.}
    \label{fig:orb_num_and_tlin_b1_set1}
\end{figure}
The dynamical evolution of the binary system following the DART impact, for parameter set 1 and momentum enhancement factor $\beta = 1$, is illustrated in Figure \ref{fig:orb_num_and_tlin_b1_set1}. The solid blue curves correspond to the numerical integration of the full, non-averaged Hamiltonian system, whereas the dashed orange curves represent the analytical linearized solution derived from Eqs. (\ref{LinSolPlan}), (\ref{LinSolz}) and (\ref{LinSolth1xy}). As seen in the numerical curves $r(t)$, $\phi_2(t)$, $\theta_{2x}(t)$, $\theta_{2y}(t)$, the post-impact behavior of the system remains regular, characterized by multi-periodic librations and small-amplitude oscillations. Furthermore, the analytical predictions by linear theory closely match the numerical curves. As regards the off-plane motion of the primary, (curve $\theta_{1x}(t)$), since in linear theory the motion in the angles $\theta_{1x}(t)$, $\theta_{2x}(t)$ is decoupled from the rest of the system, the analytical solution is a superposition of linear oscillator modes obtained by setting an arbitrary initial amplitude, and exhibit no coupling with the orbit of the system (see Subsection \ref{subsec:lineqmo}). The numerical curves, instead, show this coupling which is through nonlinear terms in the equations of motion. To compare the linear with nonlinear modes behavior, in the curves shown in Fig. \ref{fig:orb_num_and_tlin_b1_set1} we introduce an ad hoc initial orientation  $\delta \theta_{1x}(0) = \delta \theta_{1y}(0) = 0.0025^\circ$, corresponding to the maximum amplitude of libration attained by the numerical solution. We then see that the linear solution exhibits a fast oscillation with frequency close to the fast frequency of the numerical solution, but fails to reproduce the slow-frequency modulation of the latter solution. 

\begin{figure}
  \centering
    \includegraphics[scale=0.35]{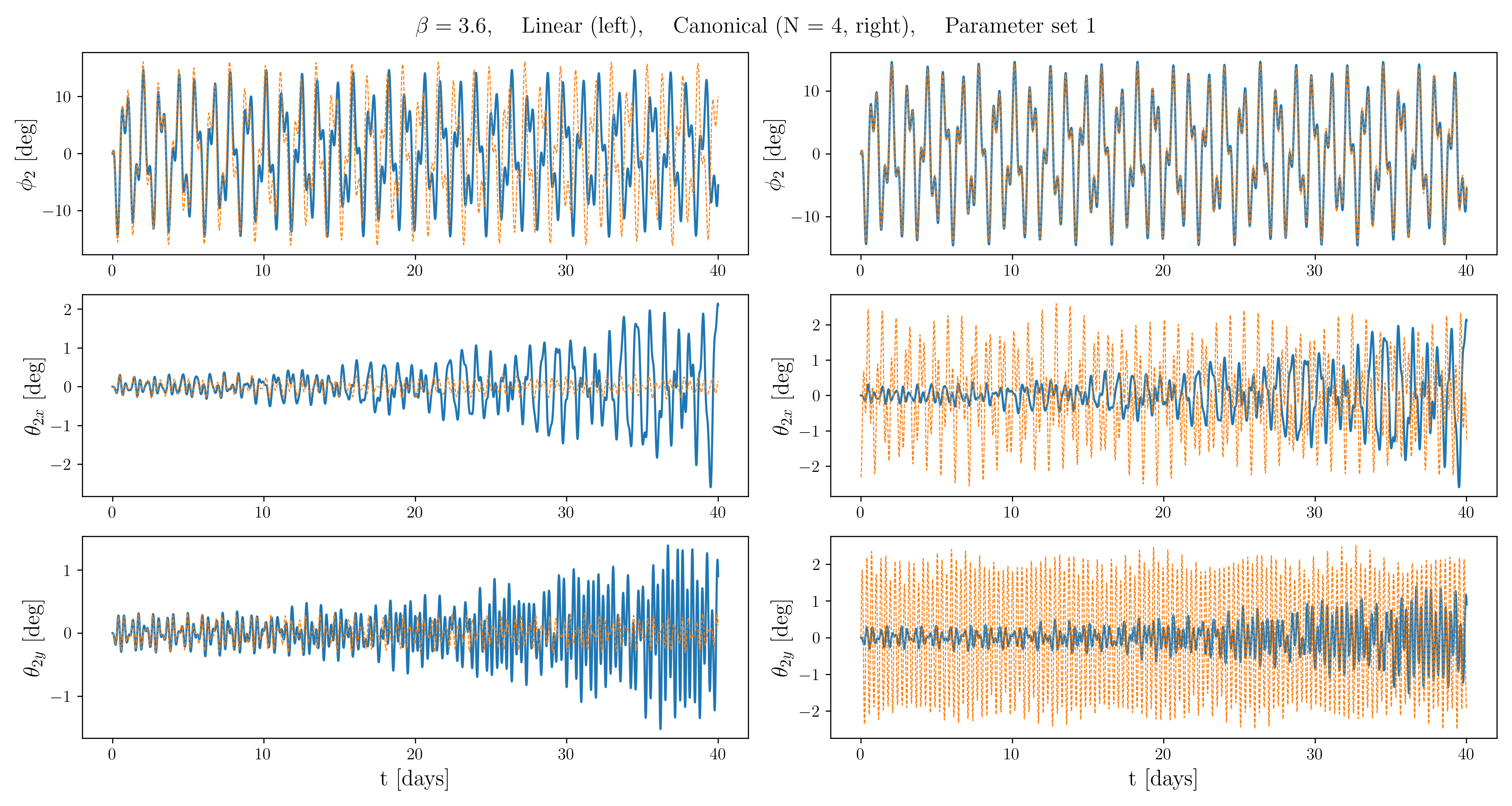}
    \caption{Left: comparison between the numerical (blue) and the linear (orange) solution for $\beta = 3.6$. Right: comparison of the numerical solution with the nonlinear solution, truncated at order $N=4$.}
    \label{fig:orb_num_and_tlin_and_can_b36_set1_short}
\end{figure}
\begin{figure}
  \centering
    \includegraphics[scale=0.35]{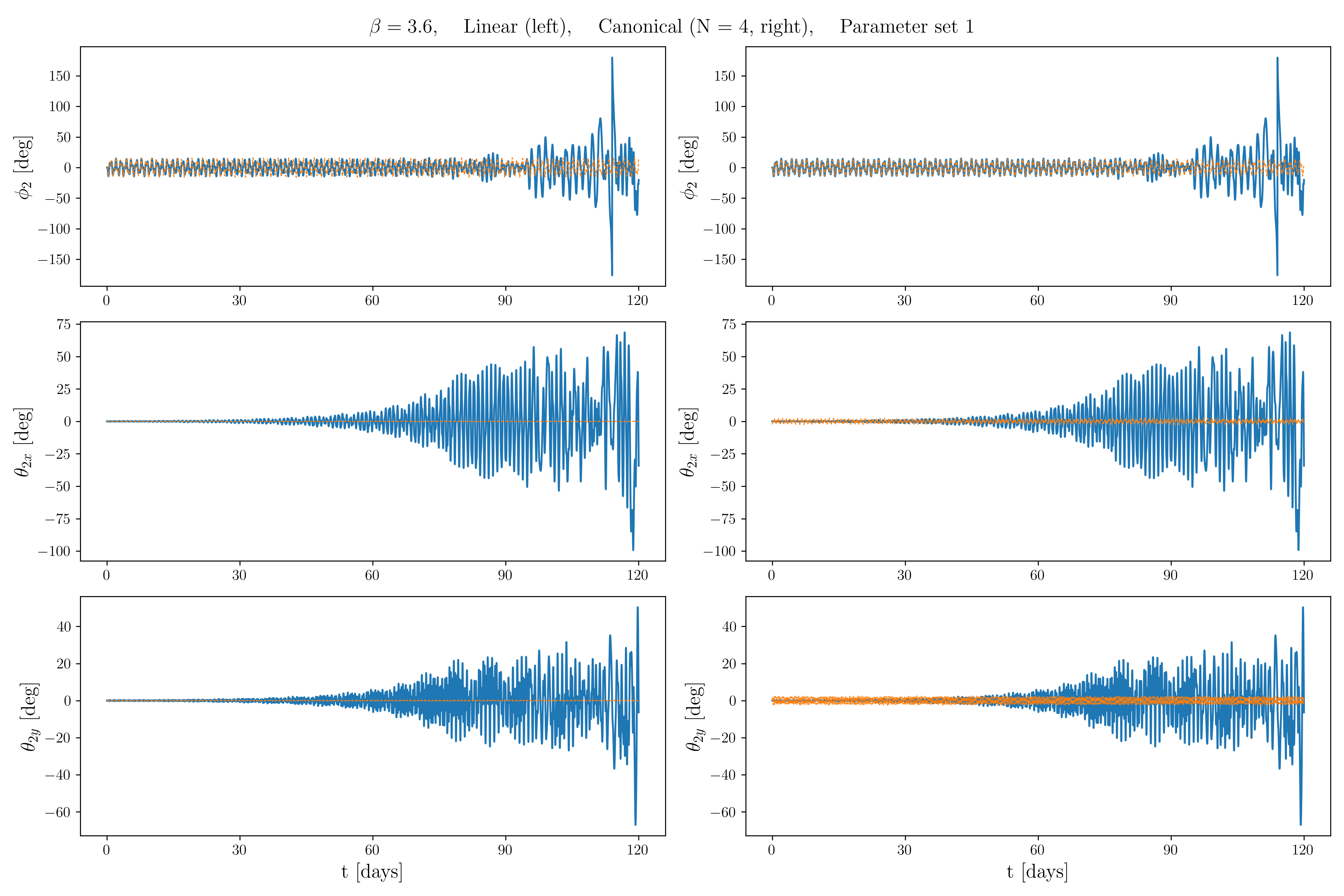}
    \caption{Same as in Figure \ref{fig:orb_num_and_tlin_and_can_b36_set1_short}, but for long term propagation.}
    \label{fig:orb_num_and_tlin_and_can_b36_set1_long}
\end{figure}
\begin{figure}
  \centering
    \includegraphics[scale=0.35]{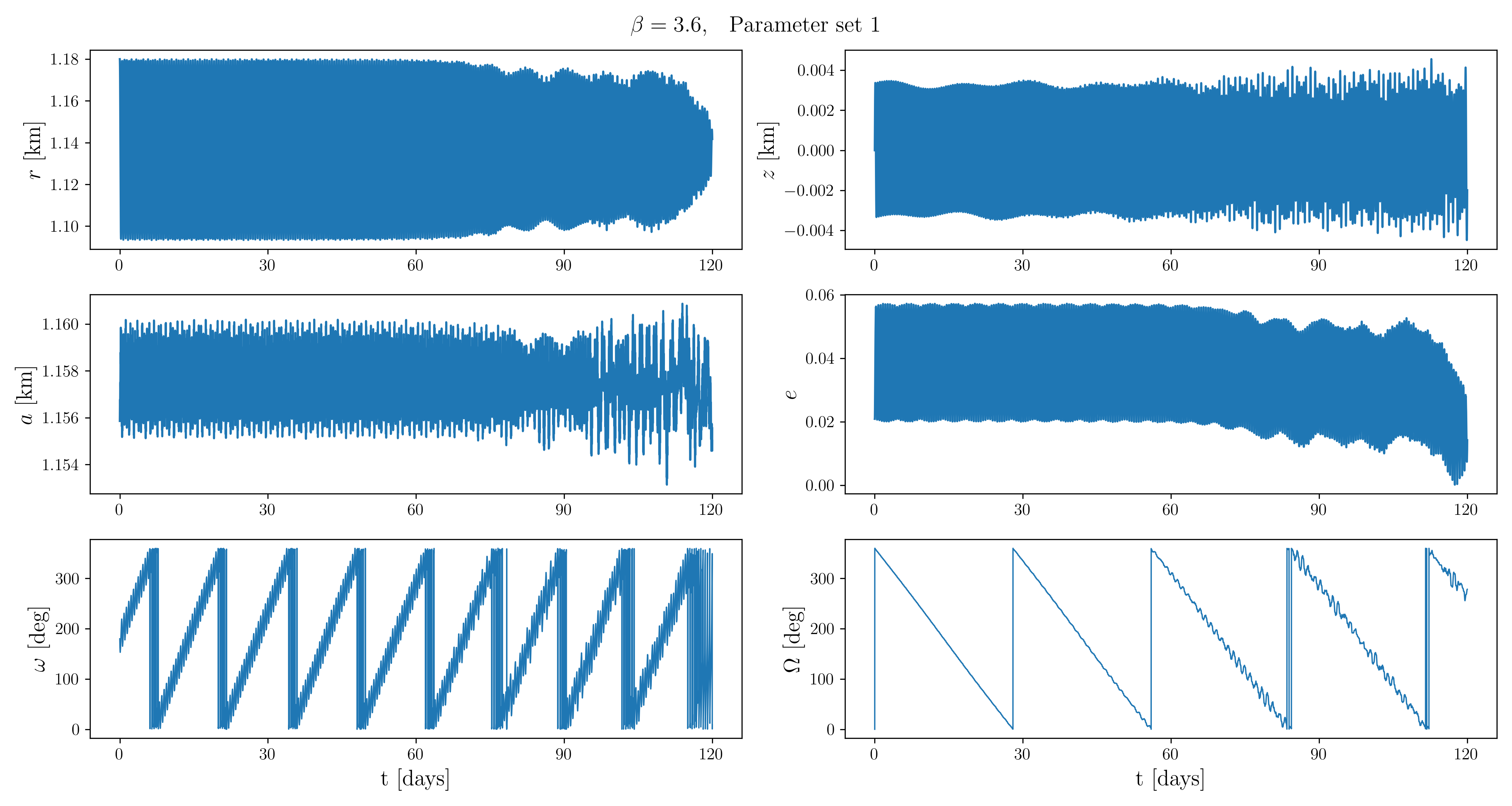}
    \caption{Top: numerical Cartesian states $r(t)$ and $z(t)$. Mid and bottom: numerical Keplerian elements $a(t), e(t),\omega(t)$ and $\Omega(t)$ - all for $\beta = 3.6$, corresponding to chaotic evolution.}
    \label{fig:orb_kep_num_b36_set1}
\end{figure}

Figures \ref{fig:orb_num_and_tlin_and_can_b36_set1_short}, \ref{fig:orb_num_and_tlin_and_can_b36_set1_long} and \ref{fig:orb_kep_num_b36_set1}, now,  highlight the transition to chaotic behavior of the system under more energetic impact conditions, with $\beta = 3.6$. The left-column panels in Fig. \ref{fig:orb_num_and_tlin_and_can_b36_set1_short} compare the full numerical solution with the linear analytical approximation for a relatively short-time propagation (40 days). 
We observe that the libration angle $\phi_2(t)$ retains a quasi-regular behavior, and it is fitted qualitatively by the linear analytical solution, the difference in the two solutions being due mostly to the de-phasing of the analytical solution due to the error in the linear estimate of the libration frequency. In fact, correcting for this error by the nonlinear theory (see next subsection) makes the analytical and numerical solutions to coincide (compare the top-left and top-right panels in Fig. \ref{fig:orb_num_and_tlin_and_can_b36_set1_short}. On the other hand, the numerical solution in the off-plane variables $\theta_{2x}(t)$, $\theta_{2y}(t)$ shows that the corresponding modes are unstable from the beginning, and cannot be fitted neither by the linear nor the nonlinear analytical solution. 

On the other hand, as the instability leads the secondary's off-plane oscillations to grow in amplitude, the nonlinear couplings between in-plane and off-plane degrees of freedom eventually destabilize the in-plane regular libration as well (Fig. \ref{fig:orb_num_and_tlin_and_can_b36_set1_long}). In the chosen parameter set the destabilization takes place at a time $t\approx 90~$days. Note that the timescales involved imply possible consequences in the interpretation of observational data of the system post-impact. In particular, as shown in Fig. \ref{fig:orb_kep_num_b36_set1}, after the loss of partial stability the osculating semi-major axis and eccentricity of the system undergo variations, leading, in particular, to noticeable change in the value of the mean eccentricity and rate of apsidal precession. 

\subsubsection{Solution of the nonlinear theory}
\label{subsec:dartnonlin}
To obtain the corresponding nonlinear analytical solution for a DART-like perturbation, we set the same post-impact initial conditions as in the linear theory, given by Eqs. (\ref{PostimpactICVel})-(\ref{AvgPostmpactICMom}). However, we assign no initial offset in the orientation of the primary, i.e., we set $\delta\theta_{1x}(0) = \delta\theta_{1y}(0) = 0$. This choice is justified by the fact that, within the full (non-averaged) Hamiltonian system, all degrees of freedom are coupled. Thus, even in the absence of a direct perturbation to the primary's attitude angles, the inclined impact is sufficient to excite all relevant dynamical modes - an effect clearly illustrated in the numerical results of Figure \ref{fig:orb_num_and_tlin_b1_set1}, where the blue curve of $\theta_{1x}(t)$ displays non-trivial evolution despite zero initial displacement.

Under these assumptions, the nonlinear analytical solution computed as in Eqs. (\ref{ZNormalFormSol}), back transformed in the original variables is
\begin{equation}\label{CanSol}
\begin{gathered}
r_{can}(t;\beta) = r^{\ast} + P_{r,0}(\beta) + \sum_{i=1}^{N_{r,\text{tot}}}\Bigg[C_{r,i}\beta^{k_{r,i}}\cos\big[\Omega_i(\beta) t \big] \Bigg]\\
z_{can}(t;\beta) = \sum_{i=1}^{N_{z,\text{tot}}}\Bigg[C_{z,i}\beta^{k_{z,i}}\sin\big[\Omega_i(\beta) t \big]\Bigg]\\
\theta_{1x,can}(t;\beta) = \sum_{i=1}^{N_{\theta_{1x},\text{tot}}}\Bigg[C_{\theta_{1x,i}}\beta^{k_{\theta_{1x,i}}}\cos\big[\Omega_i(\beta) t \big]\Bigg]\\
\theta_{1y,can}(t;\beta) = \sum_{i=1}^{N_{\theta_{1y},\text{tot}}}\Bigg[C_{\theta_{1y,i}}\beta^{k_{\theta_{1y,i}}}\sin\big[\Omega_i(\beta) t \big]\Bigg]\\
\theta_{2x,can}(t;\beta) = \sum_{i=1}^{N_{\theta_{2x},\text{tot}}}\Bigg[C_{\theta_{2x,i}}\beta^{k_{\theta_{2x,i}}}\cos\big[\Omega_i(\beta) t \big]\Bigg]\\
\theta_{2y,can}(t;\beta) = \sum_{i=1}^{N_{\theta_{2y},\text{tot}}}\Bigg[C_{\theta_{2y,i}}\beta^{k_{\theta_{2y,i}}}\sin\big[\Omega_i(\beta) t \big]\Bigg]\\
\phi_{2,can}(t;\beta) = \sum_{i=1}^{N_{\phi_{2},\text{tot}}}\Bigg[C_{\phi_{2,i}}\beta^{k_{\phi_{2,i}}}\sin\big[\Omega_i(\beta) t \big]\Bigg]\\
\theta_{can}(t;\beta) = \theta(0) + \omega_{\theta c}(\beta)t + \sum_{i=1}^{N_{\theta,\text{tot}}}\Bigg[C_{\theta,i}\beta^{k_{\theta,i}}\sin\big[\Omega_i(\beta) t \big]\Bigg]\\
\phi_{1,can}(t;\beta) = \phi_1(0) + \omega_{\phi_1 c}(\beta)t + \sum_{i=1}^{N_{\phi_1,\text{tot}}}\Bigg[C_{\phi_1,i}\beta^{k_{\phi_1,i}}\sin\big[\Omega_i(\beta) t \big]\Bigg]~~~.
\end{gathered}
\end{equation}
We observe that the nonlinear solution (\ref{CanSol}) is represented as trigonometric time series, whose frequencies $\Omega_i(\beta)$ are linear combinations of the canonical frequencies $\omega_{jc},\omega_{\theta c},\omega_{\phi_1 c}$, which themselves are polynomial expressions of $\beta$. Specifically, the frequencies $\Omega(\beta)$ are expressed as
\begin{equation}
\Omega_i(\beta) = \sum_{j=1}^{7} \big[n_j\omega_{jc}(\beta)\big] + n_{\phi_1}\omega_{\phi_1 c}(\beta) + n_{\theta}\omega_{\theta c}(\beta)~, ~~~ n_j,n_\theta,n_{\phi_1} \in \mathbb{Z}~~~,
\end{equation}
with amplitudes also determined by polynomial expressions in $\beta$. The coefficients $C_{r,i}, C_{z,i}, ..., C_{\phi_1,i}$ depend directly on the physical parameters inherent to the system. Each solution includes a finite series summation, indicated by indices $i$, which run from $1$ up to their respective totals  $N_{r,\text{tot}},N_{z,\text{tot}},...,N_{\phi_1,\text{tot}}$. Similar expressions (here omitted) can be derived for the conjugate momenta. 

From the structure of the solution of $r_{can}(t;\beta)$, we observe that although the starting point of the canonical normalization procedure is the idealized two-sphere single-synchronous configuration, where both bodies are assumed spherical, the canonical formalism recovers `on the go' (i.e. through iteration of the normalization steps) a more precise value for the equilibrium state $r_{eq}$, consistent with the effect on the orbit by the bodies' triaxiality. The correction is encoded in the polynomial term $P_{r,0}(\beta)$, which represents a displacement of the equilibrium radius of the orbit with respect to the radius $r^{\ast}$. As a result, $r^{\ast} + P_{r,0}(\beta)$ serves as a refined estimate of the mean orbital separation of the bodies, around which small oscillations take place. This estimate mirrors the behavior observed by `scissor averaging' in the linearized theory, while offering higher-order corrections consistent with the mutual potential expansion.

\begin{figure}
  \centering
    \includegraphics[scale=0.35]{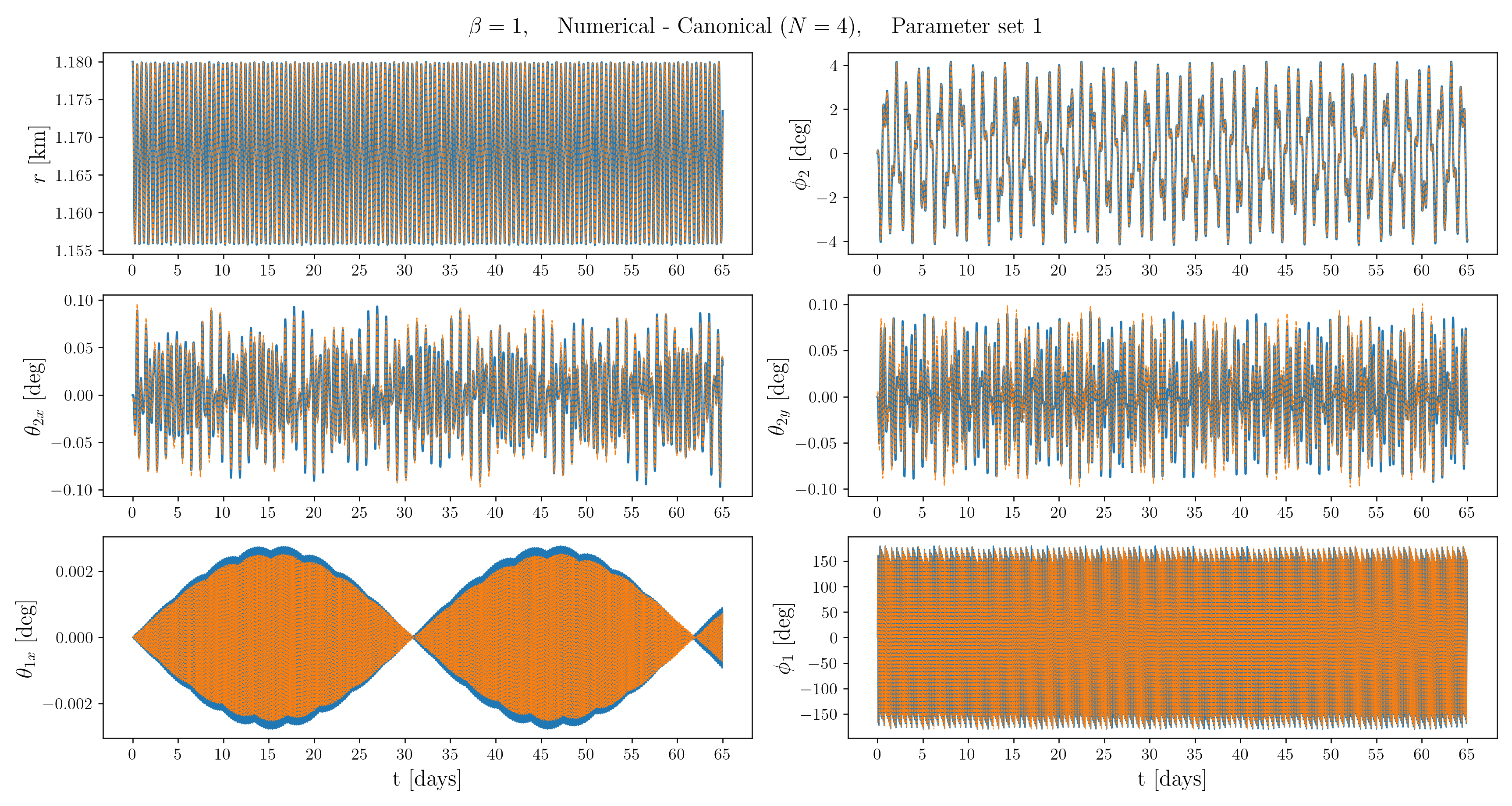}
    \caption{Comparison between the numerical (blue) and nonlinear-analytical (orange) solution for $\beta = 1$. The primary's initial disturbance is now set to $\delta\theta_{1x0}=\delta\theta_{1y0}=0$.}
    \label{fig:fig2_orb_num_and_can_b1}
\end{figure}
\begin{figure}
  \centering
    \includegraphics[scale=0.35]{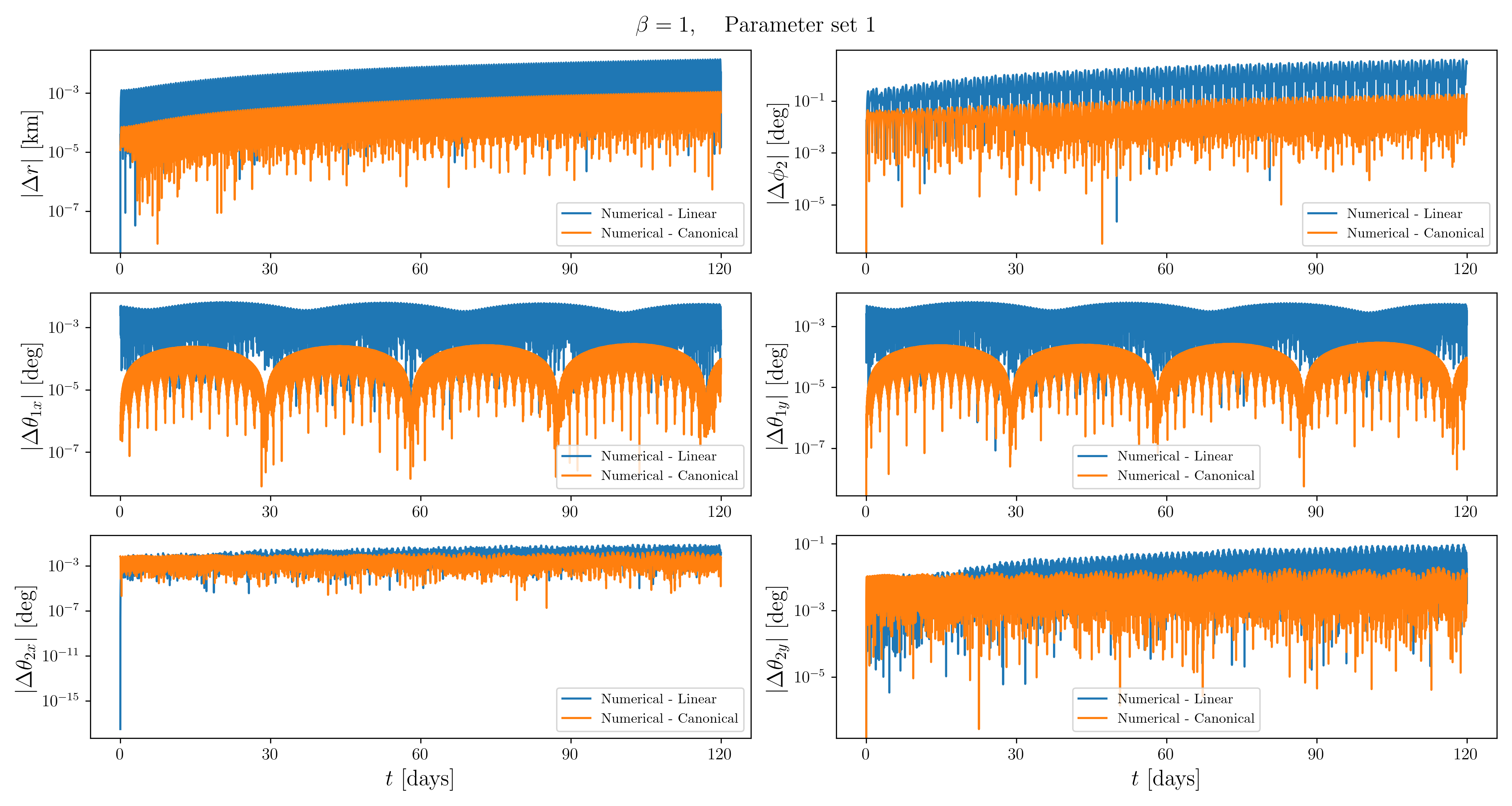}
    \caption{Absolute differences between the numerical and analytical solutions (linear and canonical) for a total duration of 120 days.}
    \label{fig:fig7_errors_num_tlin_can_b1}
\end{figure}
Figure \ref{fig:fig2_orb_num_and_can_b1} (same parameters as Fig. \ref{fig:orb_num_and_tlin_b1_set1}) shows the comparison of the full numerical solution (blue) with the nonlinear (`canonical') analytical solution obtained from the normal form iterated at order $N=4$ (orange) in the parameter set 1 for $\beta = 1$. Comparing with the linear analytic solution (Fig. \ref{fig:orb_num_and_tlin_b1_set1}), the nonlinear theory accurately captures the amplitude modulation and phase evolution in all degrees of freedom, including the primary's libration in the angle $\theta_{1x}$ (and also $\theta_{1y}$, which is similar, but not shown in the figures). Figure \ref{fig:fig7_errors_num_tlin_can_b1}, which presents the absolute difference between the numerical and linear or nonlinear analytical solution, shows that the nonlinear solution consistently improves the accuracy, i.e., produces a lower error than the linear solution.

\begin{figure}
  \centering
    \includegraphics[scale=0.33]{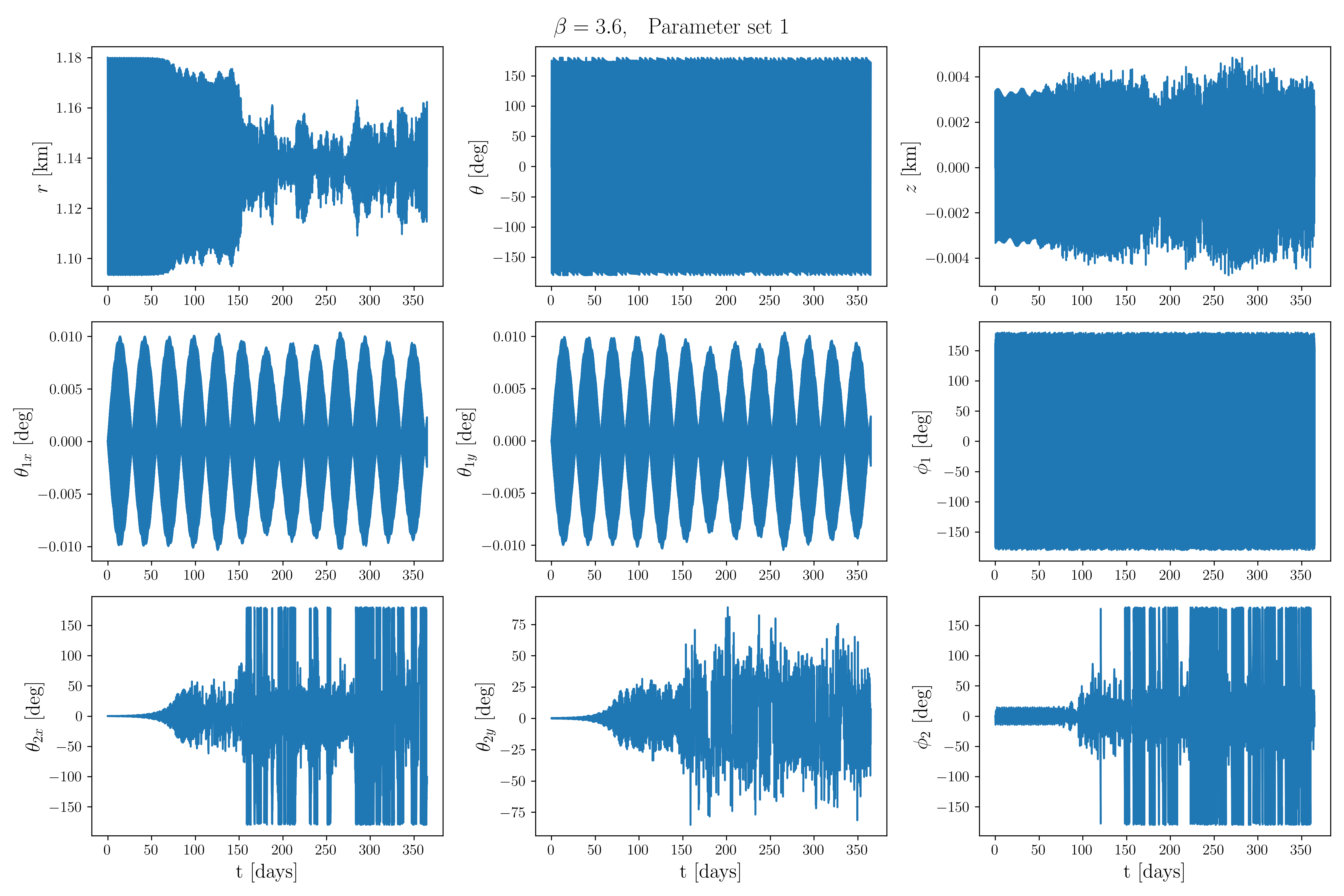}
    \caption{One year evolution of all nine coordinates of the system in the case of the parameter set 1 with $\beta=3.6$. First row: the coordinates $r(t)$, $\theta(t)$, $z(t)$ of the relative orbit. Second row: the primary's angles $\theta_{1x}(t)$, $\theta_{1y}(t)$, $\phi_1(t)$. Third row: the secondary's  angles $\theta_{2x}(t)$, $\theta_{2y}(t)$, $\phi_2(t)$.}
    \label{fig:orb_num_b36_365_set1}
\end{figure}
\begin{figure}
  \centering
    \includegraphics[scale=0.35]{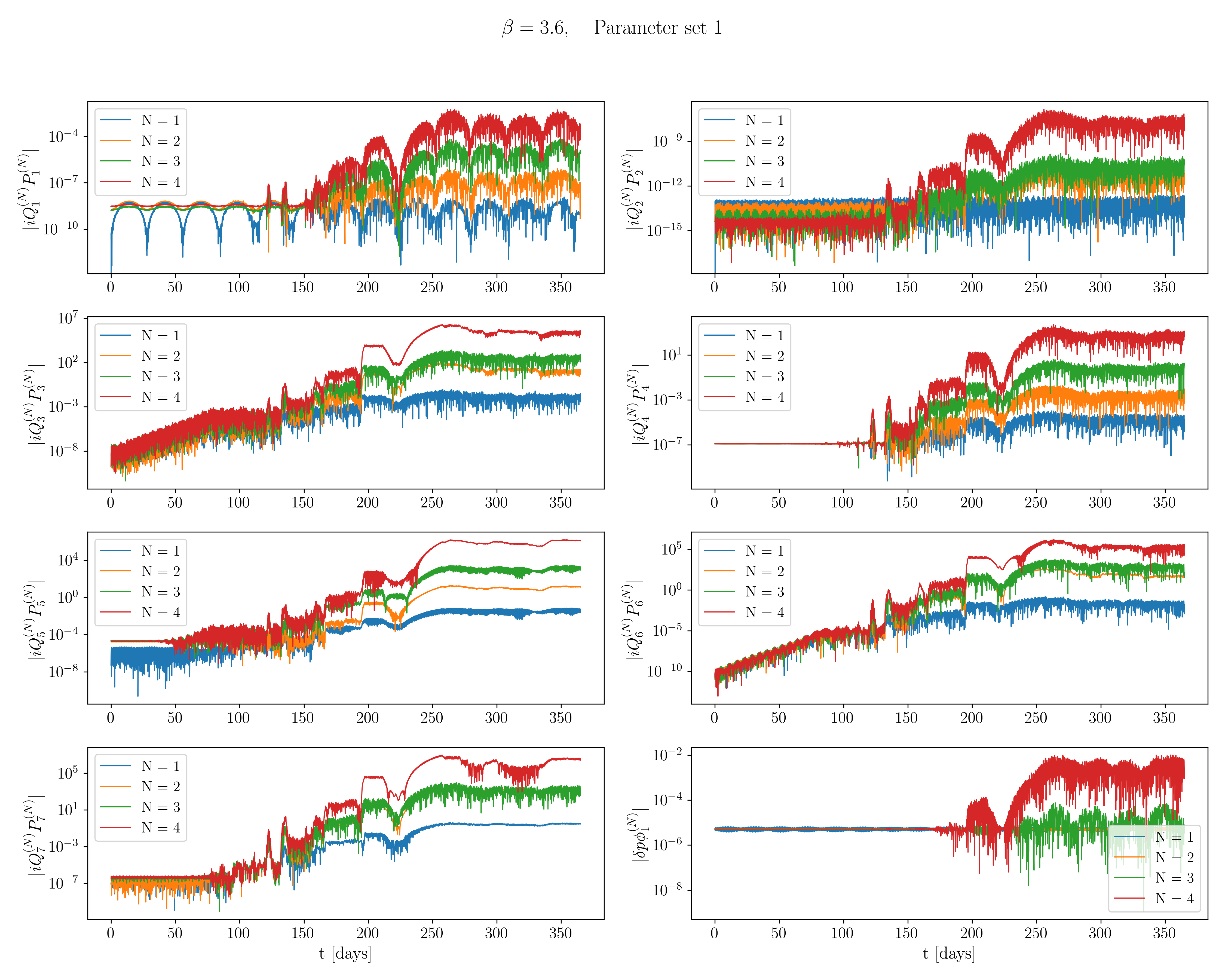}
    \caption{Quasi-integrals of motion in time, corresponding to Eq. (\ref{integrals}), but transformed via $N$ inverse Lie transformations back to the original variables $\boldsymbol{\zeta}$ of Eq. (\ref{NonLinDeltaVars}). The evolution of the quasi-integrals is shown in the case of the trajectory of parameter set 1 with $\beta=3.6$, corresponding to Figs. \ref{fig:orb_kep_num_b36_set1} and \ref{fig:orb_num_b36_365_set1}. Each panel shows the quasi-integrals computed with the normal form iterated $N$ times with $N=1$ to $4$. }
    \label{fig:fig6_integrals_b36}
\end{figure}
As already mentioned, increasing the impact parameter to $\beta=3.6$ causes a regime transition in which the system exhibits only partially regular behavior. Figure \ref{fig:orb_num_b36_365_set1} shows the one year evolution of all nine coordinates of the system. We distinguish three groups of variables depending on their observed evolution: i) the three primary's angles $(\theta_{1x},\theta_{1y},\phi_1)$ as well as the angle $\theta$ of the relative orbit exhibit a long-term regular behavior. The angles $\theta_{1x},\theta_{1y}$ undergo a slowly modulated libration of tiny amplitude, while $\phi_1$ and $\theta$ rotate with nearly constant frequencies. ii) The variables $r$, $z$, $\theta_{2y}$ initially exhibit a transient regular behavior, then converted into bounded chaotic libration-like motion. iii) the secondary's angles $\phi_2$, $\theta_{2x}$, initially exhibit a transient regular librational behavior, which then turns into a chaotic motion switching erratically between libration and rotation. From the dynamical point of view, these different types of evolution observed in the coordinates' time series can be properly characterized through the evaluation of the time series representing the quasi-integrals of Eq. (\ref{integrals}). These include the seven products $I_j^{(N)}=iQ_j^{(N)} P_j^{(N)}$, for $j=1,\dots,7$, and the quantity $\delta p_{\phi_1}^{(N)}$ which are exact invariants of the truncated flow under the normal form Hamiltonian $Z^{(N)}$, but only approximate integrals of the full Hamiltonian flow, exhibiting small and bounded variations for those variables which exhibit regular behavior. The variation of the quasi-integrals can be computed through the expressions $I^{(N)}(\boldsymbol{\zeta},\theta,\phi_1,\delta p_{\phi_1})$, $\delta p_{\phi_1}^{(N)}(\boldsymbol{\zeta},\theta,\phi_1,\delta p_{\phi_1})$ found by the transformation (\ref{nonlintra}), where the evolution of $\boldsymbol{\zeta}(t)$,$\theta(t)$,$\phi_1(t)$,$\delta p_{\phi_1}(t)$ is obtained by the numerical time series through the full equations of motion.  

Figure \ref{fig:fig6_integrals_b36} displays the evolution of the eight quasi-integrals over time for the trajectory of Fig. \ref{fig:orb_num_b36_365_set1}. A classification consistent with the decoupling observed in the linear system (see Eqs. (\ref{delta2QP57})-(\ref{delta2QP12}). Namely: i) The integrals $I_1^{(N)}=iQ_1^{(N)} P_1^{(N)}$ and $I_2^{(N)}=iQ_2^{(N)} P_2^{(N)}$ correspond to the nonlinear normal modes associated with librations in the primary's Euler angles $\theta_{1x}$, $\theta_{1y}$. ii) The integrals $I_3^{(N)}=i Q_3^{(N)} P_3^{(N)}$, $I_6^{(N)}=i Q_6^{(N)} P_6^{(N)}$, and $I_7^{(N)}=i Q_7^{(N)} P_7^{(N)}$ are related to the nonlinear normal modes associated with the secondary’s Euler angles ($\theta_{2x}$, $\theta_{2y}$, and $\phi_2$). iii) The integrals $I_4^{(N)}=i Q_4^{(N)} P_4^{(N)}$ and $I_5^{(N)}=i Q_5^{(N)} P_5^{(N)}$ are related to nonlinear normal modes associated with epicyclic oscillations of the relative orbit in the variables $r$ and $z$ with respect to the equatorial equilibrium circular orbit. iv) Finally, $\delta p_{\phi_1}$ is the angular momentum associated with the primary's rotation in the relative yaw angle $\phi_1$. 
From Figure \ref{fig:fig6_integrals_b36}, we observe that the integrals associated with the primary's Euler angles ($I_1$, $I_2$, and $\delta p_{\phi_1}$) exhibit an initial phase of stability, which lasts for $t\lesssim T_1=150~$days. This timescale is related with the modulation timescale induced by the orbit in the nearby resonances as we will see in the following. In this phase, the fluctuations on the integrals shrink in size as the order $N$ of the normal form increases. After the time $T_1$, however, the three above quasi-integrals truncated at order higher than $N=1$ show an abrupt variation, while the integrals $I_1^{(1)}$, $I_2^{(1)}$, and $\delta p_{\phi_1}^{(1)}$ (truncated at order $N=1$) keep representing quasi-constants of motion. The fact that the three quasi-integrals loose their stability beyond order $N=1$ is due to the presence in the series of small divisors (see the next subsection). Note also that the integrals $I_3^{(N)}$ and $I_6^{(N)}$, associated with the secondary's Euler angles $\theta_{2x},\theta_{2y}$ exhibit a phase of initial linear drift, lasting up to $T_2\simeq 70~$days, from where on their behavior undergoes the same qualitative transitions as for the integrals $I_1,I_2,\delta p_{\phi_1}$. Comparing with Fig. \ref{fig:orb_num_b36_365_set1}, the transition at time $T_1$ is connected to a chaotic switch of the secondary's rotational state in the angles $\theta_{2x}, \phi_2$ from libration to rotation. The transitions observed at the timescale $T_2$ are connected to resonance effects, as analyzed in the next subsection. Finally, both transitions have small effects in the orbit, which, however, remains essentially a bounded orbit with nearly constant angular velocity and mean values of the cylindrical coordinates $r$ and $z$. 

Further discussion on the behavior of the analytical series is made in the next subsection. As a general remark, the above dual approach - examining both trajectories and conserved quantities - reveals the hierarchy of stability among the degrees of freedom of the system, and illustrates how canonical normalization exposes `hidden' partial regular structure in the dynamics, even under the presence of chaos.

\subsection{Sensitivity of the solution to the system's parameters. The role of resonances}
\label{subsec:parametric}
It is possible to see that the transition of the system from regular to chaotic behavior observed in the previous figures is very sensitive to the system's parameters, which determine the proximity of the system to a single or multiple \textit{resonance} condition, i.e. a commensurability between the fundamental orbital and librational frequencies of the system. The most important resonances (see \cite{Agrusa2021}) are between the orbital frequencies of the system and the librational frequencies of the secondary. Their role is revealed through a simplified `spin-orbit' model for the secondary, obtained by considering the equations of motion for the secondary as given by the Hamiltonian (\ref{Hfullform}), with the substitution $p_{\theta}-p_{\phi_1}-p_{\phi_2}\rightarrow\omega_{\theta}(t)=\dot{\theta}(t)$, and the substitution of the relative orbit $(r(t),\theta(t),z(t))$ with the one found under the averaged, over the primary's angle $\phi_1$, axisymmetric potential
\begin{equation}\label{potaveaxi}
V_{orb}=-{G(M_1+M_2)\over(r^2+z^2)^{1/2}}
+{G(I_{1x}+I_{1y}-2I_{1z})\over 4(r^2+z^2)^{3/2}}\left(1-{3z^2\over r^2+z^2}\right)~.
\end{equation}
Epicyclic theory then yields the following approximate solution for the relative orbit under the potential (\ref{potaveaxi}):
\begin{equation}\label{epiorb}
r(t)\simeq r_c+\xi_0\cos(\kappa t),~~
z(t)=\left({v_{0z}\over\kappa_z}\right)\sin(\kappa_z(t)),~~
\omega_{\theta}(t)=\omega_c\left(1-{2\xi_0\over r_c}\cos(\kappa t)\right)
\end{equation}
where the equilibrium radius $r_{eq}=r_c$ is the positive root of the equation 
$$
r_0^2
\left(
\sqrt{r_0V_0'(r_0)}-{M_Dv_D\beta\cos\gamma\over M_2}
\right)^2=r_c^3V_0'(r_c)
$$
with $V_0(r)=V_{orb}(r,z=0)$, $r_0$ equal to the initial cylindrical radius at the moment of the impact (set equal to $r_{eq}$), $\xi_0=r_0-r_c$, $v_{0z}=M_Dv_D\beta\sin(\gamma)/M_2$, and the epicyclic frequencies given by:
\begin{align}\label{epifreq}
\kappa&=
\left({G(M_1+M_2)\over r_c^3}
+{3G(I_{1x}+I_{1y}-2I_{1z})\over 4r_c^5}\right)^{1/2} \nonumber \\
\omega_c&=
\left({G(M_1+M_2)\over r_c^3}
-{3G(I_{1x}+I_{1y}-2I_{1z})\over 4r_c^5}\right)^{1/2} \\
\kappa_z&=
\left({G(M_1+M_2)\over r_c^3}
-{9G(I_{1x}+I_{1y}-2I_{1z})\over 4r_c^5}\right)^{1/2}~. \nonumber
\end{align}

\begin{figure}
  \centering
    \includegraphics[scale=0.55]{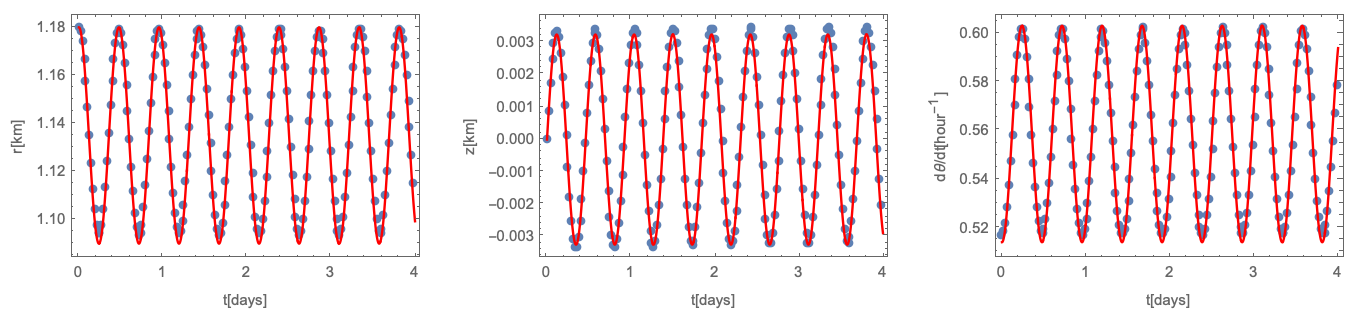}
    \caption{The epicyclic approximation of Eqs. (\ref{epiorb}) (solid line) superposed to the exact numerical solution (dotted curve) for the relative orbit for $\beta=3.6$ under parameter set 1, for a short propagation of 4 days.}
    \label{fig:epiorb}
\end{figure}
As shown in Fig. \ref{fig:epiorb}, the epicyclic approximation to the relative orbit fits excellently the numerical orbit under the full Hamiltonian as long as the latter remains regular. Inserting the approximate solution (\ref{epiorb}) to the equations of motion, linearizing with respect to the SSES solution $\theta_{2x}=\theta_{2y}=p_{\theta_{2x}}=p_{\theta_{2y}}=\phi_2=0$, $p_{\phi_2}=I_{2z}\omega_c$, considering the oscillations in $z$ as of second order, and expanding the quantity $1/(r^2+z^2)$ up to terms of first degree in the epicyclic amplitude $\xi_0=r_0-r_c$, we arrive at a simplified system of linearized equations for the Euler angles of the secondary. For the off-plane angles we find:
\begin{equation}\label{eqmoSOoffplane}
\left(
\begin{array}{c}
\dot{\delta\theta_{2x}} \\
\dot{\delta\theta_{2y}} \\
\dot{\delta P_{\theta_{2x}}} \\
\dot{\delta P_{\theta_{2y}}} 
\end{array}
\right)
=(\mathbf{B_1}+\mathbf{B_2}(t))
\left(
\begin{array}{c}
\delta\theta_{2x} \\
\delta\theta_{2y} \\
\delta P_{\theta_{2x}} \\
\delta P_{\theta_{2y}} 
\end{array}
\right)
\end{equation}
where $\delta P_{\theta_{2x}}=\delta p_{\theta_{2x}}/I_{2x}$, $\delta P_{\theta_{2y}}=\delta p_{\theta_{2y}}/I_{2y}$, and  
$$
\mathbf{B_1}=
\left(
\begin{array}{cccc}
0                                              
&\omega_c         
&1                               
&0                                              \\
\left({I_{2z}\over I_{2y}}-1\right)\omega_c            
&0 
&0                                             
&1                                \\
-\left({I_{2z}\over I_{2y}}-1\right){I_{2z}\over I_{2x}}\omega_c^2   
&0 
&0                                  
&-\left({I_{2z}-I_{2y}\over I_{2x}}\right)\omega_c   \\
0  
&-\omega_c^2{I_{2z}\over I_{2y}}-{3G(I_{2z}-I_{2x})M_1\over I_{2y}r_c^3}   
&-{I_{2x}\over I_{2y}}\omega_c                                                
&0
\end{array}
\right)
$$
$$
\mathbf{B_2}(t)=
\left(
\begin{array}{cccc}
0    &0                                                      &~~~~0  &~~~~~0 \\
0    &0                                                      &~~~~0  &~~~~~0 \\
0    &0                                                      &~~~~0  &~~~~~0 \\
0    &{9G(I_{2z}-I_{2x})M_1\over I_{2y}r_c^4}\xi_0\cos(\kappa t)   &~~~~0  &~~~~~0
\end{array}
\right)~.
$$
For the libration angle $\phi_2$, instead, we obtain the forced (non-homogeneous) spin-orbit second order differential equation
\begin{equation}\label{eqmoSOinplane}
\ddot{\phi}_2+{3GM_1\over r_c^3}{(I_{2y}-I_{2x})\over I_{2z}}
(1-{3\xi_0\over r_c}\cos(\kappa t))\phi_2=
-{2\kappa\omega_c\xi_0\over r_c}\sin(\kappa t)~.
\end{equation}

\begin{figure}
  \centering
    \includegraphics[scale=0.7]{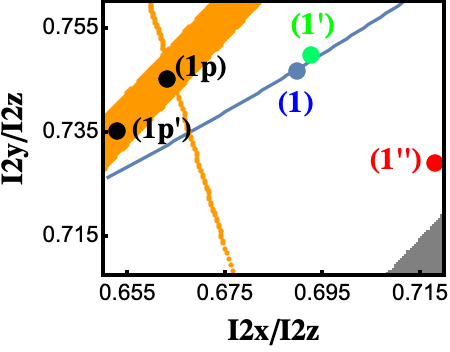}
    \caption{Plot of the moment of inertia ratios $I_{2x}/I_{2z}$, $I_{2y}/I_{2z}$ where various linear or nonlinear resonances affect the dynamics and/or the convergence of the analytical series solution. The blue line corresponds to the locus of the resonance $\kappa-3\kappa_z+2(\omega_1-\omega_3)=\kappa-3\kappa_z+2(\omega_{prec}-\omega_{lib})=0$ in the plane of the moment-of-inertia ratios $(I_{2x}/I_{2z},I_{2y}/I_{2z})$ for the system in all parameters as in parameter set 1, except for the moments of inertia $I_{2x},I_{2y}$ altered within the limits indicated in the figure. The point (1) corresponds to the exact conditions of the parameter set 1, and satisfies the resonance condition with an offset $\Delta\omega=|\kappa-3\kappa_z+2(\omega_1-\omega_3)|=2.9\cdot 10^{-4} \text{[hr}]^{-1}$. By contrast, for the parameter set $1'$ ($I_{2x}=8.23863\cdot 10^{-5}$, $I_{2y}=8.92192\cdot 10^{-5}$, $I_{2z}=1.1899\cdot 10^{-4}$ in units same as in Table \ref{tab:ParamSets} the offset is $\Delta\omega=5.09\cdot 10^{-3} \text{[hr]}^{-1}$, while for the parameter set $1''$ ($I_{2x}=8.93181\cdot 10^{-5}$, $I_{2y}=9.07161\cdot 10^{-5}$, $I_{2z}=1.24425\cdot 10^{-4}$) the offset is $\Delta\omega=2.2\cdot 10^{-1} \text{[hr}]^{-1}$. The two strips in orange indicate values where parametric resonance takes place. The points $1p$ and $1p$' correspond to the parameters $I_{2z}=1.1899\cdot 10^{-4}$ and $I_{2x}=7.88606\cdot 10^{-5}$, $I_{2y}=8.86832\cdot 10^{-5}$ (point $1p$), or $I_{2x}=7.76410\cdot 10^{-5}$, $I_{2y}=8.74755\cdot 10^{-5}$ (point $1p'$). The gray domain below the diagonal $I_{2x}=I_{2y}$ indicates initial conditions for which the secondary's libration is linearly unstable ($\omega_{lib}$ imaginary). The nonlinear analytical series solution converges for all parameter values in the white area of the plot where the numerical solution is regular.}
    \label{fig:plotres}
\end{figure}
\begin{figure}
  \centering
    \includegraphics[scale=0.33]{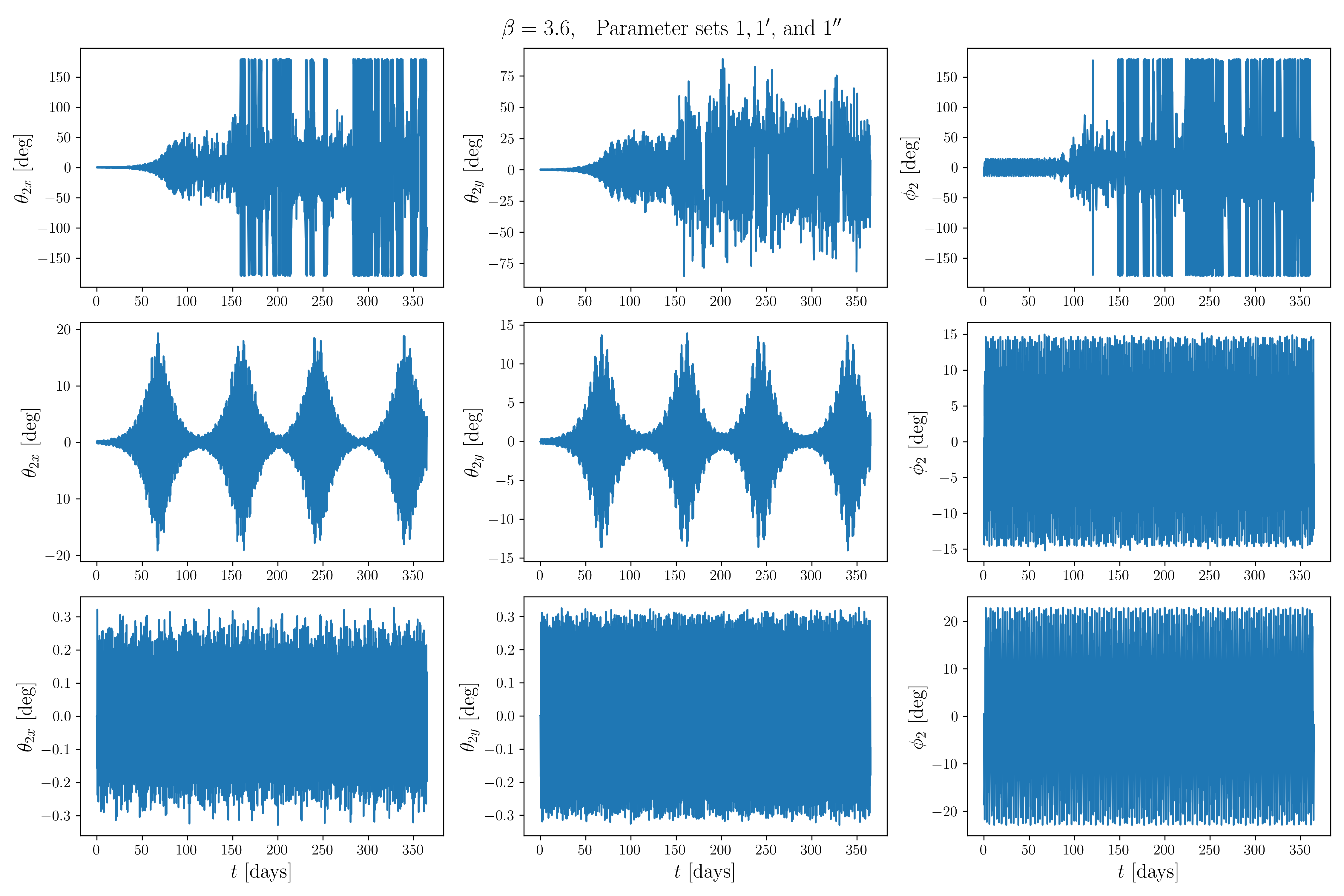}
    \caption{The evolution of the secondary's angles under the full model of the equations of motion for $\beta=3.6$ and secondary's axial ratios corresponding to the parameter set 1 (top row), 1' (middle row) or 1'' (bottom row).}
\label{fig:SecondaryEulerAngles}
\end{figure}
\begin{figure}
  \centering
    \includegraphics[scale=0.4]{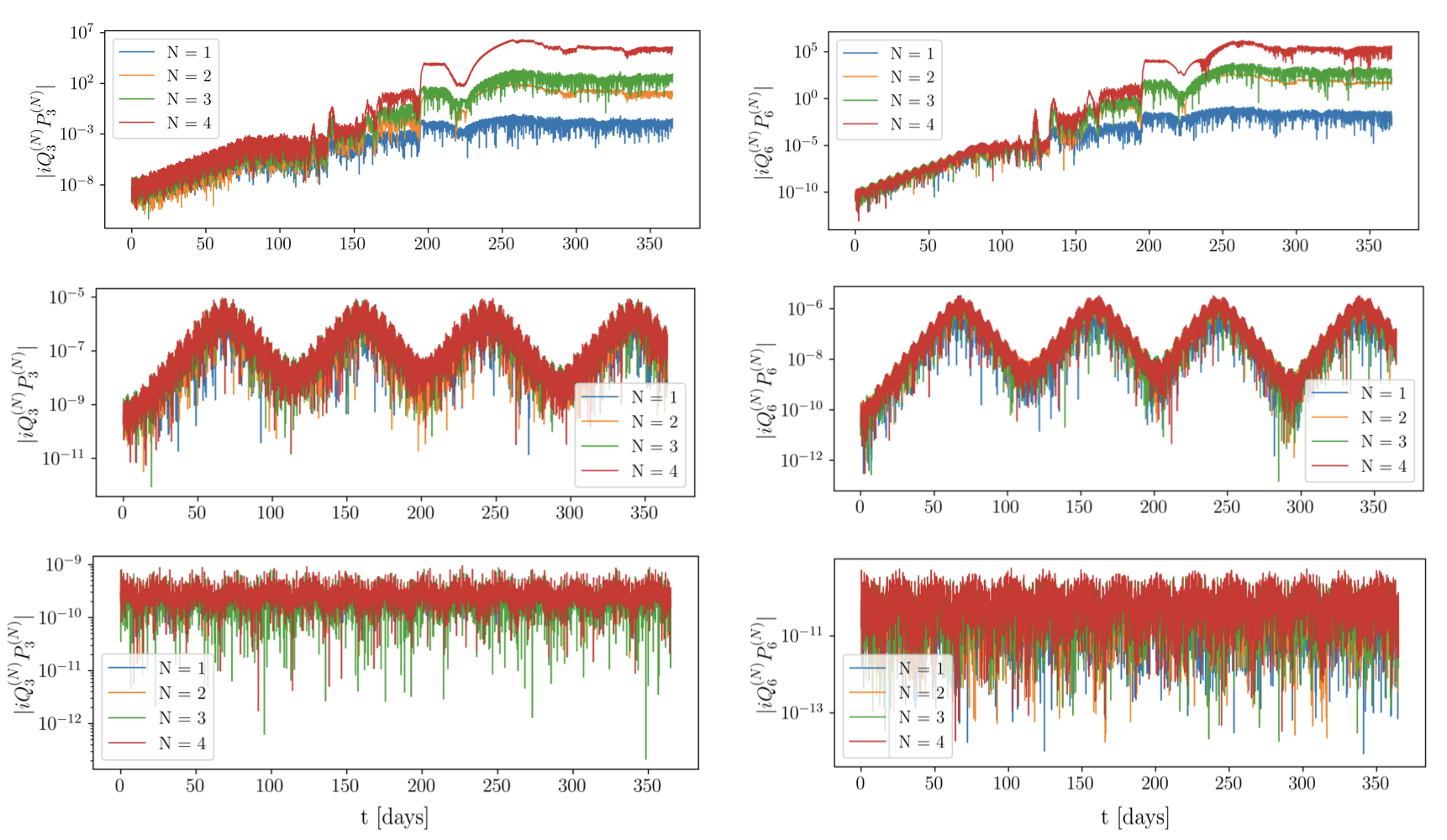}
    \caption{Evolution of the quasi-integrals of motion $I_3$ and $I_6$ for the same orbits as in Fig. \ref{fig:SecondaryEulerAngles}.}
    \label{fig:fig6_integrals_b36plus}
\end{figure}

Through the above approximation, we can now recognize two types of resonances which affect the evolution of the secondary's Euler angles: i) The linearized equations of motion (Eqs. (\ref{eqmoSOoffplane}) and (\ref{eqmoSOinplane})) exhibit parametric resonance, similar as described in \cite{Wisdom1984}. ii) Nonlinear resonances occur by small divisors involving commensurabilities between five frequencies, namely the two orbital frequencies $(\kappa,\kappa_z)$ and the three rotational frequencies $\omega_j$, $j=1,2,3$, where $\omega_1,\omega_2$ are obtained by the eigenvalues of the matrix $B_1$ and $\omega_3=\left(3GM_1(I_{2y}-I_{2x})/I_{2z}\right)^{1/2}r_c^{-3/2}$ is the classical spin-orbit libration frequency. Note that these latter resonances create multiplets around the resonances discussed in \cite{Agrusa2021}. The multiplets are due to the small frequencies of precession of the periapsis and of the nodes, $\dot{\varpi}$, $\dot{\Omega}$, of the secondary's orbit with respect to the Keplerian orbit. The latter frequencies are connected with the epicyclic frequencies through the relations $\dot{\lambda}=\omega_c$, $\dot{\varpi}=\omega_c-\kappa$, $\dot{\Omega}=\omega_c-\kappa_z$. The association of one resonance in the multiplet with the corresponding central resonance in \cite{Agrusa2021} can be done using, in addition, the relations $\omega_1=\omega_{prec}$ (frequency of precession of the secondary's spin axis), and $\omega_3=\omega_{lib}$ (secondary's libration frequency). As shown in Fig. \ref{fig:plotres}, we find that the chaotic behavior exhibited by the system for the parameter set 1 at $\beta=3.6$ is caused by the nonlinear resonance $\kappa-3\kappa_z+2(\omega_{prec}-\omega_{lib})=0$. Since all three frequencies $\omega_j$, $j=1,2,3$ depend on the moment-of-inertia ratios $I_{2x}/I_{2z}$, $I_{2y}/I_{2z}$ (or the corresponding axial ratios), keeping $\beta=3.6$ as well as all the parameters of the parameter set 1 fixed, and altering the axial ratios $I_{2x}/I_{2z}$, $I_{2y}/I_{2z}$ even slightly, as indicated in Fig. \ref{fig:plotres} (set 1'), leads the system off-resonance and restores the regular character of the motion. In particular, as shown in Fig. \ref{fig:SecondaryEulerAngles}, changing slightly the secondary's axes' length (by a few meters), suffices to set the system off-resonance and restore the regular character of the orbits. For a resonance offset $\Delta\omega=|\kappa-3\kappa_z+2(\omega_1-\omega_3)|$ of the order of $10^{-3}$, the effect of the resonance is still visible as a slow modulation of the libration in the angles $\theta_{2x},\theta_{2y}$. This creates the characteristic `zig-zag' variation of the integrals $I_3$ and $I_6$ computed through the nonlinear normal form (see Fig. \ref{fig:fig6_integrals_b36plus}). The `zig-zag' period is equal to the modulation period, and interprets the transient period $T_2$ observed in Fig. \ref{fig:fig6_integrals_b36}. On the other hand, for a resonance offset as high as $\Delta\omega\sim 10^{-1}$, the modulation disappears and the typical non-resonant behavior of all integrals of motion computed by the nonlinear normal form is restored (Fig. \ref{fig:fig6_integrals_b36plus}). In this case, the analytical predictions of the evolution of the orbit with the nonlinear normal form returns to be more precise by about two orders of magnitude with respect to the linear analytical solution. 
\begin{figure}
  \centering
    \includegraphics[scale=0.31]{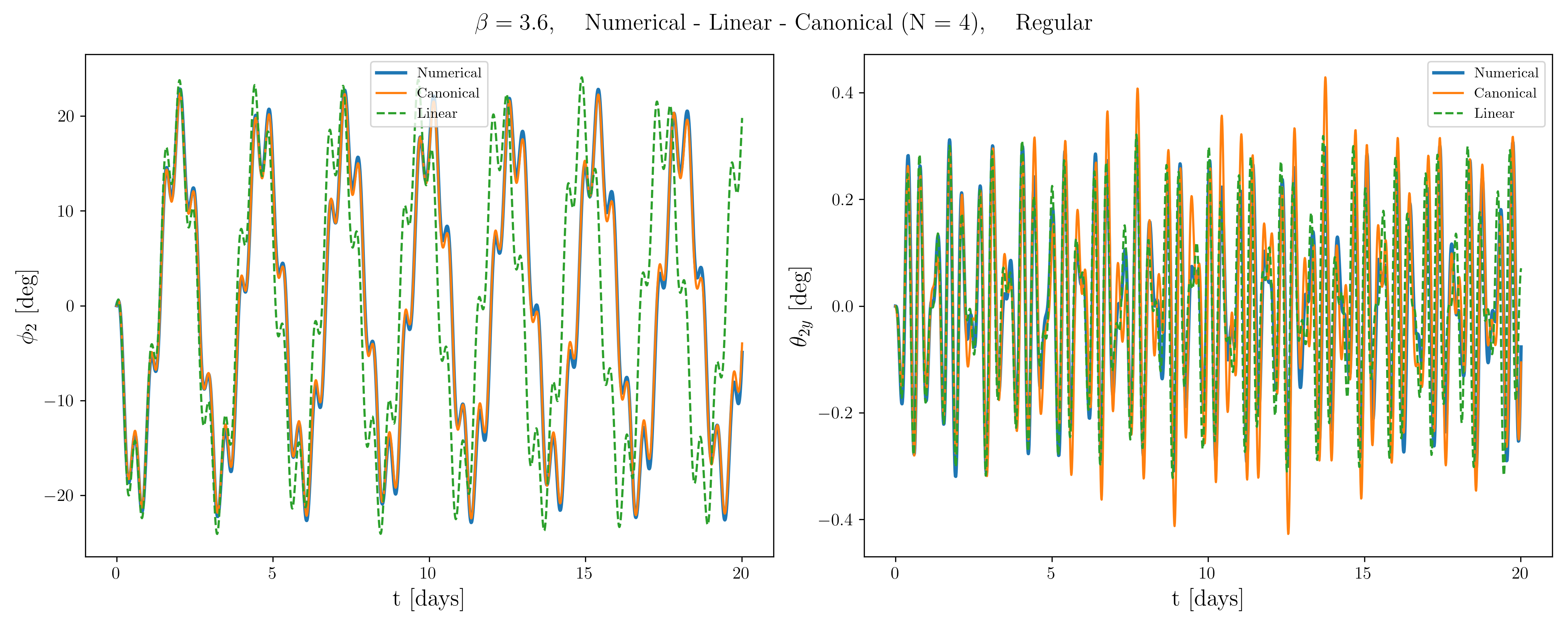}
    \caption{Comparison between numerical, linear and nonlinear solutions of $\phi_2(t)$ and $\theta_{2y}(t)$ for the orbit with $\beta=3.6$ and parameter set $1''$ (regular case).}
    \label{fig:fig10_orb_tlin_and_can_b36_short}
\end{figure}

\begin{figure}
  \centering
    \includegraphics[scale=0.42]{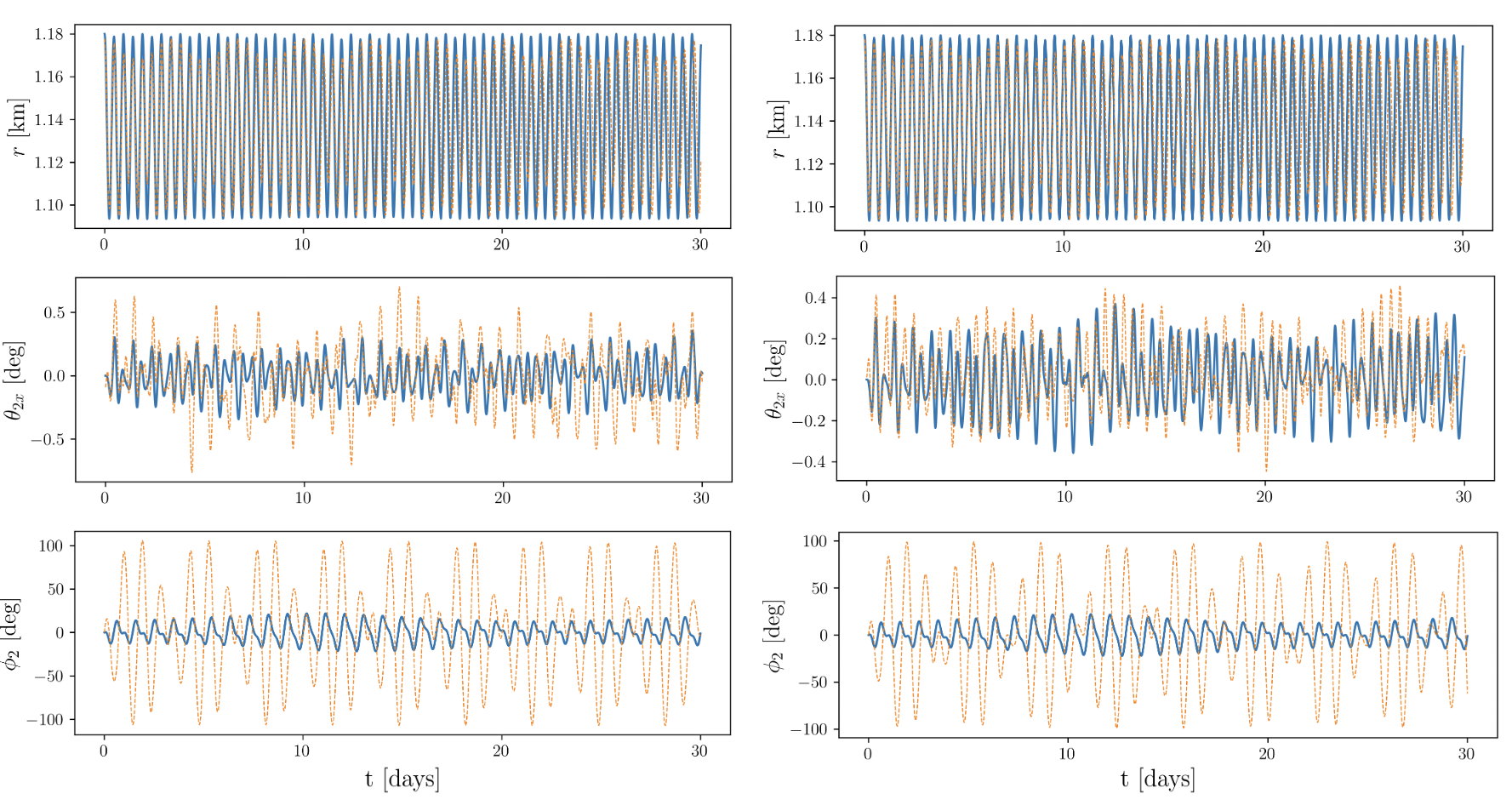}
    \caption{Comparison between numerical (blue) and nonlinear (orange) solutions in $r(t)$ and the secondary's angles $\theta_{2x}(t)$ and $\phi_2(t)$ for the orbits with $\beta=3.6$ and parameters corresponding to the parameter set 1, except for the values or $I_{2x}$, $I_{2y}$, $I_{2z}$, which are taken as indicated in Fig. \ref{fig:plotres} for the points $1p$ (left column of panels) or $1p'$ (right column of panels).}
    \label{fig:orbits_parametric}
\end{figure}
As regards the orbits in the zone of parametric resonance of Fig. \ref{fig:plotres}, we first note that, by the form of Eqs. (\ref{eqmoSOoffplane}) and (\ref{eqmoSOinplane}), parametric resonance conditions appear whenever the epicyclic frequency $\kappa$ is in near-resonance condition with one of the libration, precession or nutation frequencies of the secondary. In particular, we find that the thick orange strip in the upper corner of Fig. \ref{fig:plotres} indicates parameters leading to Floquet instability around the exact resonance $\kappa=2\omega_3=2\omega_{lib}$. The points marked ($1p$) and ($1p'$) satisfy the above resonance with an offset of $|\Delta\omega|=4\cdot 10^{-5} \text{[hr]}^{-1}$ and $|\Delta\omega| = 1.3\cdot 10^{-4} \text{[hr]}^{-1}$. Notwithstanding their close proximity to the resonance, we find that the numerical orbits (Fig. \ref{fig:orbits_parametric}) are regular, a fact which indicates that the nonlinear terms in the equations of motion restores stability in the close neighborhood of the (Floquet unstable) periodic orbit at the center of the resonance. However, the fact that these orbits are inside the parametric resonance zone implies that a nonlinear \textit{resonant} normal form is required to eliminate the corresponding small divisors and represent analytically the orbits. As a consequence, our non-resonant normal form constructions fails to reproduce correctly the evolution of all three secondary's Euler angles (see Fig. \ref{fig:orbits_parametric}). Similar results hold the the second thin line of parametric resonance in Fig. \ref{fig:plotres}, which corresponds to the resonance $3\kappa=2\omega_{prec}$. 

\begin{figure}
  \centering
    \includegraphics[scale=0.35]{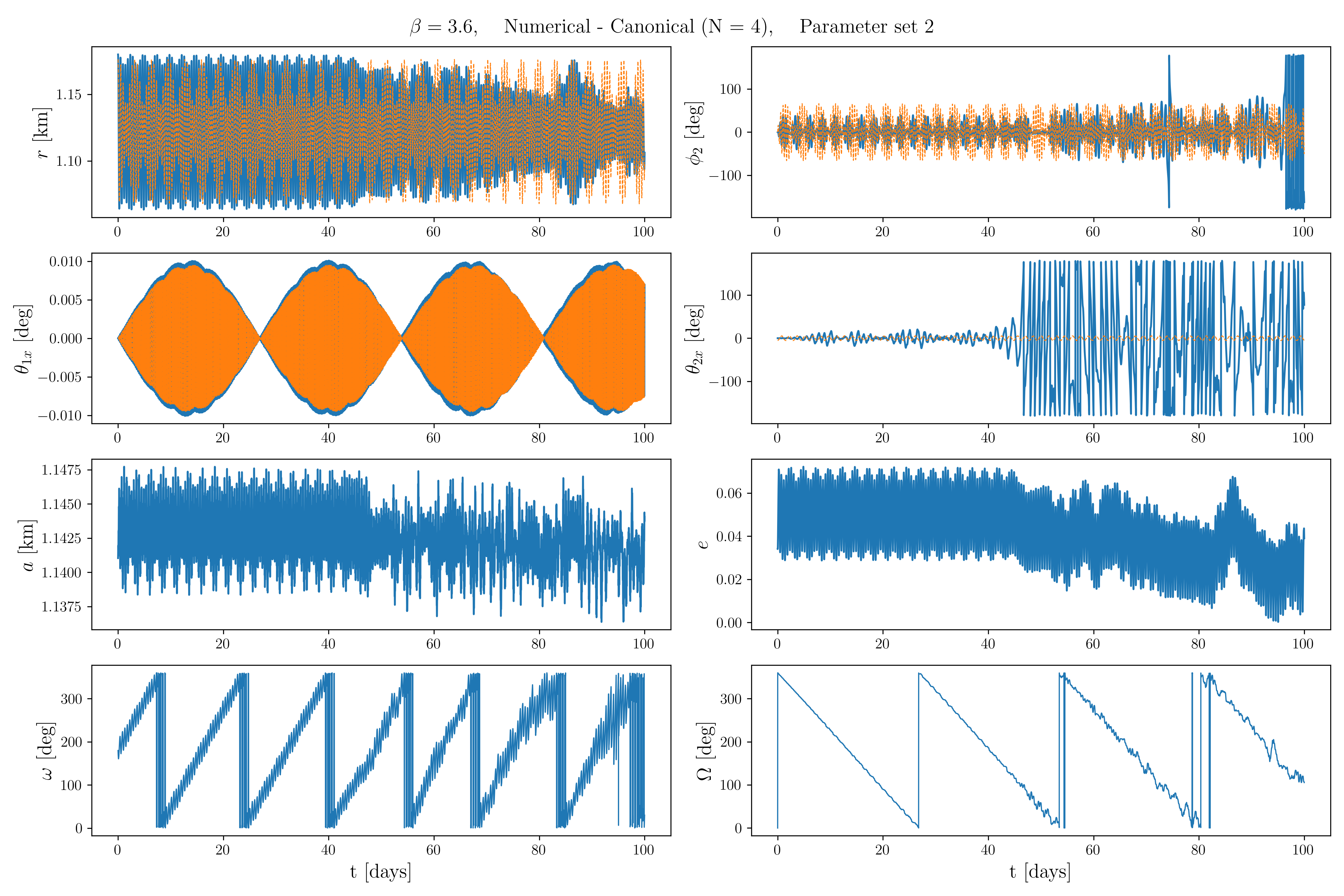}
    \caption{Comparison between numerical (blue) and nonlinear (orange) solution for the parameter set 2. The primary's attitude remains almost intact, while the rest of the variables gradually drift to chaos.}
    \label{fig:orb_num_and_tlin_b36_short_set2}
\end{figure}
Testing several simulations as those described above, in general we find that the break of near-constancy of the integrals of motion is very local, and happens only very close to nonlinear resonances as those described above. As a second example, for $\beta=3.6$, the parameter set 2 of Table \ref{tab:ParamSets} leads to chaotic behavior of the system with similar features as for the parameter set 1 (Fig. \ref{fig:orb_num_and_tlin_b36_short_set2}). In this case, a careful analysis of the frequencies shows that we are within a nonlinear \textit{doubly-resonant} domain, created by the intersection of the resonances $R=3\omega_{prec}-\omega_{lib}=0$ and $R'=2\omega_{prec}+\omega_{lib}-\omega_{nut}+2(\kappa-\kappa_z)$=0. In fact, for the orbit of Fig. \ref{fig:orb_num_and_tlin_b36_short_set2} each of the two resonances independently is satisfied by an offset of about $3\cdot 10^{-3} \text{[hr]}^{-1}$, while their linear combination is satisfied with an offset of about $1.8\cdot 10^{-4} \text{[hr]}^{-1}$. 

Most notably, as shown in Fig. \ref{fig:orb_num_and_tlin_b36_short_set2}, also in this case we find the chaotic behavior affects mostly the time series of the orbital eccentricity, and appears to lead to a circularization of the orbit, i.e., an abrupt decrease of the orbital eccentricity which takes place at about the same time ($t\approx 50~\text{[days]}$) when the onset of large scale chaotic behavior of the orbit takes place. A qualitative interpretation of this evolution of the orbital eccentricity (similar also as in Fig. \ref{fig:orb_kep_num_b36_set1}) stems from the consideration that, after the onset of large scale chaos, the stochastic evolution of all three Euler angles of the secondary implies that, as regards its contribution to the gravitational potential which determines the orbit, the secondary behaves on the average as a spherical body. That is, dynamical chaos in the rotation implies an effect similar to the physical effect caused by a secondary's re-shaping  (see \cite{Naidu2024}). It is important to recognize, however, that in our case the effect on the observable orbital parameters stems from a chaotic behavior of a conservative dynamical system, while real re-shaping is a dissipative physical process. 

\section{Conclusions}\label{sec:Conclusions}

As a summary of our results, in the previous sections we constructed two 3-dimensional, linear and nonlinear, theories able to reproduce analytically the evolution in time of the relative orbit and of the bodies' relative orientation in binary systems interacting through gravitational forces and torques being close to the `single-synchronous' equilibrium state, whose exact definition was given in Subsection \ref{subsec:StateInterest}. As a focus example of application, we reproduce analytically the post-impact state of the Didymos-Dimorphos system under various assumptions for the system's physical parameters. Our main conclusions are:\\
\\
\noindent
1) Far from i) nonlinear resonances leading to chaos, or ii) linear (parametric) resonance, we find that our theories reproduce with precision the numerically observed evolution of the perturbed system. In the case of the Didymos-Dimorphos system, our nonlinear theory, based on a Hamiltonian normal form of the fourth order with respect to a suitably-defined book-keeping parameter, leads to predictions whose errors are of the order of centimeters in the relative orbit and arcminutes in the post-impact evolution of the secondary's libration angles. On the other hand, linear theory has errors growing linearly in time as a result of the gradual de-phasing of the linear analytical from the numerical solution, due to the difference between the linear and nonlinear normal mode frequencies of the system. The cummulative error with the linear theory raises to about two orders of magnitude larger than with nonlinear theory after about one year of evolution of the system. \\
\\
\noindent
2) Linear theory, however, offers insight into the dynamics through the block structure of the variational matrix of the linearized equations of motion. In particular, through the analysis of Subsection \ref{subsec:lineqmo}, three blocks are identified: i) the block of the in-plane variables (epicyclic difference of the cylindrical radius $\delta r=r-r_c$ coupled with the secondary's libration angle $\delta\phi_2$), ii) the block of the off-plane variables ($z$ of the relative orbit and the secondary's angles $\theta_{2x},\theta_{2y}$), and iii) the primary's angles $\theta_{1x},\theta_{1y}$, which, in linear theory, are coupled only between themselves through Euler's free-torque equations. The nonlinear theory, however, introduces extra couplings through all of the above three blocks.  \\
\\
\noindent
3) General precision tests on the validity of the nonlinear theory are provided through the computation of approximate integrals of motion associated with the nonlinear normal modes of the systems. Such integrals break partially close to resonances, or completely inside resonance zones, where the evolution of the system becomes chaotic. We identify two types of resonances: `linear resonances' are parametric resonances qualitatively similar as described in \cite{Wisdom1984}, but with a forced frequency equal to the epicyclic frequency $\kappa$ of the relative orbit. Nonlinear resonances, instead, appear as harmonics in the series of perturbation theory involving both the radial and vertical epicyclic frequencies $\kappa,\kappa_z$ of the relative orbit, as well as all three librational frequencies of the secondary.  \\
\\
\noindent
4) In the case of the Didymos-Dimorphos system, we examine two different parameter sets suggested in literature by the analysis of post-impact observational data (see Section \ref{sec:dart}). Besides probing the accuracy of our analytical theories, we examined also states close to a nonlinear resonance. As regards the application of our theories to the analysis of real data based on the DART experiment, we emphasize two aspects: i) our time series of the evolution of the system are obtained through simple analytical formulas explicitly containing the physical parameters and beta, which practically eliminate any need of numerical integrations once the system's post impact state is assumed to be non-chaotic. For example, our formulas can be inserted directly in any algorithm aimed to parmeter-fit the post-impact light curves, radar data etc. of the system. ii) For chaotic orbits, instead, we emphasized the existence of transient timescales of several months (see Subsection \ref{subsec:dartnonlin}) corresponding to the system's stickiness close to a regular state before the onset of chaos. \\
\\
\noindent
5) Inspecting the time series of the evolution of orbital elements for chaotic orbits, we find that the timescale corresponding the onset of large-scale chaotic evolution of the secondary's Euler angles is imprinted in observable time series mostly as a sudden drop of the system's orbital eccentricity. We interpret this effect by the fact that the stochastic evolution of all three secondary's Euler angles after the onset of chaos implies that the secondary affects the orbit on the average as a spherical body. Hence, as regards the orbit, this dynamical effect in an otherwise conservative dynamical system mimics the dissipative effect of real re-shaping of the secondary. The comparison between these two different scenaria leading to circularization of the orbit presents interest as regards the interpretation of real post-impact data for the Didymos-Dimorphos system, and it is proposed for further study.   

\backmatter

\bmhead{Conflicts of interest}

M.G., I.G. and G.V. have no conflicts of interest to declare that are relevant to the content of this article. C.E. and K.T. are Editorial Board Members of Celestial Mechanics and Dynamical Astronomy.

\begin{appendices}

\section{Nonzero elements of the Jacobian matrix $\boldsymbol{J}$ }\label{app:JacElements}
\begin{equation}
    j_{1,8}  = \frac{1}{m}
\end{equation}
\begin{equation}
    j_{3,4}  = \dot{\theta}_{1z}(0) = \nu_{1,obs}
\end{equation}
\begin{equation}
    j_{3,10} = \frac{1}{I_{1x}}
\end{equation}
\begin{equation}
    j_{4,3}  = \frac{\dot{\theta}_{1z}(0)(I_{1z} - I_{1y})}{I_{1y}}
\end{equation}
\begin{equation}
    j_{4,11} = \frac{1}{I_{1y}}
\end{equation}
\begin{equation}
    j_{5,6}  = \dot{\theta}_{eq}
\end{equation}
\begin{equation}
    j_{5,12}  = \frac{1}{I_{2x}}
\end{equation}
\begin{equation}
    j_{6,5}  = \frac{\dot{\theta}_{eq}(I_{2z} - I_{2y})}{I_{2y}} 
\end{equation}
\begin{equation}
    j_{6,13}  = \frac{1}{I_{2y}} 
\end{equation}
\begin{equation}
    j_{7,1}  = \frac{2\dot{\theta}_{eq}}{r_{eq}} 
\end{equation}
\begin{equation}
    j_{7,14}  = \frac{1}{I_{2z}} + \frac{1}{mr_{eq}^2} 
\end{equation}
\begin{equation}
\begin{split}
    j_{8,1} = & -3m\dot{\theta}_{eq}^2 \\ +
              & \frac{G[ -3M_2( I_{1x} + I_{1y} - 2I_{1z} ) + 2M_1( -6I_{2x} + 3(I_{2y} + I_{2z}) + M_2r_{eq}^2 ) ]}{r_{eq}^5}
\end{split}
\end{equation}
\begin{equation}
    j_{9,2} = \resizebox{0.82\hsize}{1.4\baselineskip}{$ \frac{-G[ -9M_2( I_{1x} + I_{1y} - 2I_{1z} ) + 2M_1( -12I_{2x} + 3(I_{2y} + 3I_{2z}) + 2M_2r_{eq}^2 ) ]}{4r_{eq}^5}
    $}
\end{equation}
\begin{equation}
    j_{9,6} = \frac{3GM_1(I_{2x} - I_{2z})}{r_{eq}^4}
\end{equation}
\begin{equation}
    j_{10,3} = \frac{(I_{1y} - I_{1z})(3GM_2I_{1y} + 2\dot{\theta}_{1z}(0)^2I_{1z}r_{eq}^3)}{2I_{1y}r_{eq}^3}
\end{equation}
\begin{equation}
    j_{11,4} = -\dot{\theta}_{1z}^2(0)I_{1z} + \frac{3GM_2(I_{1x} - I_{1z})}{2r_{eq}^3}
\end{equation}
\begin{equation}
    j_{12,5} = \frac{\dot{\theta}_{eq}^2(I_{2y} - I_{2z})I_{2z}}{I_{2y}}
\end{equation}
\begin{equation}
    j_{13,6} = -\dot{\theta}_{eq}^2I_{2z} + \frac{3GM_1(I_{2x} - I_{2z})}{r_{eq}^3}
\end{equation}
\begin{equation}
    j_{14,7} = \frac{3GM_1(I_{2x} - I_{2y})}{r_{eq}^3}
\end{equation}

\section{Coefficients and parameters of the linear solution}
\label{app:SysLinSol}
The parameters $u_0$, $u_2$ of the characteristic polynomial $P_1(\lambda)$ (Eq. (\ref{P1Morph})) are given by
\begin{equation}
u_0 = j_{1,8}j_{14,7}( j_{7,1}^2 + j_{7,14}j_{8,1} )
\end{equation}
\begin{equation}
u_2 = -j_{1,8}j_{8,1} - j_{7,14}j_{14,7}~.
\end{equation}
The linear amplitudes $A_r$,$A_{\phi_2}$,$A_{p_r}$,$A_{p_{\phi_2}}$ of the solution 
(\ref{P1LinVar}) are given by
\begin{equation}
    A_r(\omega_\ell, \omega_m) =
    \frac{ \delta p_{\phi_2}(0)j_{1,8}j_{7,1} - \delta r(0)(\omega_m^2 + j_{1,8}j_{8,1}) }{\omega_\ell^2-\omega_m^2}
\end{equation}
\begin{equation}
    A_{\phi_2}(\omega_\ell, \omega_m) =
    \frac{\omega_\ell^2 j_{7,14} + j_{1,8}(j_{7,1}^2 + j_{7,14}j_{8,1})}{\omega_\ell j_{1,8}j_{7,1}}A_r(\omega_\ell, \omega_m)
\end{equation}
\begin{equation}
    A_{p_r}(\omega_\ell, \omega_m) =
    \frac{-\omega_\ell}{j_{1,8}}A_r(\omega_\ell, \omega_m)
\end{equation}
\begin{equation}
    A_{p_{\phi_2}}(\omega_\ell, \omega_m) =
    \frac{\omega_\ell^2 + j_{1,8},j_{8,1}}{j_{1,8}j_{7,1}}A_r(\omega_\ell,\omega_m)~~~.
\end{equation}

\noindent
The coefficients of the characteristic polynomials $P_2(\lambda)$, $P_3(\lambda)$ (Eqs. (\ref{P2Morph}), (\ref{P3Morph}))) are given by
\begin{equation}
\mathrm{v}_0 = -j_{1,8}\left( j_{6,5}^2 + j_{6,13}j_{12,5} \right)
                  \left[ j_{9,2}j_{5,6}^2 + j_{5,12} \left( j_{9,2}j_{13,6} - j_{9,6}^2 \right) \right]
\end{equation}
\begin{equation}
\begin{split}
\mathrm{v}_2 & = 2j_{1,8}j_{5,6}j_{6,5} j_{9,2} \\
      & + j_{5,6}^2\left( j_{6,5}^2 + j_{6,13}j_{12,5}\right ) \\
      & + j_{5,12}j_{13,6}\left( j_{6,5}^2 + j_{6,13}j_{12,5}\right ) \\
      & + j_{1,8}\left[ j_{5,12}j_{9,2}j_{12,5} + j_{6,13}\left( j_{9,2}j_{13,6} - j_{9,6}^2 \right) \right]
\end{split}
\end{equation}
\vspace{0.2cm}
\begin{equation}
\mathrm{v}_4 = -2j_{5,6}j_{6,5} - j_{1,8}j_{9,2} - j_{5,12}j_{12,5} - j_{6,13}j_{13,6}
\end{equation}
\vspace{0.2cm}
\begin{equation}\label{P2longexpr}
    \kappa = -27\mathrm{v}_0 + 9\mathrm{v}_2\mathrm{v}_4 - 2\mathrm{v}_4^3 + 3\sqrt{3}\sqrt{27\mathrm{v}_0^2 + 4\mathrm{v}_2^3 - 18\mathrm{v}_0\mathrm{v}_2\mathrm{v}_4 - \mathrm{v}_2^2\mathrm{v}_4^2 + 4\mathrm{v}_0\mathrm{v}_4^3}
\end{equation}

\vspace{0.2cm}

\begin{equation}
\mathrm{w}_0 = (j_{4,3}^2 + j_{4,11}j_{10,3})(j_{3,4}^2 + j_{3,10}j_{11,4})
\end{equation}

\begin{equation}
\mathrm{w}_2 = -2j_{3,4}j_{4,3} - j_{3,10}j_{10,3} - j_{4,11}j_{11,4}
\end{equation}

\vspace{0.2cm}

\noindent
The corresponding linear amplitudes of the non-planar solutions (\ref{P2LinVar}), (\ref{P3LinVar}) are:

\begin{equation}
    A_z(\omega_{\ell},\omega_m,\omega_n) = \resizebox{0.69\hsize}{1.2\baselineskip}{$
    \frac{\delta p_z(0)j_{1,8}(\omega_m^2 + j_{1,8}j_{9,2})(\omega_n^2 + j_{1,8}j_{9,2})[\omega_{\ell}^2j_{6,13} + j_{5,12}(j_{6,5}^2 + j_{6,13}j_{12,5})]}
    {\omega_{\ell}(\omega_{\ell}^2-\omega_{m}^2)(\omega_{\ell}^2-\omega_{n}^2)[-j_{1,8}j_{6,13}j_{9,2} + j_{5,12}(j_{6,5}^2 + j_{6,13}j_{12,5})]}
    $}
\end{equation}
\begin{equation}
    A_{\theta_{2x}}(\omega_{\ell},\omega_m,\omega_n) = \resizebox{0.67\hsize}{1.1\baselineskip}{$
    \frac{\delta p_z(0)(j_{5,12}j_{6,5} - j_{5,6}j_{6,13})(\omega_{\ell}^2 + j_{1,8}j_{9,2})(\omega_m^2 + j_{1,8}j_{9,2})(\omega_n^2 + j_{1,8}j_{9,2})}
    {(\omega_{\ell}^2-\omega_m^2)(\omega_m^2-\omega_n^2)j_{9,6}[-j_{1,8}j_{6,13}j_{9,2} + j_{5,12}(j_{6,5}^2 + j_{6,13}j_{12,5})]}
    $}
\end{equation}
\begin{equation}
    A_{\theta_{2y}}(\omega_{\ell},\omega_m,\omega_n) = \resizebox{0.67\hsize}{1.0\baselineskip}{$ -
    \frac{\delta p_z(0)(\omega_{\ell}^2 + j_{1,8}j_{9,2})(\omega_m^2 + j_{1,8}j_{9,2})(\omega_n^2 + j_{1,8}j_{9,2})[\omega_{\ell}^2j_{6,13} + j_{5,12}(j_{6,5}^2 + j_{6,13}j_{12,5})]}{\omega_{\ell}(\omega_{\ell}^2 - \omega_m^2)(\omega_{\ell}^2 - \omega_n^2)j_{9,6}[-j_{1,8}j_{6,13}j_{9,2}+j_{5,12}(j_{6,5}^2 + j_{6,13}j_{12,5})]}
    $}
\end{equation}
\begin{equation}
    A_{p_z}(\omega_{\ell},\omega_m,\omega_n) = \frac{\omega_\ell}{j_{1,8}}A_z(\omega_\ell,\omega_m,\omega_n)
\end{equation}
\begin{equation}
    A_{p_{\theta_{2x}}}(\omega_{\ell},\omega_m,\omega_n) =  \resizebox{0.46\hsize}{1.2\baselineskip}{$ \frac{(\omega_\ell^2 + j_{1,8}j_{9,2})[\omega_{\ell}^2j_{6,5} + j_{5,6}(j_{6,5}^2 + j_{6,13}j_{12,5})]}{j_{1,8}j_{9,6}[\omega_{\ell}^2j_{6,13} + j_{5,12}(j_{6,5}^2 + j_{6,13}j_{12,5})]}$}A_z(\omega_\ell,\omega_m,\omega_n)
\end{equation}
\begin{equation}
    A_{p_{\theta_{2y}}}(\omega_{\ell},\omega_m,\omega_n) =  \resizebox{0.46\hsize}{1.2\baselineskip}{$ \frac{-\omega_\ell(\omega_\ell^2 + j_{1,8}j_{9,2})(\omega_{\ell}^2 + j_{5,6}j_{6,5} + j_{5,12}j_{12,5})}{j_{1,8}j_{9,6}[\omega_{\ell}^2j_{6,13} + j_{5,12}(j_{6,5}^2 + j_{6,13}j_{12,5})]}$}A_z(\omega_\ell,\omega_m,\omega_n)
\end{equation}
\begin{equation}
\begin{aligned}
    A_{1_{\theta_{1x}}}(\omega_{\ell},\omega_m) & =  \frac{\delta \theta_{1x}(0)(\omega_\ell^2 + j_{3,4}j_{4,3} + j_{3,10}j_{10,3})}{\omega_\ell^2 - \omega_m^2} \\
    A_{2_{\theta_{1x}}}(\omega_{\ell},\omega_m) & =  \frac{\delta \theta_{1y}(0)\omega_\ell[\omega_m^2j_{4,3} + j_{3,4}(j_{4,3}^2 + j_{4,11}j_{10,3})]}{(\omega_\ell^2 - \omega_m^2)(j_{4,3}^2 + j_{4,11}j_{10,3})}
\end{aligned}
\end{equation}
\begin{equation}
\begin{aligned}
    A_{1_{\theta_{1y}}}(\omega_{\ell},\omega_m) & = \frac{\omega_m^2j_{4,11} + j_{3,10}(j_{4,3}^2 + j_{4,11}j_{10,3})}{\omega_m(j_{3,10}j_{4,3}-j_{3,4}j_{4,11})}A_{1_{\theta_{1x}}}(\omega_{\ell},\omega_m) \\
    A_{2_{\theta_{1y}}}(\omega_{\ell},\omega_m) & = \frac{-A_{1_{\theta_{1y}}}(\omega_{\ell},\omega_m)A_{2_{\theta_{1x}}}(\omega_m,\omega_\ell)}{A_{1_{\theta_{1x}}}(\omega_{\ell},\omega_m)}
\end{aligned}
\end{equation}
\begin{equation}
\begin{aligned}
    A_{1p_{\theta_{1x}}}(\omega_{\ell},\omega_m) & = \frac{-[\omega_m^2j_{4,3} + j_{3,4}(j_{4,3}^2 + j_{4,11}j_{10,3})]}{\omega_m(j_{3,10}j_{4,3} - j_{3,4}j_{4,11})}A_{1_{\theta_{1x}}}(\omega_{\ell},\omega_m) \\
    A_{2p_{\theta_{1x}}}(\omega_{\ell},\omega_m) & = \frac{-A_{1p_{\theta_{1x}}}(\omega_m,\omega_\ell)A_{2_{\theta_{1x}}}(\omega_{\ell},\omega_m)}{A_{1_{\theta_{1x}}}(\omega_m,\omega_\ell)}
\end{aligned}
\end{equation}
\begin{equation}
\begin{aligned}
    A_{1p_{\theta_{1y}}}(\omega_{\ell},\omega_m) & = \frac{\omega_m^2 + j_{3,4}j_{4,3} + j_{3,10}j_{10,3}}{j_{3,10}j_{4,3} - j_{3,4}j_{4,11}}A_{1_{\theta_{1x}}}(\omega_{\ell},\omega_m) \\
    A_{2p_{\theta_{1y}}}(\omega_{\ell},\omega_m) & = \frac{A_{1p_{\theta_{1y}}}(\omega_m,\omega_\ell)A_{2_{\theta_{1x}}}(\omega_{\ell},\omega_m)}{A_{1_{\theta_{1x}}}(\omega_m,\omega_\ell)}
\end{aligned}
\end{equation}

\section{Full form of $Z_0$}\label{app:Z0}
\begin{equation}
\begin{split}
    Z_0  & = \frac{\nu_1^2I_{1z}}{2} + \frac{G\big( 2M_1(2I_{2x}-I_{2y} - M_2r^{\ast 2}) + M_2(I_{1x} + I_{1y} - 2I_{1z} + 2I_{2z}) \big)}{4r^{\ast 3}} \\
    & + \frac{G(6M_1(2I_{2x}-I_{2y}-I_{2z})+3M_2(I_{1x}+I_{1y}-2I_{1z})+M_1M_2r^{\ast 2})}{2r^{\ast 5}} \delta r^2 \\
    & + \frac{G(-9M_2(I_{1x}+I_{1y}-2I_{1z})+2M_1(-12I_{2x}+3I_{2y}+9I_{2z}+2M_2r^{\ast 2}))}{8r^{\ast 5}}\delta z^2 \\
    & + \frac{(I_{1z}-I_{1y})(3GM_2I_{1y}+2\nu_1^2I_{1z}r^{\ast 3})}{4I_{1y}r^{\ast 3}}\delta\theta_{1x}^2 \\ 
    & + \frac{3GM_2(I_{1z}-I_{1x})+2\nu_1^2I_{1z}r^{\ast 3}}{4r^{\ast 3}}\delta\theta_{1y}^2 \\
    & + \frac{G(M_1+M_2)(I_{2z}-I_{2y})I_{2z}}{2I_{2y}r^{\ast 3}}\delta \theta_{2x}^2 \\
    & + \frac{G(-3I_{2x}M_1+I_{2z}(4M_1+M_2))}{2r^{\ast 3}}\delta \theta_{2y}^2 \\
    & + \frac{3GM_1(I_{2y}-I_{2x})}{2r^{\ast 3}}\delta \phi_2^2 \\
    & + \frac{\delta p_r^2}{2m} + \frac{\delta p_z^2}{2m} +
        \frac{\delta p_{\theta_{1x}}^2}{2I_{1x}} + \frac{\delta p_{\theta_{1y}}^2}{2I_{1y}} + \frac{\delta p_{\theta_{2x}}^2}{2I_{2x}} + \frac{\delta p_{\theta_{2y}}^2}{2I_{2y}} + \frac{I_{2z} + mr^{\ast 2}}{2I_{2z}mr^{\ast 2}}\delta p_{\phi_2}^2 \\
    & + \nu_\theta^{\ast}\delta p_\theta + (\nu_1-\nu_\theta^{\ast})\delta p_{\phi_1} \\
    & + \frac{2\nu_\theta^{\ast}}{r^{\ast}}\delta r \delta p_{\phi_2} +\frac{\nu_1(I_{1z} - I_{1y})}{I_{1y}}\delta \theta_{1x}\delta p_{\theta_{1y}} + \nu_1\delta \theta_{1y}\delta p_{\theta_{1x}} \\
    & + \frac{\nu_{\theta}^{\ast}(I_{2z}-I_{2y})}{I_{2y}}\delta \theta_{2x}\delta p_{\theta_{2y}} + \nu_{\theta}^{\ast} \delta \theta_{2y} \delta p_{\theta_{2x}}+ \frac{3GM_1(I_{2z}-I_{2x})}{r^{\ast 4}}\delta z \delta \theta_{2y}
\end{split}
\end{equation}

\end{appendices}

\bibliography{sn-bibliography}

\end{document}